\documentclass[1p,final]{elsarticle}
\usepackage{lineno,hyperref}
\usepackage{subfigure}
\usepackage{placeins}
\usepackage{algorithm}
\usepackage{algorithmic}
\usepackage{amsmath}
\usepackage{amsmath,amsfonts,amssymb}
\usepackage{setspace}
\usepackage{xcolor}
\usepackage{booktabs}
\journal{Journal of Computational Physics}
\mathchardef\Gamma="7100 % Maiuscole greche inclinate (modo matematico)
\mathchardef\Delta="7101
\mathchardef\Theta="7102
\mathchardef\Lambda="7103
\mathchardef\Xi="7104
\mathchardef\Pi="7105
\mathchardef\Sigma="7106
\mathchardef\Upsilon="7107
\mathchardef\Phi="7108
\mathchardef\Psi="7109
\mathchardef\Omega="710A

\newcommand{\qeq}{\quad \textrm{and} \quad}

\newcommand{\dts}{\partial_t}
\newcommand{\dxs}{\partial_x}

\newcommand{\trasp}{^{\scriptstyle\textrm{T}}}

\newcommand{\cref}{c_\ast}

\newcommand{\FirstRef}[1]{ \color{black}{#1}}
\newcommand{\SecondRef}[1]{ \color{black}{#1}}
\begin{document}
\begin{frontmatter}
\title{A pressure-based diffuse interface method for low-Mach multiphase flows with mass transfer}
\author[kth]{Andreas D. Demou\corref{mycorrespondingauthor}}
\ead{demou@mech.kth.se}
\cortext[mycorrespondingauthor]{Corresponding author}
\author[kth]{Nicol\`o Scapin}
\ead{nicolos@mech.kth.se}
\author[ensta]{Marica Pelanti}
\ead{marica.pelanti@ensta-paris.fr}
%
%\author[zur]{Barbara Re}
%\ead{barbara.re@math.uzh.ch}
%
%\author[zur]{R\'emi Abgrall}
%\ead{remi.abgrall@math.uzh.ch}
%
\author[kth,ntnu]{Luca Brandt}
\ead{luca@mech.kth.se}
\address[kth]{Department of Engineering Mechanics, Royal Institute of Technology (KTH), SE-10044 Stockholm, Sweden}
\address[ensta]{ENSTA Paris - Institut Polytechnique de Paris, 91120 Palaiseau, France}
%\address[zur]{Institute of Mathematics, University of Z\"urich, CH-8057 Z\"urich, Switzerland}
\address[ntnu]{Department of Energy and Process Engineering, Norwegian University of Science and Technology (NTNU), Trondheim, Norway}
\begin{abstract}
This study presents a novel pressure-based methodology for the efficient numerical solution of a four-equation two-phase diffuse interface model. The proposed methodology has the potential to simulate %{\lb not weakly??} weakly compressible 
low-Mach flows with mass transfer. In contrast to the classical conservative four-equation model formulation, the adopted set of equations features volume fraction, temperature, velocity and pressure as the primary variables. The model includes the effects of viscosity, surface tension, thermal conductivity and gravity, and has the ability to incorporate complex equations of state. Additionally, a Gibbs free energy relaxation procedure is used to model mass transfer. A key characteristic of the proposed methodology is the use of high performance and scalable solvers for the solution of the Helmholtz equation for the pressure, which drastically reduces the computational cost compared to analogous density-based approaches. We demonstrate the capabilities of the methodology to simulate flows with large density and viscosity ratios  through extended verification against a range of different test cases. Finally, the potential of the methodology to tackle challenging phase change flows is demonstrated with the simulation of three-dimensional nucleate boiling.
\end{abstract}
\end{frontmatter}
%
%\linenumbers
%
%===============================
\section{Introduction}
%===============================
Many flows of engineering interest exhibit compressibility effects, even if the flow velocities are relatively small and much smaller than the speed of sound in extended regions of the domain. Labelled as low-Mach flows, these types of flows share characteristics with both the fully compressible regime (pressure-density coupling dominates) and the fully incompressible regime (pressure-velocity coupling dominates), making their numerical simulation challenging~\cite{van1986segregated}.  
%in the transonic regime, the pressure is strongly coupled with both velocity and density (Journal of Computational Physics 367 (2018) 192–234)
To further complicate things, real-life applications usually involve two or more fluids that interact dynamically, with the possibility of mass transfer between different phases. 

A typical example of this type of flow is boiling; the phenomenon occurring when a heated liquid reaches or exceeds its saturation temperature at a specific pressure. At that point, the vaporisation process is initiated and bubbles are formed either on the heated surface or in an adjacent liquid layer, due to the presence of microscopic surface imperfections or other impurities~\cite{dhir1998boiling}. With continuous heat transfer, the bubbles grow and eventually detach from the heated surface, and rise inside the liquid. Even though the liquid phase can be considered approximately incompressible, the large heat and mass transfer rates induce compressibility effects in the gas phase that cannot be ignored. Additional characteristics such as the sensitivity on wetting and surface roughness render boiling as a very complex physical phenomenon, one whose physical interpretation is yet to be fully revealed. Nonetheless, boiling is utilised in steam generators~\cite{zhao2003convective}, heat exchangers~\cite{amalfi2016flow} and electronics cooling~\cite{narumanchi2008numerical} amongst other applications. Other phase transition processes such as cavitation and evaporation can also exhibit weak or strong compressibility effects. Even though these phase transition processes are seemingly different to boiling, they are all governed by the same physical mechanism, the equilibrium of the local Gibbs free energy between the two phases~\cite{saurel2016general}. 

One of the most important aspects of the numerical modeling of multiphase flows is the method used to describe the movement of the liquid-gas interface. This is usually done with the use of a marker function that acquires different values for each fluid, and helps identify the interface. These methods can be broadly categorised as~\cite{tryggvason2011direct} (i) surface tracking methods, where the marker function is reconstructed by marker points on the interface that are advected~\cite{unverdi1992front,tryggvason2001front}, and (ii) surface capturing methods, where the marker function is advected directly. Prominent examples of this second category are the level--set methods~\cite{dervieux1980finite}, the volume--of--fluid methods~\cite{hirt1981volume} and the diffuse interface (DI) methods~\cite{cahn1958free,saurel1999multiphase}. Even though DI methods have the drawback of adopting an interface thickness that is significantly larger compared to the physical thickness, these methods have important advantages when used for multiphase compressible flows. First of all, the thermodynamic consistency is retained everywhere, even at the interface where the averaging of the properties of each phase takes place. Also the diffused shape of the interface allows the numerical resolution of the property gradients, which is very beneficial on the overall accuracy and stability of the solution methodology. Moreover, the dynamic creation and disappearance of interfaces emerges naturally, a feature that is of great importance in boiling simulations~\cite{lemartelot2013liquid}. 

The most general two-phase DI model is the Baer--Nunziato model~\cite{baer1986two} (and the variant of Saurel--Abgrall~\cite{saurel1999multiphase}), consisting of seven equations: two equations for the mass conservation in each phase, two equations for the momentum conservation, two equations for the total energy conservation and one for the evolution of the volume fraction. This model is often characterised as a non-equilibrium model, meaning that in the regions where both phases coexist, there is no requirement for kinetic equilibrium (same velocity), mechanical equilibrium (same pressure),  thermal equilibrium (same temperature) or chemical equilibrium (same Gibbs free energy). From this parent model, a hierarchy of models arises via relaxation processes that drive the system to specific equilibrium states~\cite{linga2019hierarchy}, such as,
\begin{itemize}
    \item the six-equation model of Saurel et al.~\cite{saurel2009simple}, a kinetic equilibrium model where stiff pressure relaxation is applied (see also~\cite{yeom2013modified,pelanti2014mixture}),
    \item the five-equation model of Kapila et al.~\cite{kapila2001two}, with kinetic and  mechanical equilibrium (see also~\cite{allaire2002five,murrone2005five,perigaud2005compressible,shukla2010interface,jain2020conservative}),
    \item the four-equation model of Abgrall~\cite{abgrall1996prevent}, with kinetic, mechanical and thermal equilibrium 
    (see also~\cite{saurel1999simple,johnsen2012preventing,lund_proc,le2014towards,saurel2016general}). 
\end{itemize}
Amongst these models, the five-equation model provides a good compromise between physical complexity and performance, and it is the most widely used model for compressible two-phase simulations. Nonetheless, in the presence of conductive heat transfer, the additional simplification of stiff thermal relaxation is justified~\cite{le2014towards}. More specifically, when the thermal boundary layers on either side of the liquid-gas interface are properly resolved, the temperature at the interface should be continuous and a four-equation model becomes appropriate. Since the numerical methodology proposed in this study aims to simulate boiling flows where conduction heat transfer is prominent, a four-equation model will be adopted.

To numerically solve the adopted model in the low-Mach regime, two broad categories of solution strategies exist, namely (i) the \textit{density-based} approach, originating directly from methodologies for compressible flows (e.g.~\cite{murrone2008behavior,lemartelot2013liquid,pelanti-amc}), and (ii) the \textit{pressure-based} approach, originating from methodologies for incompressible flows (e.g.~\cite{jemison2014compressible,denner2018pressure}). Even though density-based approaches have been shown to perform well in a range of different multiphase flows, they rely on preconditioning techniques to overcome the stiffness problem in the low-Mach limit~\cite{weiss1995preconditioning}. Preconditioning techniques continue to develop and become more sophisticated, but there is still much room for improvement, especially for the simulation of unsteady flows~\cite{turkel2005local} or three-dimensional cases where the computational cost of preconditioning becomes hardly feasible for practical applications. As commented in the \textit{Future Issues} section of the recent review paper of Saurel and Pantano~\cite{saurel2018diffuse}, preconditioning techniques for two-phase low-Mach simulations need to become more efficient. On the other hand, the pressure-based approach can achieve good performance in terms of computational cost. Moreover, it has the advantage of preventing pressure oscillations at interfaces, since the pressure is solved for and not retrieved from the energy.{\FirstRef This approach was mainly employed in single-phase cases, with the first true all-Mach number flow solver presented in~\cite{park2005multiple}. Numerical schemes specific to single-phase low-Mach number flows were also presented in~\cite{klein1995semi,klein2001asymptotic,munz2003extension}, inspiring further development of novel pressure-based low- and all-Mach number methods~\cite{dumbser2016conservative,bermudez2020staggered,busto2021semi}}. Only a small number of studies were devoted to weakly compressible multiphase flows, such as~\cite{re2018non,re2021pressure} based on the Baer--Nunziato model,~\cite{kuhn2021all} based on the six-equation model, and other sharp interface methods~\cite{denner2018pressure,fuster2018all,jemison2014compressible,dalla2021interface}. These few studies showed promising results, both in terms of numerical efficiency and overall performance for several test cases such as bubble oscillations, explosion shocks, oscillating water column, etc. In addition, few studies demonstrated the possibility to add phase transition to pressure-based methods, using either a sharp interface approach~\cite{juric1998computations,sato2013sharp,tanguy2014benchmarks,scapin2020volume} or the Cahn--Hilliard phase-field method~\cite{jafari2016numerical,wang2021phase}.

This study presents the development of a novel pressure-based methodology for the solution of a four-equation DI model. The methodology is capable of simulating low-Mach flows, where weak compressibility effects cannot be ignored even though the flow velocities are much smaller than the speed of sound. Mass transfer is also taken into account via a Gibbs free energy relaxation procedure that is activated whenever the proper thermodynamic conditions are locally met. The physical description is enriched with additional terms, which account for viscous stresses, surface tension, heat conduction, and gravity. Moreover, the method is able to incorporate complex equations of state, specific to each phase, and can handle large density and viscosity ratios. The rest of this paper is organised as follows: Section~\ref{sec:math} presents the mathematical model, followed by the description of the proposed solution methodology in Section~\ref{sec:algorithm}. The verification of the methodology against benchmark single-phase and multiphase cases, with and without mass transfer, is presented in Section~\ref{sec:verification}, where the conservation of mass and total energy are also evidenced. Furthermore, the potential of this method to simulate the computationally demanding nucleate boiling flow is demonstrated in Section~\ref{sec:demonstration}. Finally, Section~\ref{sec:future} lists some possible extensions and improvements to be addressed and Section~\ref{sec:conclusions} concludes the study with a summary of the key findings.

%===============================
\section{Mathematical model}\label{sec:math}
%===============================
%===============================
\subsection{Governing equations}\label{sec:equations}
%===============================

As indicated in the Introduction, the diffuse interface model presented here uses a  four-equation model describing a two-phase flow in kinetic, mechanical and thermal equilibrium \cite{lund_proc, le2014towards,saurel2016general}. This two-phase model results from the velocity, pressure and temperature relaxation of the full Baer--Nunziato model~\cite{baer1986two}. For completeness, the main steps of the derivation of the relaxed model are illustrated in~\ref{BNrelax}. 
%[The diffused interface model presented here results from the pressure and temperature relaxation of %the full Baer--Nunziato model~\cite{baer1986two}. For completeness, the derivation of the relaxed %model  is described in~\ref{BNrelax}. Additional terms are added to the relaxed set of equations to %account for viscous stresses, surface tension, heat conduction, and gravity effects. {\color{red}(It %would be nice to prepare a few SymPy scripts as supplementary material so that the equations used %don't cause any issues.)}] 
As a convention, phasic quantities are identified with subscript $k$ or explicitly with \{1,2\}, while mixture quantities bare no such identification.
%The mixture density per unit volume ($X$) are related to the corresponding phasic quantities ($X_k$) %with $X=a_1X_1+a_2X_2$, where $a_k$ is the volume fraction of phase $k$. The governing equations read:
We will denote with $a_k$ the volume fraction of phase $k$, with $\rho$ the mixture density,
$\rho=a_1\rho_1 +a_2\rho_2$, with $\vec{u}$ the velocity field, and with $\mathcal{E}$ the
internal energy per unit volume.
The pressure and temperature equilibrium four-equation two-phase model in the literature is commonly written in terms of the conserved variables  $a_1\rho_1$, $a_2\rho_2$, $\rho \vec{u}$
and $E = \mathcal{E} +\rho \frac{|\vec{u}|^2}{2}$ (mixture total energy per unit volume).
The formulation in terms of these variables, including here viscous stresses, surface tension, heat conduction, and gravity effects reads:
\begin{subequations}
\label{eq:sysTboilc}
\begin{eqnarray}
  \partial_t \left(a_1 \rho_1  \right) + \vec{\nabla} \cdot
  \left(a_1 \rho_1 \vec{u}\right)&=& M,\\
 \partial_t \left(a_2\rho_2\right) +\vec{\nabla} \cdot  \left(a_2\rho_2 \vec{u}\right) &=& -M,\\
 \partial_t \left(\rho \vec{u}\right) +  \vec{\nabla}\cdot \left(\rho \vec{u}\otimes \vec{u}\right) +  \vec{\nabla}p  &=& \vec{D}_u  + \vec{\Sigma} + \vec{G}, \label{eq:u2} \\
 \dts E +\vec{\nabla} \cdot ((E+p)\vec{u}) &=& 
 \vec{D}_u \cdot \vec{u} + D_{\mathcal{E}} +K
 +\vec{\Sigma}  \cdot \vec{u}+ \vec{G} \cdot{\vec{u}}.
\end{eqnarray}
\end{subequations}
Alternatively, one equation for one partial density (e.g.\ $a_2\rho_2$) can be  replaced by the
equation for the mixture density $\rho$:
\begin{equation}
\label{eq:partr}
\partial_t \rho + \vec{\nabla} \cdot (\rho \vec{u}) = 0.  
\end{equation}
The source terms on the right hand side of Eqs.~(\ref{eq:sysTboilc}), are defined as follows:
\begin{itemize}
    \item Mass transfer $M$
        \begin{equation}
            M=\nu(g_2-g_1),\label{eq:M_define}
        \end{equation}
        where $\nu$ is the chemical relaxation parameter, and $g$ is the Gibbs free energy.
    \item Viscous stress $\vec{D}_u$
        \begin{equation}
            \vec{D}_u=\vec{\nabla}\cdot \vec{\vec{\tau}},
        \end{equation}
        \begin{equation}
             \mathrm{\ with \ } \vec{\vec{\tau}}=\mu\left(\vec{\nabla}\vec{u}+\left(\vec{\nabla}\vec{u}\right)^T-\frac{2}{3}\left(\vec{\nabla}\cdot\vec{u}\right)\vec{\vec{I}}\right),
        \end{equation}
        where $\mu$ is the dynamic viscosity.
    \item Surface tension $\vec{\Sigma}$
        \begin{equation}
            \vec{\Sigma} = \sigma \  \vec{\nabla}\cdot\left(\frac{\vec{\nabla}a_1}{\left|\vec{\nabla}a_1\right|}\right)\vec{\nabla}a_1,
        \end{equation}
        where $\sigma$ is the surface tension coefficient. This modeling follows the continuum surface force model (CSF) proposed by Brackbill et al.~\cite{brackbill1992continuum}.%, and $Y_1=a_1\rho_1/\rho$ is the mass fraction of phase 1. 
    \item Gravity $\vec{G}$
        \begin{equation}
            \vec{G} = \rho \vec{g},
        \end{equation}
        where $\vec{g}$ is the acceleration of gravity.
    \item Viscous dissipation $D_{\mathcal{E}}$
        \begin{equation}
            D_{\mathcal{E}}= \vec{\nabla}\vec{u} : \vec{\vec{\tau}}\label{eq:DE_define}
        \end{equation}
    \item Heat conduction $K$
        \begin{equation}
            K= \vec{\nabla} \cdot \left(\lambda \vec{\nabla}T\right),\label{eq:K_define}
        \end{equation}
        where $\lambda$ is the heat conduction coefficient and $T$ is the temperature.
\end{itemize}

In the present work we adopt a different choice of primary variables for reasons related to the 
numerical discretization method which will be discussed in the next section.
Our four-equation model formulation uses the governing equations for
the volume fraction $a_1$ of one phase, the temperature $T$,{\SecondRef the velocity $\vec{u}$},
and the pressure $p$:
\begin{subequations}
\label{eq:sysprim}
\begin{eqnarray}
\partial_t a_1 + \vec{\nabla} \cdot \left( a_1 \vec{u} \right) + \left(S_{a}^{(3)}-a_1\right) \vec{\nabla}\cdot \vec{u} &=& S_{a}^{(1)}M  + S_{a}^{(2)}(D_{\mathcal{E}}+K), \label{eq:a1} \\
%
%\partial_t \left(\rho_1 a_1\right) + \vec{\nabla}\cdot \left(\rho_1 a_1 \vec{u} \right) &=& M, \label{eq:a1rho1} \\ 
\partial_t T + \vec{\nabla} \cdot \left( T \vec{u} \right) + \left(S_{T}^{(3)}-T\right) \vec{\nabla}\cdot \vec{u} &=& S_{T}^{(1)}M +S_{T}^{(2)}(D_{\mathcal{E}}+K), \label{eq:T} \\
%
%\partial_t \left(\rho_2 a_2\right) + \vec{\nabla}\cdot \left(\rho_2 a_2 \vec{u} \right) &=& -M, \label{eq:rho2a2} \\ 
%
\partial_t  \vec{u} +  \vec{\nabla}\cdot \left( \vec{u}\otimes \vec{u}\right) - \vec{u}\left(\vec{\nabla}\cdot \vec{u}\right) &=& \frac{1}{\rho}\left(\vec{D}_u  + \vec{\Sigma} + \vec{G}\right), \label{eq:u} \\ 
\partial_t p+ \vec{u} \cdot \vec{\nabla} p +\rho c^2 \vec{\nabla}\cdot \vec{u} &=& S_{p}^{(1)}M+S_{p}^{(2)}(D_{\mathcal{E}}+K).\label{eq:p}
%&&	\frac{\partial P}{\partial t}+ u_j \frac{\partial P}{\partial x_j} +\rho c^2 \frac{\partial u_j}{\partial x_j} = S_p \nu(g_2-g_1)+S\left(\frac{\partial u_i}{\partial x_j} \tau_{ij}+ \frac{\partial}{\partial x_j} \left(\lambda \frac{\partial T}{\partial x_j} \right)\right),\label{eq:p}
\end{eqnarray}
\end{subequations}
%along with an equation of state (EOS) which will be presented in Section~\ref{sec:eos}. 
%In the above equations $T$ the temperature, $\rho$ the density, $\kappa_{p}$ the specific heat capacity %at constant pressure, $u$ the velocity field, $p$ the pressure and $c$ the speed of sound, given by,

{\SecondRef The momentum Eq.~\eqref{eq:u} is recast in a different form from Eq.~\eqref{eq:u2} using the conservation of mass, Eq.~\eqref{eq:partr}. This is done to avoid the need of the updated density since, as shown in the algorithmic part of this study (see Algorithm~\ref{tab:algorithm}), the density is updated after the numerical solution of Eq.~\eqref{eq:u}.} In the above equations $c$ is the speed of sound, given by \cite{flatten-lund:rel},
\begin{equation}
\dfrac{1}{c^2}= \rho\left(\dfrac{a_1}{\rho_1 c_1^2}+\dfrac{a_2}{\rho_2 c_2^2}\right)+\dfrac{\rho T C_{p1}C_{p2}}{C_{p1}+C_{p2}}\left( \dfrac{\Gamma_2}{\rho_2 c_2^2}-\dfrac{\Gamma_1}{\rho_1 c_1^2} \right)^2. \label{eq:speedofsound}
\end{equation}
The quantities $S_{a}^{(1)}$, $S_{a}^{(2)}$, $S_{T}^{(1)}$, $S_{T}^{(2)}$, $S_{p}^{(1)}$ and $S_{p}^{(2)}$ are defined as:
\begin{subequations}
\begin{eqnarray}
S_{a}^{(1)} &=&\frac{1}{\bar{D}}\left[
\left(\frac{\chi_1}{\Gamma_1}-\frac{\chi_2}{\Gamma_2}\right)(\phi\zeta)^T+\left(\frac{a_1}{\Gamma_1}+\frac{a_2}{\Gamma_2}\right)\phi_v
\right], \\ \label{eq:Sa1}
S_{T}^{(1)} &=&\frac{1}{\bar{D}}\left[
\left(\frac{\chi_2}{\Gamma_2}-\frac{\chi_1}{\Gamma_1}\right)(\zeta\rho)^T+\left(\frac{\rho_1c_1^2}{\Gamma_1}-\frac{\rho_2c_2^2}{\Gamma_2}\right)\zeta_v+\left(\frac{a_1}{\Gamma_1}+\frac{a_2}{\Gamma_2}\right)\Delta\rho
\right], \\ \label{eq:ST1}
S_{p}^{(1)} &=&\frac{1}{\bar{D}}\left[\left(\frac{\chi_1}{\Gamma_1}-\frac{\chi_2}{\Gamma_2}\right)(\phi\rho)^T +
\left(\frac{\rho_2 c^{2}_2}{\Gamma_2}-\frac{\rho_1 c^{2}_1}{\Gamma_1}\right)\phi_v\right], \label{eq:Sp1} \\
S_{a}^{(2)}&=&\frac{1}{\bar{D}}(\phi \zeta)^T, \label{eq:Sa2}\\
S_{T}^{(2)}&=&-\frac{1}{\bar{D}}(\zeta\rho)^T, \label{eq:ST2}\\
S_{p}^{(2)}&=&\frac{1}{\bar{D}}(\phi\rho)^T, \label{eq:Sp2}\\
S_{a}^{(3)} &=& 
\rho c^2 \left[\left(\frac{a_1 a_2}{\rho_2 c_2^2}-\frac{a_1 a_2}{\rho_1 c_1^2}\right)+\frac{T C_{p1}C_{p2}}{C_{p1}+C_{p2}}\left(\frac{\Gamma_2}{\rho_2 c_2^2}-\frac{\Gamma_1}{\rho_1 c_1^2}\right)\left(\frac{a_1\Gamma_2}{\rho_2 c_2^2}+\frac{a_2\Gamma_1}{\rho_1 c_1^2}\right)\right], \label{eq:Sa3} \\
S_{T}^{(3)} &=& \frac{\rho c^2 T}{C_{p1}+C_{p2}}\left(\frac{C_{p1}\Gamma_1}{\rho_1 c_1^2}+\frac{C_{p2}\Gamma_2}{\rho_2 c_2^2}\right), \label{eq:ST3}
\end{eqnarray}
\end{subequations}
where, 
\begin{subequations}
\begin{eqnarray}
(\phi\rho)^T&=& a_1\phi_1\rho_2+a_2\phi_2\rho_1, \hspace{0.575 cm} \phi_v= a_1\phi_1+a_2\phi_2, \\
(\zeta\rho)^T &=& a_1\zeta_1\rho_2+a_2\zeta_2\rho_1, \hspace{0.75 cm} \zeta_v = a_1\zeta_1+a_2\zeta_2,\\
(\phi \zeta)^T&=&a_1 a_2(\phi_1\zeta_2-\phi_2\zeta_1), \hspace{0.28 cm} \Delta\rho = \rho_2-\rho_1, \\
\bar{D}&=& \left(\frac{\rho_1 c_1^2}{\Gamma_1}-\frac{\rho_2 c_2^2 }{\Gamma_2}\right)(\phi\zeta)^T+\left(\frac{a_1}{\Gamma_1}+\frac{a_2}{\Gamma_2}\right)(\phi\rho)^T,\\
\phi_k&=& \left(\frac{\partial \rho_k}{\partial T}\right)_{p}=-\rho_k \dfrac{\Gamma_k \kappa_{pk}}{c_k^2},\\ \label{eq:phi}
\zeta_k&=& \left(\frac{\partial \rho_k}{\partial p}\right)_{T}= \dfrac{1}{c_k^2}+\dfrac{\Gamma_k^2 \kappa_{pk} T}{ c_{k}^4},\\ \label{eq:zeta}
\chi_k&=& \left(\frac{\partial p}{\partial \rho_k}\right)_{\mathcal{E}_k}= c^2_k-\Gamma_k h_k.
\end{eqnarray}
\end{subequations}
In the above expressions,  $\Gamma = \left(\frac{\partial p}{\partial \mathcal{E}} \right)_{\rho}$ is the Gr\"uneisen coefficient,  $\kappa_p$ is the specific heat capacity at constant pressure and $h$ is the specific enthalpy. The extensive heat capacity at a constant pressure for each phase is given by $C_{pk}=a_k\rho_k\kappa_{pk}$.
The derivation of the  form of the source terms appearing in the model formulation
(\ref{eq:sysprim}) {\SecondRef from the conservative formulation (\ref{eq:sysTboilc})} is presented in~\ref{extraPhysics}. {\SecondRef Let us note that in the equations (\ref{eq:sysprim}) the source terms $S_a^{(1)} M$, $S_T^{(1)}M$ and $S_p^{(1)}M$ 
model the effect of mass transfer on the evolution of the volume fraction $a_1$, the temperature $T$ and the pressure $p$; 
$S_a^{(2)} (D_{\mathcal{E}} +K)$, $S_T^{(2)}(D_{\mathcal{E}} +K)$ and $S_p^{(2)}(D_{\mathcal{E}} +K)$ model viscous dissipation and heat conduction associated to the evolution of the same variables,  $a_1$, $T$ and $p$. Note also that the contribution
$S_a^{(3)}\vec{\nabla} \cdot \vec{u}$ in the volume fraction equation for liquid-gas mixtures  accounts for mechanical cavitation, leading to an increase of the gaseous volume fraction in expansion regions  and its decrease in compression regions.    The term $S_T^{(3)}\vec{\nabla} \cdot \vec{u}$ 
in the temperature equation leads to an increase of the temperature in compression regions and a decrease in expansion regions.}

Let us remark that even though the model is fully compressible, discretizations based on the  adopted non-conservative formulation (\ref{eq:sysprim}) are not appropriate for the numerical solution of flows with strong compressibility effects.{\SecondRef Since the total energy and mass are not solved for, and momentum is solved in a non-conservative form, the conservation of these quantities is not rigorously guaranteed at the discrete level. As a consequence, highly compressible flows such as those involving shock waves cannot be captured accurately.} Nonetheless, as shown in the following sections, the model is appropriate for the simulation of weakly compressible flows with phase change, which are the applications targeted by the present study.

\subsection{Equation of state}\label{sec:eos}
%===============================
To close the system of equations above we need to specify an equation of state (EOS) for each phase, 
for instance by providing a pressure law $p_k(\mathcal{E}_k,\rho_k)$ and a temperature law 
$T_k(p_k,\rho_k)$. 
Given the equation of state of each phase, the equation of state for the mixture is determined by 
the pressure and temperature equilibrium conditions $p_1=p_2=p$, $T_1=T_2=T$, by the mixture density relation $\rho=\alpha_1\rho_1+\alpha_2\rho_2$ and
by the mixture energy relation $\mathcal{E} = \alpha_1\mathcal{E}_1+\alpha_2\mathcal{E}_2$.

The thermodynamic closure used in this study is the Noble--Abel stiffened--gas (NASG) EOS~\cite{le2016noble}. The use of this specific EOS is not in any way mandatory, since the proposed method can be coupled with any complex EOS. The NASG EOS is expressed by,
\begin{subequations}
\begin{eqnarray}
T_k(p_k,\rho_k)&=&\frac{(1-\rho_k b_k)(p_k+p_{\infty k})}{\kappa_{vk}\rho_k(\gamma_k-1)}, \label{eq:T_EOS}\\
p_k(\mathcal{E}_k,\rho_k) &=& \frac{\gamma_k-1}{1-\rho_k b_k}\left(\mathcal{E}_k-\eta_k\rho_k\right)-\gamma_k p_{\infty k}. \label{eq:P_EOS}
\end{eqnarray} \label{eq:eos1}
\end{subequations}
{\SecondRef Since the pressure and temperature fields are solved for using Eqs.~\eqref{eq:p} and~\eqref{eq:T}, Eq.~\eqref{eq:T_EOS} is used to calculate the phasic densities $\rho_k$, by setting $p_k=p$ and $T_k=T$ (pressure and temperature relaxation).} Following the above relations, the phasic specific entropy $s_k$, specific enthalpy $h_k$, Gibbs free energy $g_k$, speed of sound $c_k$ and Gr\"uneisen coefficient $\Gamma_k$ become:
\begin{subequations}
\begin{eqnarray}
s_k(p,T)&=& \kappa_{vk} \log\left(\frac{T^{\gamma_k}}{(p+p_{\infty k})^{\gamma_k -1}}\right)+\Tilde{\eta}_k, \label{eq:s_EOS}\\
h_k(p,T)&=& \kappa_{vp} T+b_k p+\eta_k, \label{eq:h_EOS}\\
g_k(p,T)&=& (\gamma_k \kappa_{vk} -\Tilde{\eta}_k)T-\kappa_{vk}T \log\left(\frac{T^{\gamma_k}}{(p+p_{\infty k})^{\gamma_k -1}}\right)+\eta_k+b_k p, \label{eq:g_EOS}\\
c_k(p,\rho_k)&=& \sqrt{\gamma_k\frac{p+p_{\infty k}}{\rho_k(1-\rho_k b_k)}}, \label{eq:c_EOS}\\
\Gamma_k(\rho_k)&=&\frac{\gamma_k-1}{1-\rho_k b_k}, \label{eq:gamma_EOS}
\end{eqnarray} \label{eq:eos2}
\end{subequations}
where $\kappa_{vk}$ (phasic specific heat capacity at constant pressure), $\gamma_k$, $\eta_k$, $\Tilde{\eta}_k$, $p_{\infty}$ and $b_k$ are the phasic parameters of the EOS. It is noted that $\kappa_{pk}=\gamma_k\kappa_{vk}$. The set of adopted values will be presented separately for each test case.

Given the equations of state for the liquid and vapor phases of a species, the theoretical
pressure-temperature saturation curve is determined by the Gibbs free energy equilibrium 
condition $g_1=g_2$. For liquid and vapor phases governed by the NASG EOS \cite{le2016noble}
this gives  the following equation defining the $p$-$T$ saturation curve: 
\begin{equation}
    A_s+\frac{B_s}{T}+C_s\log T+D_s\log(p+p_{\infty 1})-\log(p+p_{\infty 2})+\frac{pE_s}{T}=0, \label{eq:sat_tmp}
\end{equation}
where,
\sloppy
\[
A_s=\frac{\kappa_{p1}-\kappa_{p2}-\Tilde{\eta}_1+\Tilde{\eta}_2}{\kappa_{p2}-\kappa_{v2}}
\mathrm{, \ }
B_s=\frac{\eta_1-\eta_2}{\kappa_{p2}-\kappa_{v2}}
\mathrm{, \ }
C_s=\frac{\kappa_{p2}-\kappa_{p1}}{\kappa_{p2}-\kappa_{v2}}
\mathrm{, \ } 
\]
\begin{equation}
D_s=\frac{\kappa_{p1}-\kappa_{v1}}{\kappa_{p2}-\kappa_{v2}}
\mathrm{, \ and \ }
E_s=\frac{b_1-b_2}{\kappa_{p2}-\kappa_{v2}}.
\end{equation}
The parameters of the NASG EOS of the two phases are defined so that the theoretical saturation curves
fit the experimental ones of the chosen material in a certain temperature range \cite{le2016noble}.

%===============================
\section{Numerical methodology}\label{sec:algorithm}
%===============================
This section details the solution of Eqs.~\eqref{eq:a1}--\eqref{eq:p} on a staggered (marker and cell) Cartesian grid, where all scalar fields are defined on cell centers while the velocity is defined on cell faces.{\SecondRef To handle numerically the mass transfer source term $M$ we use an operator splitting technique,
which is commonly employed to treat relaxation terms in Baer--Nunziato type models 
(e.g.\ \cite{saurel1999multiphase,sa-lem:multi}): the equations are first solved without the mass transfer term $M$,  and then a relaxation procedure is applied to integrate this source term accounting for phase transition.} 

\subsection{{\SecondRef Pressure-based solution method for the system without mass transfer term}}

The equations are integrated in time using an explicit, $3^{rd}$ order Runge--Kutta (RK3) method~\cite{wesseling2009principles}. Within the context of the RK3 method, each time-step $n$ is split into three sub-steps $m={1,2,3}$. With this notation, the following convention is followed to represent any quantity $q$ at the beginning and the end of each time-step,
\begin{equation*}
    q^{n,m=1} = q^{n} \mathrm{ \ \ and \ \ } 
    q^{n,m=4} = q^{n+1}.
\end{equation*}

%===============================
\subsubsection{Volume fraction}\label{solve_a1}
%===============================

First, Eq.~\eqref{eq:a1} (without the mass transfer term) is solved to obtain the volume fraction at the new sub-step $a_1^{n,m+1}$. Following the RK3 method, the updated $a_1^{n,m+1}$ is calculated as:
\begin{equation}
\begin{aligned}
     a_1^{n,m+1} = a_1^{n,m} -\Delta t \ \Bigg[ &\tilde{\alpha}^{m}\left(\vec{\nabla} \cdot \left( a_1 \vec{u} \right) + \left(S_a^{(3)}-a_1\right) \vec{\nabla}\cdot \vec{u}\right)^{n,m} + \\ &\tilde{\beta}^{m}\left(\vec{\nabla} \cdot \left( a_1 \vec{u} \right) + \left(S_a^{(3)}-a_1\right) \vec{\nabla}\cdot \vec{u}\right)^{n,m-1}-\\
     &\tilde{\gamma}^{m}\left( S_a^{(2)} \left( D_{\mathcal{E}} + K \right)\right)^{n,m} \Bigg], \label{eq:solve_a1}
\end{aligned}
\end{equation}
where $\Delta t$ is the time step, $\tilde{\alpha}^{m} = \{8/15,5/12,3/4\}$, $\tilde{\beta}^{m} = \{0,-17/60,-5/12\}$ and $\tilde{\gamma}^{m}=\tilde{\alpha}^{m}+\tilde{\beta}^{m}$, in accordance to~\cite{wesseling2009principles}. The convection terms $\vec{\nabla} \cdot \left( a_1 \vec{u} \right)$ are discretised using the van Leer flux limiter~\cite{van1977towards,prosperetti2009computational}, while the velocity divergence $\vec{\nabla}\cdot \vec{u}$, heat conduction and viscous dissipation terms are discretised with central differences.

Consequently, the viscosity and thermal conductivity fields can be updated as,
\begin{equation}
\mu^{n,m+1} = a_1^{n,m+1}\mu_1+a_2^{n,m+1}\mu_2 \mathrm{ \ \ and \ \ } \lambda^{n,m+1} = a_1^{n,m+1}\lambda_{c1}+a_2^{n,m+1}\lambda_{c2},
\end{equation}
where $a_2=1-a_1$ and the phasic properties $\mu_1$, $\mu_2$, $\lambda_{c1}$, $\lambda_{c2}$ are constant.

%===============================
\subsubsection{Temperature}\label{solve_T}
%===============================
Similar to the volume fraction, the solution of Eq.~\eqref{eq:T} (without the mass transfer term) is advanced in time as,
\begin{equation}
\begin{aligned}
     T^{n,m+1} = T^{n,m} -\Delta t \ \Bigg[ &\tilde{\alpha}^{m}\left(\vec{\nabla} \cdot \left( T \vec{u} \right) + \left(S_T^{(3)}-T\right) \vec{\nabla}\cdot \vec{u}
     \right)^{n,m} + \\ &\tilde{\beta}^{m}\left(\vec{\nabla} \cdot \left( T \vec{u} \right) + \left(S_T^{(3)}-T\right) \vec{\nabla}\cdot \vec{u}
     \right)^{n,m-1} - \\
     &\tilde{\gamma}^{m}\left( S_T^{(2)} \left( D_{\mathcal{E}} + K \right)\right)^{n,m} \Bigg]. \label{eq:solve_T}
\end{aligned}
\end{equation}
For consistency, the same spatial discretisation schemes used for Eq.~\eqref{eq:solve_a1} are also applied to Eq.~\eqref{eq:solve_T}.

%===============================
\subsubsection{Predicted velocity}\label{solve_u*}
%===============================
To decouple velocity and pressure, a fractional-step approach is adopted (in the spirit of~\cite{amsden1970simplified}), where a predicted velocity field $\vec{u}^{n,m*}$ is first calculated without considering the pressure gradient term. In this form, the predicted velocity can be obtained as,
\begin{equation}
\begin{aligned}
     u^{n,m*} = u^{n,m} -\Delta t \ \Bigg[ &\tilde{\alpha}^{m}\left(\vec{\nabla}\cdot \left( \vec{u}\otimes \vec{u}\right) - \vec{u}\left(\vec{\nabla}\cdot \vec{u}\right) -\frac{\vec{D}_u}{\rho}
     \right)^{n,m} + \\ 
     &\tilde{\beta}^{m}\left(\vec{\nabla}\cdot \left( \vec{u}\otimes \vec{u}\right) - \vec{u}\left(\vec{\nabla}\cdot \vec{u}\right) -\frac{\vec{D}_u}{\rho}
     \right)^{n,m-1} + \\
     &\tilde{\gamma}^{m}\left(\frac{\vec{\Sigma} + \vec{G}}{\rho}\right)^{n,m}\Bigg].
     \label{eq:solve_u}
\end{aligned}
\end{equation}
Following the discretisation used for the volume fraction and temperature equations, the convection term $\vec{\nabla}\cdot \left( \vec{u}\otimes \vec{u}\right)$ is discretised using the van Leer flux limiter, while all other terms are discretised with central differences.

Once the updated pressure field $p^{n,m+1}$ becomes available, the corrected velocity field $\vec{u}^{n,m}$ can be obtained as,
\begin{equation}
 \vec{u}^{n,m+1}=\vec{u}^{n,m*}-\tilde{\gamma}^{m}\Delta t \frac{\vec{\nabla}p^{n,m+1}}{\rho^{n,m}}. \label{eq:correction}
\end{equation}

%===============================
\subsubsection{Pressure solution}\label{solve_p}
%===============================
Eq.~\eqref{eq:p} (without the mass transfer term) is discretised in time as,
\begin{equation}
\begin{aligned}
   p^{n,m+1} = p^{n,m} -&\Delta t \ \Big[ \tilde{\alpha}^{m}\left(\vec{\nabla} \cdot \left( p \vec{u} \right) -p\vec{\nabla}\cdot \vec{u} \right)^{n,m} +
   \tilde{\beta}^{m}\left(\vec{\nabla} \cdot \left( p \vec{u} \right) -p\vec{\nabla}\cdot \vec{u} \right)^{n,m-1}\Big] + \\
   \tilde{\gamma}^{m}&\Delta t \left[S_p^{(2),n,m}\left(D_{\mathcal{E}}+K \right)^{n,m}-(\rho c^2)^{n,m}\vec{\nabla}\cdot \vec{u}^{n,m+1}\right]
   \label{eq:solve_p}
\end{aligned}
\end{equation}
Term $\vec{\nabla}\cdot \vec{u}^{n,m+1}$ is replaced by the divergence of Eq.~\eqref{eq:correction},
\begin{equation}
    \vec{\nabla}\cdot \vec{u}^{n,m+1} = \vec{\nabla}\cdot \vec{u}^{n,m*} -\tilde{\gamma}^{m}\Delta t \vec{\nabla}\cdot\left(\frac{\vec{\nabla}p^{n,m+1}}{\rho^{n,m}}\right), \label{eq:velo_div}
\end{equation}
yielding the following Helmholtz equation for the pressure,
\begin{equation}
\begin{aligned}
   p^{n,m+1} -&(\tilde{\gamma}^{2}\Delta t^2\rho c^2)^{n,m}\vec{\nabla}\cdot \left(\dfrac{\vec{\nabla}p^{n,m+1}}{\rho^{n,m}} \right) = \\
    p^{n,m}-&\Delta t \ \Big[ \tilde{\alpha}^{m}\left(\vec{\nabla} \cdot \left( p \vec{u} \right) -p\vec{\nabla}\cdot \vec{u} \right)^{n,m} +
   \tilde{\beta}^{m}\left(\vec{\nabla} \cdot \left( p \vec{u} \right) -p\vec{\nabla}\cdot \vec{u} \right)^{n,m-1}\Big] + \\
   \tilde{\gamma}^{m}&\Delta t \left[S_p^{(2),n,m}\left(D_{\mathcal{E}}+K \right)^{n,m}-(\rho c^2)^{n,m}\vec{\nabla}\cdot \vec{u}^{n,m*}\right].
   \label{eq:helmholtz}
\end{aligned}
\end{equation}
In the present study, the Helmholtz equation is solved using the parallel semicoarsening multigrid (PFMG) solver combined with a Red-Black (RB) preconditioner, both available in the HYPRE library~\cite{falgout2002hypre}. Once $p^{n+1}$ is obtained, the corrected velocity field $u^{n+1}$ is updated using Eq.~\eqref{eq:correction}. 

%===============================
\subsection{Phase transition solver}\label{phase_transition_solver}
%===============================
{\SecondRef As mentioned above the mass transfer is treated via a (first-order) operator  splitting method: we first
solve the system without the source term $M$ and then we solve a system of ordinary differential equations accounting for the mass  transfer. Hence we consider: 
$\partial_t [a_1, T, \vec{u},p]^T = \Phi_{M}$, where here 
$\Phi_M =[S_a^{(1)} M, S_T^{(1)} M,0, S^{(1)}_p M]^T$ is the vector of the source terms in Eq.~\eqref{eq:sysprim} containing the mass transfer term $M= \nu(g_2-g_1)$.}
As one can easily see from this
system of ODEs, during this chemical relaxation process the mixture density, mixture energy and velocity
remain constant. To determine the state after the mass transfer step we need to determine three independent variables,
for instance $a_1$, $T$ and $p$.
We have implemented two different relaxation techniques proposed in the literature
to determine the updated values of $a_1$, $T$ and $p$ after mass transfer. 

The first one 
\cite{le2014towards,pelanti2014mixture} assumes
instantaneous chemical relaxation, $\nu\rightarrow \infty$, so that thermodynamic equilibrium
is instantaneously attained. In this case we do not need  to solve a system of ODEs, but instead 
impose directly the equilibrium condition $g_1(p,T)=g_2(p,T)$. This gives an algebraic system of equations to be solved for the thermodynamic equilibrium state with the mixture relations
$\rho=a_1\rho_1(p,T)+a_2\rho_2(p,T)$ and $\mathcal{E} = a_1\mathcal{E}_1(p,T)+a_2\mathcal{E}_2(p,T)$.

The second relaxation procedure, based on \cite{de2019hyperbolic,pelanti2019numerical,pelanti2021arbitrary}, allows the modelling of chemical relaxation of arbitrary rate, finite-rate
(for instance with a given function to define $\nu$) or instantaneous.
It is based on the idea of approximating the relaxation process toward the
equilibrium $g_1=g_2$ by an exponential behavior. Within this approximation, a semi-exact exponential
solution of the system of ODEs   $\partial_t [a_1, T,  \vec{u},p]^T = \Phi_{M}$ can be found.
This approximate solution is used to define the solution after the mass transfer step.
This second approach is simpler since we update the variables using explicit formulas, whereas 
in the first approach we need the solution of a non-linear system of algebraic equations. 
We typically use this second relaxation technique, nonetheless the first method was also tested within the context of the present study (where we assume instantaneous mass transfer in all the tests) with very similar results.

Let us note that the temperature is equal to its saturation value $T_{sat}(p)$ at the equilibrium $g_1(p,T)=g_2(p,T)$.
Denoting here with subscript $1$ the liquid and with subscript $2$ the vapor, if $g_1>g_2$ then $T>T_{sat}(P)$ and liquid-to-vapor transition occurs (evaporation), whereas if $g_1<g_2$ then $T<T_{sat}(P)$ 
and vapor-to-liquid transition occurs (condensation). This mass transfer processes modelled by the chemical relaxation term 
may be activated and deactivated in the
numerical model depending on the desired criteria. 
 In the present study, mass transfer  is activated only in the presence of a two phase mixture, i.e.\ when both $a_1>\epsilon$ and $a_2>\epsilon$, with $\epsilon=10^{-8}$. Moreover, in some numerical tests we may activate chemical relaxation only
 if the condition for evaporation  $T>T_{sat}(P)$ is met, as done for various tests for instance in 
 \cite{saurel2008modelling,zein2010modeling,pelanti2014mixture}.
 This criterion is applied in the test taken from these references in Section \ref{sec:cavtube}. In such case, for the NASG EOS we solve the equation (\ref{eq:sat_tmp}) using $p^{n,m+1}$ as an independent variable, which provides $T_{sat}^{n,m+1}$.

\subsection{Algorithm overview and additional remarks}
\sloppy
%===============================
To make the proposed methodology as clear as possible, a step by step description of the overall solution procedure is presented in Algorithm~\ref{tab:algorithm}. For the purposes of the present study, the algorithm was implemented using the framework already available in~\cite{costa2018fft}, with the substitution of the fast Fourier transform library with the \textit{Hypre} library.
%AAAAAAAAAAAAAAAAAAAAAAAAAAAAAAAAAAAAAAAAAAAAa

\begin{algorithm}
\caption{Overall solution procedure of the proposed methodology.}
\begin{algorithmic}[1]
\STATE $a_1, \ T, \ \vec{u}, \ p$ are initialised.
\STATE $\rho_k$ are calculated using Eq.~\eqref{eq:T_EOS}. $\rho$, $\mu$ and $\lambda_c$ are calculated using the corresponding phasic quantities and $c$ is calculated from Eq.~\eqref{eq:speedofsound}.
\STATE  $S_a^{(2)}$, $S_T^{(2)}$, $S_p^{(2)}$, $S_a^{(3)}$ and $S_T^{(3)}$ are calculated from Eqs.~\eqref{eq:Sa2}--\eqref{eq:ST3} using the EOS Eqs.~\eqref{eq:eos1} and~\eqref{eq:eos2}.
\STATE $n=0$ is set.
\WHILE{$t<t_{tot}$}
    \STATE $n=n+1$, $m=0$ are set.
    \STATE $\Delta t$ is calculated using Eq.~\eqref{eqn:max_dt}.
	\WHILE{$m<3$} 
	    \STATE $m=m+1$ is set.
	    \STATE $a_1^{n,m+1}$ is calculated from Eq.~\eqref{eq:solve_a1}.
	    \STATE $T^{n,m+1}$ is calculated from Eq.~\eqref{eq:solve_T}.
	    \STATE $\vec{u}^{n,m*}$ is calculated from Eq.~\eqref{eq:solve_u}.
	    \STATE Helmholtz Eq.~\eqref{eq:helmholtz} is solved and $p^{n,m+1}$ is obtained.	    \STATE $\vec{u}^{n,m+1}$ is calculated from equation~\eqref{eq:correction}.
	    \STATE $\rho_k$, $\rho$, $\mu$, $\lambda_c$, $c$, $S_a^{(2)}$, $S_T^{(2)}$, $S_p^{(2)}$, $S_a^{(3)}$ and $S_T^{(3)}$ are updated.
	   % \STATE $T_{sat}$ is calculated from Eq.~\eqref{eq:sat_tmp}.
	    \IF{phase change conditions} 
	      \STATE $a_1^{n,m+1}$, $T^{n,m+1}$ and $p^{n,m+1}$ are locally modified following the relaxation procedure of~\cite{de2019hyperbolic,pelanti2019numerical,pelanti2021arbitrary}.
	      \STATE $\rho_k$, $\rho$, $\mu$, $\lambda_c$, $c$, $S_a^{(2)}$, $S_T^{(2)}$, $S_p^{(2)}$, $S_a^{(3)}$ and $S_T^{(3)}$ are updated.
	    \ENDIF
    \ENDWHILE
\ENDWHILE
\STATE End of simulation.
\end{algorithmic}
\label{tab:algorithm}
\end{algorithm}

%AAAAAAAAAAAAAAAAAAAAAAAAAAAAAAAAAAAAAAAAAAAAa
%

Since the adopted four-equation model and the proposed solution methodology are significantly different from what was previously used in multiphase diffuse interface studies, a few key points on various aspects of the methodology are discussed below:  
\begin{itemize}
    \item \underline{Set of equations}: 
    Common formulations of the four-equation model in the literature adopt either $(a_1 \rho_1, a_2 \rho_2, \rho u, E)$~\cite{abgrall1996prevent,lund_proc} or $(a_1 \rho_1, \rho, \rho u, E)$~\cite{le2014towards} as the set of primary variables (see the set of Eqs.~\eqref{eq:sysTboilc},~\eqref{eq:partr}). In the present study, the choice to follow a pressure-based methodology was made to avoid pressure oscillations at interfaces when the pressure is retrieved from the energy. To construct a pressure-based methodology using either of these sets of equations, the total energy equation can be transformed to a pressure equation, and the temperature can be calculated from the EOS.{\FirstRef This approach was followed in~\cite{bermudez2020staggered} for single-phase problems, using a pressure equation similar to Eq.~\eqref{eq:p}.} Using this approach, the authors of the present study noted that the calculation of the temperature field was not very accurate, especially at the interface. Even a small error in the calculation of the temperature field in phase transition simulations could cause the temperature to artificially exceed or fall below the saturation temperature, severely affecting the results. For this reason, we adopted $(a_1, T, \rho u, p)$ as the primary variables, so to have a more accurate calculation of the temperature field. As mentioned earlier, this model is not appropriate for the numerical solution of fully compressible flows, because there are no discrete equations that guarantee the conservation of the total energy and mass,{\SecondRef while the momentum is solved using a non-conservative form.} Nevertheless, the following sections demonstrate the validity of this model in simulations of compressible flows with phase change at low speeds.
    \item \underline{Mass and energy conservation}: 
    Since the adopted four-equation model does not consider equations for the conservation of mass and energy, these quantities are not automatically conserved. Nonetheless, with proper spatial and temporal resolution, mass and energy are indeed conserved over long simulation times. This observation will be demonstrated and quantified in the verification cases presented in Section~\ref{sec:verification}.
    \item \underline{Momentum discretisation}: 
    As described in Section~\ref{solve_u*}, the discretised momentum equations adopt a non-conservative form of the advection term. The reason behind this is to avoid invoking the updated density field which is yet to be computed at this point. An alternative treatment (not used in the present study) would be to over-constrain the system of equations by solving an extra equation for the mass conservation before solving the momentum equation. In that case, mass will be conserved by definition and the updated density field would be available to be used in a fully conservative momentum equation. This of course would add the cost of having to solve an additional equation and encounter some loss of consistency between the various thermodynamic quantities because in that case $\rho\neq a_1\rho_1(p,T)+a_2\rho_2(p,T)$. A similar treatment of over-constraining the system for algorithmic purposes was adopted in sharp-interface formulations~\cite{jemison2014compressible,fuster2018all}.
    \item \underline{Time step restrictions}: 
    It is generally accepted that explicit pressure-based methods bare overwhelming time-step restrictions. In the proposed methodology, this is overcome by using the updated pressure field in the momentum equation, that results in a Helmholtz equation (Eq.~\eqref{eq:helmholtz}) for the pressure field. In addition, the adoption of the RK3 method for numerical integration improves the overall time step restrictions~\cite{wesseling2009principles}. The time-step restriction employed in this study is~\cite{kang2000boundary},
    \begin{equation}
	\Delta t=C_{\Delta t}\min(\Delta t_c,\Delta t_{\sigma},\Delta t_{\mu},\Delta t_{\lambda})\mathrm{,}
	\label{eqn:max_dt}
    \end{equation}
    where $\Delta t_c$, $\Delta t_{\sigma}$, $\Delta t_{\mu}$ and $\Delta t_{\lambda}$ are the maximum allowable time steps due to convection, surface tension, momentum and thermal energy diffusion. These are determined as suggested in~\cite{kang2000boundary}:
    \begin{equation}
	\begin{aligned}
	\Delta t_c & =\left(\dfrac{|u_{x,\max}|}{\Delta x}+\dfrac{|u_{y,\max}|}{\Delta y}+\dfrac{|u_{z,\max}|}{\Delta z}\right)^{-1}\mathrm{,} \\
	\Delta t_{\sigma}&=\sqrt{\dfrac{(\rho_{1,min}+\rho_{2,min})\min(\Delta x^3,\Delta y^3,\Delta z^3)}{4\pi\sigma}}\mathrm{,} \\
	\Delta t_{\mu}&=\left[\max\left(\dfrac{\mu_1}{\rho_{1,min}},\dfrac{\mu_2}{\rho_{2,min}}\right)\left(\dfrac{2}{\Delta x^2}+\dfrac{2}{\Delta y^2}+\dfrac{2}{\Delta z^2}\right)\right]^{-1}\mathrm{,} \\
	\Delta t_{\lambda}&=\left[\max\left(\dfrac{\lambda_1}{\rho_{1,min}C_{p,1}},\dfrac{\lambda_2}{\rho_{2,min}C_{p,2}}\right)\left(\dfrac{2}{\Delta x^2}+\dfrac{2}{\Delta y^2}+\dfrac{2}{\Delta z^2}\right)\right]^{-1}\mathrm{,}
	\end{aligned}
	\label{eqn:diff_con_t}
    \end{equation}
    where $|u_{i,\max}|$ is an estimate of the maximum value of the $i$th component of the flow velocity, $\rho_{k,min}$ is the minimum density of phase $k$ in the domain and $\Delta x,\ \Delta y,\ \Delta z$ are the grid spacings along the $x,\ y,\ z$ directions. Since this study considers only weakly compressible flows, the acoustic time-step restrictions are not taken into account. Setting $C_{\Delta t}=0.25-1.0$ was seen to be sufficient for a stable and accurate time integration. For a more accurate and fair comparison against reference results, a constant time step was adopted in some of the test cases. This is clearly specified in each test case.
    %
    %\item \underline{Reduced memory version}: 
     %For the implementation of the algorithm presented in Sections~\ref{solve_a1}--\ref{phase_transition_solver}, a number of three-dimensional arrays need to be allocated in the memory, including the volume fraction and the temperature at the $(m)$ sub-step. These arrays are necessary because, after $a_1^{n,m+1}$ and $T^{n,m+1}$ are calculated, the older values $a_1^{n,m}$ and $T^{n,m}$ should remain saved in memory since these are needed for the calculation of thermodynamic terms that appear in the momentum and pressure equations. A small adjustment to this algorithm that reduces the memory cost of this method would be to use $a_1^{n,m+1}$ and $T^{n,m+1}$  whenever $a_1^{n,m}$ and $T^{n,m}$ are needed for the calculation of the aforementioned thermodynamic quantities. Preliminary simulations carried out during this study revealed almost negligible differences when the lower memory implementation was used.
    %\item Advantages over similar density based approaches: complexity as admitted in~\cite{lemartelot2013liquid}: 
\end{itemize}

%===============================
\section{Verification}\label{sec:verification}
%===============================
The methodology presented in the previous sections will be verified in a number of different test cases, under incompressible and compressible conditions, with and without mass transfer. When mass transfer is activated it is always assumed as an instantaneous process (chemical
relaxation parameter $\nu \rightarrow +\infty$). In the following sections, wherever a two-phase mixture is present, subscript (1) refers to the liquid phase while subscript (2) refers to the gas phase. 

%=============================
\subsection{Gresho vortex}\label{sec:Gresho}
%=============================
The proposed numerical algorithm is first tested against the Gresho vortex benchmark~\cite{liska2003comparison}, a rotating vortex which is a time-independent solution of the incompressible Euler equations. The aim of this test is to assess the accuracy of the method against an exact solution and its ability to preserve the vortex structure for different Mach numbers. To this purpose, we employ a variant of the original benchmark, where the analytical solution, reported in Eqs.~\eqref{eqn:gre_vortex_u} depends on a reference Mach number~\cite{miczek2013simulation} and it is a continuous differentiable function~\cite{thomann2020all}: 
 \begin{equation}
     u_{\phi}(r) = \dfrac{1}{u_r}\begin{cases}
                    75r^2-250r^3 &0\leq r\leq 0.2l_r\mathrm{,} \\
                    -4+60r-225r^2+250r^3 &0.2l_r< r\leq 0.4l_r\mathrm{,} \\
                    0\mathrm{,} &r> 0.4l_r
                   \end{cases}
                   \label{eqn:gre_vortex_u}
 \end{equation}
 where $u_{\phi}$ is the angular velocity and $u_r$ a reference velocity. In this steady configuration, the pressure gradient is balanced by the centrifugal force and the density is uniform and equal to a reference value, i.e. $\rho=\rho_r$. Therefore the radial momentum balance reads,
 \begin{equation}
     \dfrac{1}{Ma_r^2}\dfrac{\partial p}{\partial r} = \dfrac{u^2_{\phi}}{r}\mathrm{.}
     \label{eqn:mom_bal}
 \end{equation}
 By splitting the pressure $p$ into a reference pressure $p_r$ and a second order pressure $p^{(2)}$, i.e., $p=p_r+p^{(2)}Ma_r^2$ and using Eq.~\eqref{eqn:mom_bal}, $p_r$ results to be a uniform and constant field, while $p^{(2)}$ can be computed as,
 \begin{equation}
   p^{(2)}Ma_r^2=\int_{r_1}^{r_2}\dfrac{u^2_{\phi}(s)}{s}ds\mathrm{,}
   \label{eqn:p2_exp}
 \end{equation}
 where $Ma_r=\sqrt{\rho_r/(p_r\gamma_r)}$ is the reference Mach number. The EOS parameters used for the calculation of the reference Mach number and all other necessary quantities are listed in Table~\ref{table:gre_eos}, modelling an ideal gas. Using Eq.~\eqref{eqn:p2_exp} finally provides an expression for $p$,
 \begin{equation}
     p=p_r+\int_{r_1}^{r_2}\dfrac{u^2_{\phi}(s)}{s}ds\mathrm{.}
     \label{eqn:gre_vortex_p}
 \end{equation}
 The integrals in Eq.~\eqref{eqn:gre_vortex_p} are evaluated with a Gaussian quadrature method for each interval where $u_{\phi}$ is defined, i.e. $[r_1,r_2]=[0,0.2]l_r$, $[0.2,0.4]l_r$ and $[0.4,1]l_r$. Note that in Eq.~\eqref{eqn:gre_vortex_p} the reference pressure field $p_r$ is given by $p_r=\rho_ru_r^2/(\gamma_rMa_r)$ (with $\rho_r=1$, $u_r=1$), $r$ is the radial coordinate, given by $r=\sqrt{(x-l_r/2)^2+(y-l_r/2)^2}$ and $l_r$ a reference length. Since Eq.~\eqref{eqn:gre_vortex_u} is formulated in a polar reference frame, a coordinate transformation is performed to obtain the Cartesian velocities components, i.e. $u(x,y)=u_\phi\sin(\theta)$ and $v(x,y)=v_\phi\cos(\theta)$ with $\theta=\arctan2(y-l_r/2,x-l_r/2)$. The governing equations are solved in a two-dimensional square domain $\Omega=[0,l_r]\times[0,l_r]$, discretized with four different grid spacings $[\Delta x,\Delta y]=[l_r/N_x,l_r/N_y]$ with $N_x\times N_y=[16\times 16, 32\times 32, 64\times 64, 128\times 128]$. Periodic boundary conditions are prescribed in both directions. The analytical solution given by Eqs.~\eqref{eqn:gre_vortex_u} and~\eqref{eqn:gre_vortex_p}, is prescribed as initial condition for three different Mach numbers, $Ma_r=10^{-1},\ 10^{-2}$ and $10^{-3}$. Simulations are conducted up to $tu_{r}/l_r=2$ (i.e., one complete revolution of the vortex) using a constant time-step $\Delta tu_r/l_r=2.5\times 10^{-3}$. Note that this value represents the maximum allowable time-step to ensure a stable time integration for the highest grid resolutions cases (i.e., $128\times 128$) and is employed for the coarser cases, irrespective of $Ma_r$. 
 
 %TTTTTTTTTTTTTTTTTTTTTTTTTTTTTTTTTTTTTTTTTTTTT
\begin{table}[ht]
\centering
\begin{tabular}{lcccccccc}
\hline
 $\gamma$ & $\eta$ & $\tilde{\eta}$ & $p_{\infty}$ & $b$ & $\kappa_v$\\
\hline
%&- & - & - & - & $\mathrm{N/m}$ & $\mathrm{J/(kg\cdot K)}$ & $\mathrm{Pa}$ & $\mathrm{m/s^2}$ \\
%\hline
1.4 & 0 & 0 & 0 & 0 & 717.5 \\
\hline
\end{tabular}
\caption{EOS parameters adopted for (a) the "Gresho vortex" test case (Section~\ref{sec:Gresho}) and (b) the thermally driven flow in a differentially heated cavity (Section~\ref{sec:DHC}).}
\label{table:gre_eos}
\end{table}
%TTTTTTTTTTTTTTTTTTTTTTTTTTTTTTTTTTTTTTTTTTTTT

  %FFFFFFFFFFFFFFFFFFFFFFFFFFFFFFFFFFFFFFFFFFFF
 \begin{figure}[ht!]
 \centering
 \includegraphics[width=\textwidth]{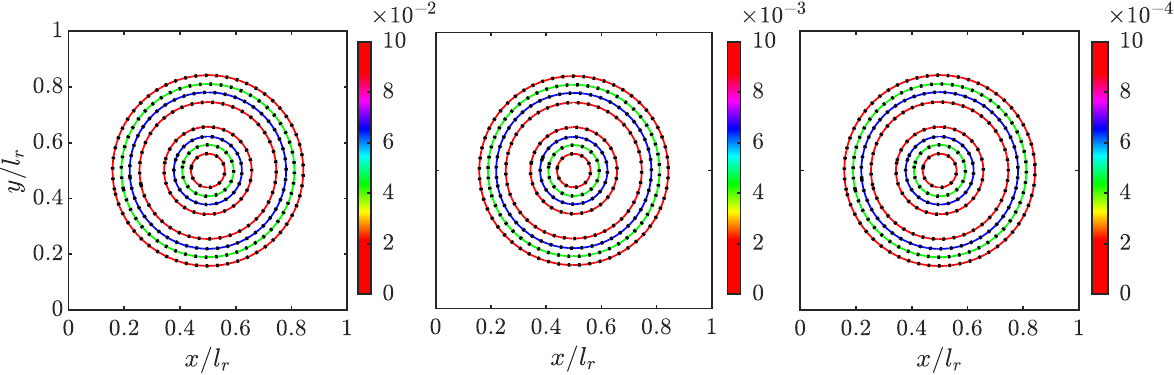}
 %\includegraphics[width=\textwidth]{figures/vortex/mach_t0.pdf}\vspace{-0.0625cm}
 %\includegraphics[width=\textwidth]{figures/vortex/mach_t1.pdf}%\vspace{-1.015625cm}
 %\caption{Mach number distribution in the Gresho vortex case for $Ma_r=10^{-1},\ 10^{-2}$ and $10^{-3}$ (from the left to the right) using $128\times 128$ grid points. The contours at the top refer to the initial condition (i.e., $tu_r/l_r=0$), while the ones at the bottom to the solution after one revolution of the vortex (i.e., $tu_r/l_r=2$).}
 \caption{Mach number distribution in the Gresho vortex case for $Ma_r=10^{-1},\ 10^{-2}$ and $10^{-3}$ (from the left to the right) using $128\times 128$ grid points.{\SecondRef{The dotted black lines refer to the initial condition, while the solid lines to the solution after one revolution of the vortex (i.e., $tu_r/l_r=2$)}} .}
 \label{fig:vortex}
 \end{figure}
%FFFFFFFFFFFFFFFFFFFFFFFFFFFFFFFFFFFFFFFFFFFF

\begin{figure}[ht!]
 \centering
 \includegraphics[width=13.0 cm, height=5.0 cm]{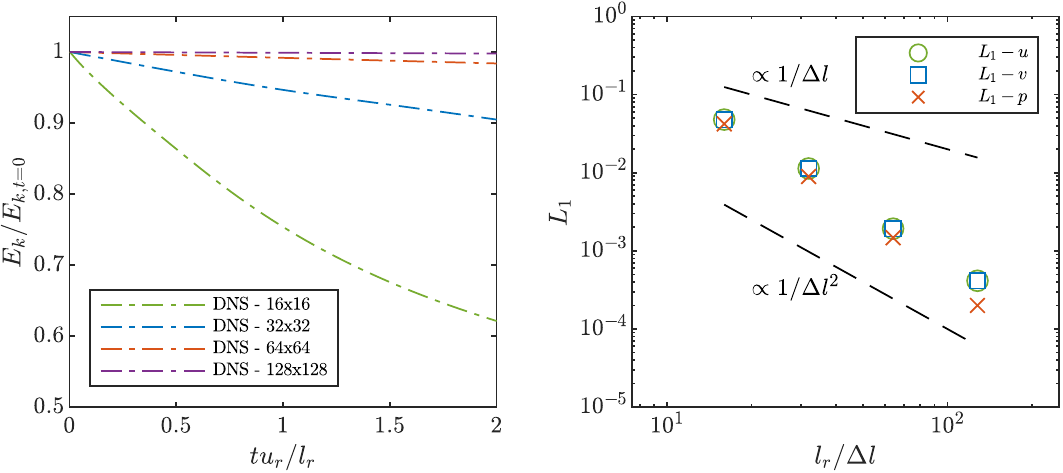}
    \put(-369,132){\small(\textit{a})}
    \put(-179,132){\small(\textit{b})}
 \caption{Gresho vortex with $Ma_r=0.001$: (a) evolution of the normalized kinetic energy $E_k/E_{k,t=0}$ in the time interval $tu_r/l_r=0-2$ for different grid spacings, (b): $L_1$ norm for $(u,v,p)$ at $tu_r/l_r=1$.}
 \label{fig:ekin}
 \end{figure}

 Fig.~\ref{fig:vortex} shows the Mach number distribution for the various cases considered. It is clear that the proposed numerical methodology is able to preserve the vortex shape regardless of the employed $Ma_r$. This result is possible thanks to the implicit treatment of the acoustic part of the pressure field (using a prediction-correction approach~\cite{kwatra2009method}), which ensures a stable and bounded solution of the pressure equation, even for $Ma_r\rightarrow 0$. The excellent ability of method in preserving the vortex shape is reflected also in the good conservation property of the kinetic energy $E_{k}=1/2\int_V\rho\vec{u}\cdot\vec{u}dV$, whose temporal evolution is reported in Fig.~\ref{fig:ekin}(a) for different grid resolutions. Note that $E_k$ is conserved at the highest resolution cases with $Ma_r=0.001$ and a similar behaviour has been observed also for $Ma_r=0.01$ and $0.1$ (not shown).
 
 We conclude the analysis with the accuracy assessment in terms of the $L_1$-norm, evaluated as: 
\begin{equation}
    L_1 = \dfrac{1}{N_xN_y}\sum_{i=1}^{N_x}\sum_{j=1}^{N_y}|s(i,j)-s_{ex}(i,j)|\mathrm{,}
    \label{eqn:l1_def}
\end{equation}
where $s$ represents either one of the Cartesian velocity components, $u$ or $v$ or the pressure, $p$, while $s_{ex}$ represents the corresponding exact fields computed with the analytical solution. If $L_{1,N}$ is the $L_1$-error using $N_x\times N_y$ grid point and $L_{1,2N}$ is the $L_1$-error evaluated with $2N_x\times 2N_y$ grid points, the order of accuracy $n_{L1}$ is computed as:
\begin{equation}
    n_{L1} = \dfrac{\log\left(\dfrac{L_{1,2N}}{L_{1,N}}\right)}{\log(2)}\mathrm{.}
\end{equation}
Both $L_1$-error and $n_{L1}$ are evaluated at $tu_r/l_r=1$ and the results are reported in Fig.~\ref{fig:ekin}(b) for $Ma_r=0.001$. As expected, a second-order accurate solution for $(u,v,p)$ is achieved for all cases (a similar trend has been observed also for $Ma_r=0.01$ and $0.1$), confirming the correct behaviour of the proposed method irrespective of $Ma_r$.
%FFFFFFFFFFFFFFFFFFFFFFFFFFFFFFFFFFFFFFFFFFFF
%FFFFFFFFFFFFFFFFFFFFFFFFFFFFFFFFFFFFFFFFFFFF

%=============================
\subsection{Thermally driven flow in a differentially heated cavity}\label{sec:DHC}
%=============================
In this section, the flow of air in a closed two-dimensional square cavity with heated and cooled side walls and adiabatic horizontal walls is considered. The ascending and descending buoyant currents next to the heated and cooled walls form a circulation current, while the central region of the cavity features an almost stagnant fluid with stratified temperature. In thermally driven flows, the compressibility effects are directly related to the temperature difference between the thermally active side walls. For temperature differences less than approximately 30 K the flow of air is considered incompressible~\cite{gray1976validity}, and the Oberbeck--Boussinesq approximation is typically adopted~\cite{de1983natural,hortmann1990finite,le1991accurate}. Outside the limits of the Oberbeck--Boussinesq approximation,  weak compressibility effects appear, and different low-Mach methodologies were utilised to study these effects~\cite{le2005modelling, armengol2017effects,demou2019low}. For the purposes of this study, this test case helps to assess the ability of the methodology to accurately incorporate the effects of viscosity and thermal conductivity.

The case simulated here follows the setup presented in~\cite{le2005modelling,armengol2017effects} and involves a temperature difference $\Delta T=720$~K, around a reference temperature of $T_r = 600$~K. The reference thermophysical quantities and the height of the cavity $L$ are chosen such that the Rayleigh and Prandtl numbers are equal to,
\begin{equation}
\mathrm{Ra}=\frac{g\rho_r^2\kappa_p\beta\Delta T L^3}{\mu \lambda}=10^6
\mathrm{, \ and \ }
\mathrm{Pr}=\frac{\mu \kappa_p}{\lambda}=0.71,
\end{equation}
where $\beta$ is the thermal expansion coefficient. The ideal gas EOS is used as the thermodynamic closure (parameters listed in Table~\ref{table:gre_eos}) and the reference density value $\rho_r$ is calculated using the reference temperature $T_r=$600~K and the reference pressure $p_r$=101325~Pa. All other thermophysical quantities are considered constant. Furthermore, no-slip boundary conditions are applied to the walls. Initially the air in the cavity is stagnant and isothermal with $T(t=0,x,y)=T_r$, and a hydrostatic pressure field with $p(t=0,x=L/2,y=L/2)=p_r$ is applied. The time step is dynamically adjusted according to Eq.~\eqref{eqn:max_dt}, with $C_{\Delta t}=0.5$.
% TTTTTTTTTTTTTTTTTTTTTTTTTTTTTTTTTTTTTT
%\begin{table}[b]
%\centering
%\begin{tabular}{lcccccccc}
%\hline
% $\gamma$ & $\eta$ & $\tilde{\eta}$ & $p_{\infty}$ & $b$ & $\kappa_v$\\
%\hline
%1.4 & 0 & 0 & 0 & 0 & 717.5 \\
%\hline
%\end{tabular}
%\caption{EOS parameters that are adopted for the study of thermally driven flow of air in a differentially heated cavity.}
%\label{table:DHC_eos}
%\end{table}
% TTTTTTTTTTTTTTTTTTTTTTTTTTTTTTTTTTTTTT

\begin{figure}[t]
 \centering
 \includegraphics[width=0.49\textwidth]{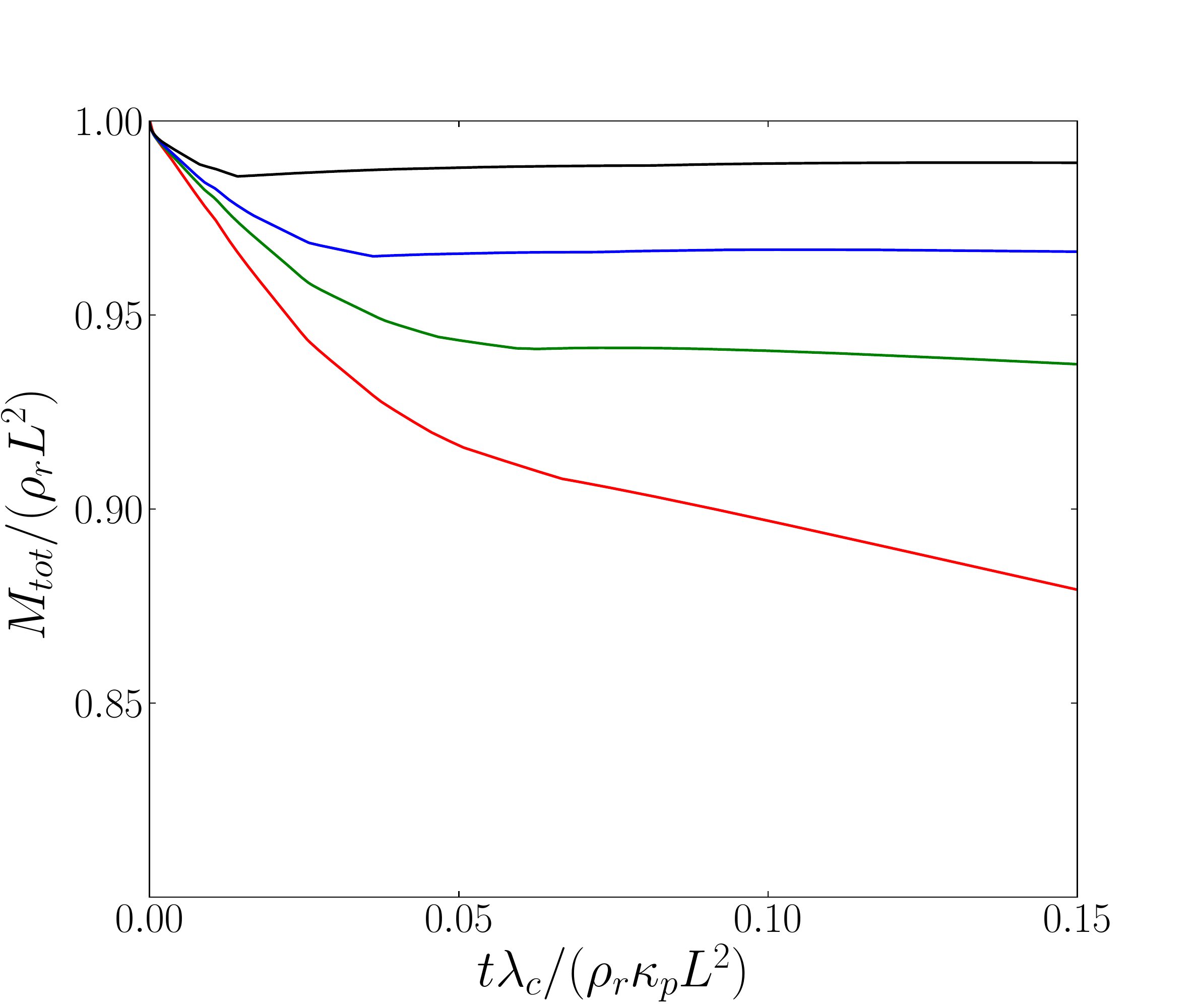}
 \includegraphics[width=0.49\textwidth]{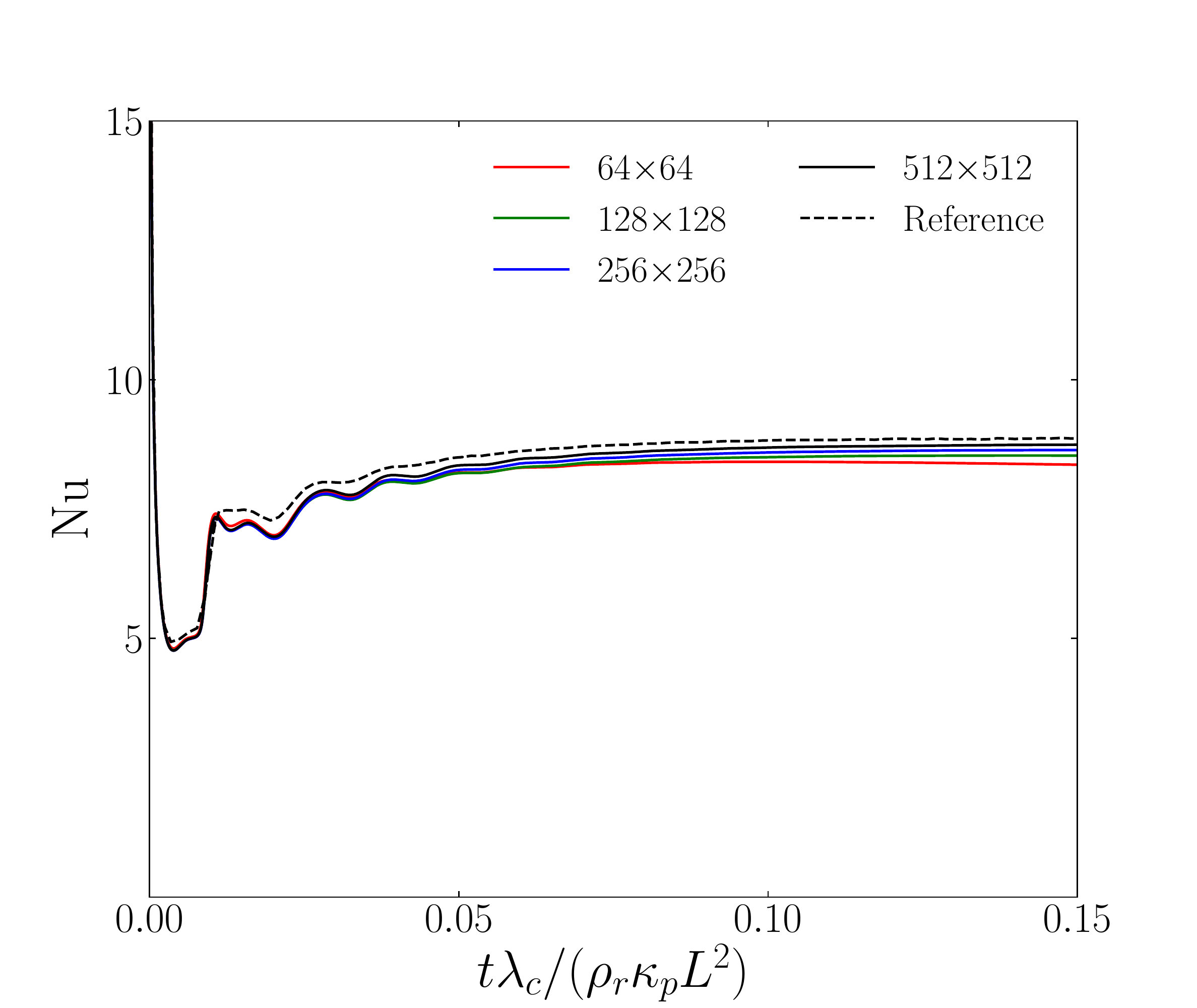}
 \\
  \begin{picture}(0,0)(0,0)
    \put(-195,140){(\textit{a})}
    \put(5,140){(\textit{b})}
  \end{picture}
 \caption{\SecondRef Temporal evolution of (\textit{a}) the dimensionless total mass in the cavity and (\textit{b}) the Nusselt number at the heated wall, for different grid sizes. The reference solution for the Nusselt number is reported in~\cite{armengol2017effects}.}
 \label{fig:heated_cavity_nusselt}
 \end{figure}
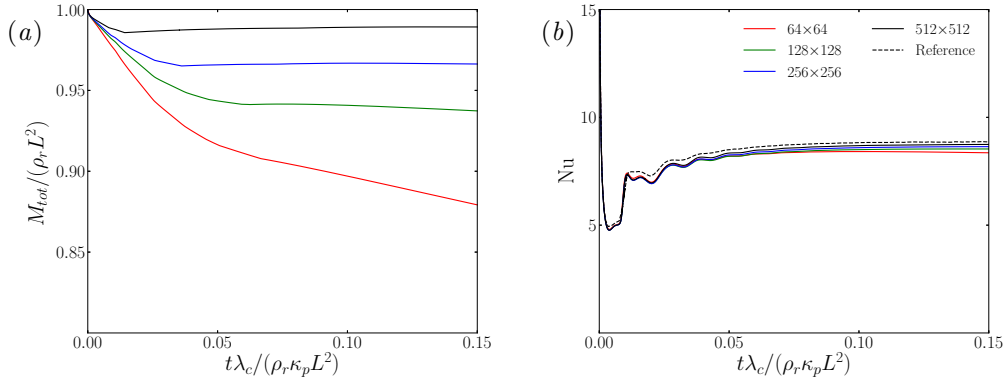

{\SecondRef Even though this is a simple configuration, the introduction of large temperature difference increases the resolution requirements due to the thinning of the thermal boundary layer at the cooled wall~\cite{demou2020variable}. In~\cite{armengol2017effects}, a non-uniform grid of $256\times256$ was used, with clustering of grid points next to the walls, while in~\cite{le2005modelling} a uniform grid of size $512\times512$ was chosen. In the present study, four different uniform grid sizes were used, namely $64\times64$, $128\times128$, $256\times256$ and $512\times512$ grid nodes. As shown in Fig.~\ref{fig:heated_cavity_nusselt}(\textit{a}), the highest resolution considered is able to achieve conservation of the total mass with approximately 1\% error. }

The heat transfer rate inside the cavity is expressed through the  Nusselt number, defined as,
\begin{equation}
\mathrm{Nu}=\frac{h L}{\lambda_c}=\frac{L}{\Delta T}\vec{\nabla}T\big|_w\cdot\hat{n}_w
\end{equation}
where $h$ is the heat transfer coefficient, $\vec{\nabla}T\big|_w$ is the temperature gradient on any of the thermally active vertical walls and $\hat{n}_w$ is the corresponding unit normal vector on the wall. The temporal evolution of the Nusselt number at the heated wall is plotted in Fig.~\ref{fig:heated_cavity_nusselt}, for different grid sizes. After a steep drop during the initial stages of the simulation, the Nusselt number increases gradually to a steady state value.{\SecondRef All grid sizes considered capture the evolution of the Nusselt number fairly well, converging to the reference solution from~\cite{armengol2017effects} with increasing resolution.} More specifically, the steady state solution of the $512\times512$ grid differs by 1.3\% with respect to the reference solution. Noting that the solution methodology followed in the reference study is based on the low-Mach number asymptotic expansion of the Navier-Stokes equations, and therefore is significantly different to the present method, the agreement between the two solutions is considered satisfactory. 
%FFFFFFFFFFFFFFFFFFFFFFFFFFFFFFFFFFFFFFFFFFFF
%FFFFFFFFFFFFFFFFFFFFFFFFFFFFFFFFFFFFFFFFFFFF

%=============================
\subsection{Rising bubble}\label{sec:rising_bubble}
%=============================
The well-established "Rising bubble" test case is employed here as a two-phase numerical benchmark~\cite{hysing2009quantitative}. This test case helps to assess the ability of the proposed numerical methodology to capture topological changes of a moving interface, in the presence of surface tension. More specifically, the evolution of the shape, position and velocity of the center of mass of a rising bubble in a two dimensional liquid column will be compared against the reference solution in~\cite{hysing2009quantitative}. 

By introducing a reference length $l_r$, a velocity $u_r$, gas and liquid densities $\rho_{g,r}$ and $\rho_{l,r}$ and gas and liquid dynamic viscosity $\mu_{g,r}$ and $\mu_{l,r}$, we can define the five dimensionless groups which governs the problem: the Reynolds number $Re=\rho_{g,r}u_rl_r/\mu_{g,r}$, the Weber number $We=\rho_{g,r}u_{r}^2l_r/\sigma$ with $\sigma$ equal to the surface tension, the Froude number $Fr=u_r^2/(|\vec{g}|d_0)$, the density ratios $\lambda_{\rho}=\rho_{l,r}/\rho_{g,r}$ and the viscosity ratio $\lambda_{\mu}=\mu_{l,r}/\mu_{g,r}$. The liquid column has a dimension $l_x=2d_0$ and $l_y=4d_0$ where $d_0$ is the bubble diameter, whose initial center of mass position is $(x_{c,0},y_{c,0})=(d_0,d_0)$. 

This section considers two different rising bubble test cases. For the first test case, simulations are conducted setting $l_r=d_0$, $u_r=\sqrt{|\vec{g}|d_0}$, $Re=35$, $We=1$, $Fr=1$, $\lambda_{\rho}=10$ and $\lambda_{\mu}=10$ in a domain discretized with $N_x\times N_y=[32\times 64, 64\times 128, 128\times 256]$ grid points. Note that the EOS parameters reported in Table~\ref{table:ris_bub} are chosen to match the specific $\lambda_{\rho}$. The top and bottom boundaries are no-slip non-moving walls, while periodic conditions are prescribed in the horizontal directions. The initial velocity field is zero and the pressure is uniform. A constant time step $\Delta t\sqrt{|\vec{g}|/d_0}=3\times 10^{-4}$ is used. This value is the maximum allowable time-step to ensure a stable time integration for the highest grid resolutions cases (i.e., $128\times 256$) and is employed for the coarser cases in order to ensure that the same time discretization error is introduced in all the cases.

%TTTTTTTTTTTTTTTTTTTTTTTTTTTTTTTTTTTTTTTTTTTTT
\begin{table}[ht]
\centering
\begin{tabular}{lcccccccc}
\hline
 & $\gamma$ & $\eta$ & $\tilde{\eta}$ & $p_{\infty}$ & $b$ & $\kappa_v$\\
\hline
%&- & - & - & - & $\mathrm{N/m}$ & $\mathrm{J/(kg\cdot K)}$ & $\mathrm{Pa}$ & $\mathrm{m/s^2}$ \\
\multicolumn{7}{c}{Test case (1)} \\
\hline
 liquid (1) & 1.187 & -1.178$\times10^6$ & -1.178$\times10^6$ & $10^5$ & 0 & 3610 \\
 gas   (2) & 1.400 & 0                  & 0                  & $10^5$ & 0 & 717.5 \\
 \hline
%&- & - & - & - & $\mathrm{N/m}$ & $\mathrm{J/(kg\cdot K)}$ & $\mathrm{Pa}$ & $\mathrm{m/s^2}$ \\
\multicolumn{7}{c}{Test case (2)} \\
\hline
 liquid (1) & 1.187 & 0 & 0 & 1.013$\times10^9$ & 6.61$\times10^{-4}$ & 3610 \\
 gas   (2) & 1.400 & 0  & 0 & 0                 & 0                 & 717.5 \\
\hline
\end{tabular}
\caption{EOS parameters adopted for the two rising bubble test cases.}
\label{table:ris_bub}
\end{table}
%TTTTTTTTTTTTTTTTTTTTTTTTTTTTTTTTTTTTTTTTTTTTT

\begin{figure}[t]
 \centering
 %\hspace{3.5 cm}
 \includegraphics[width=10.0 cm, height=4.680 cm]{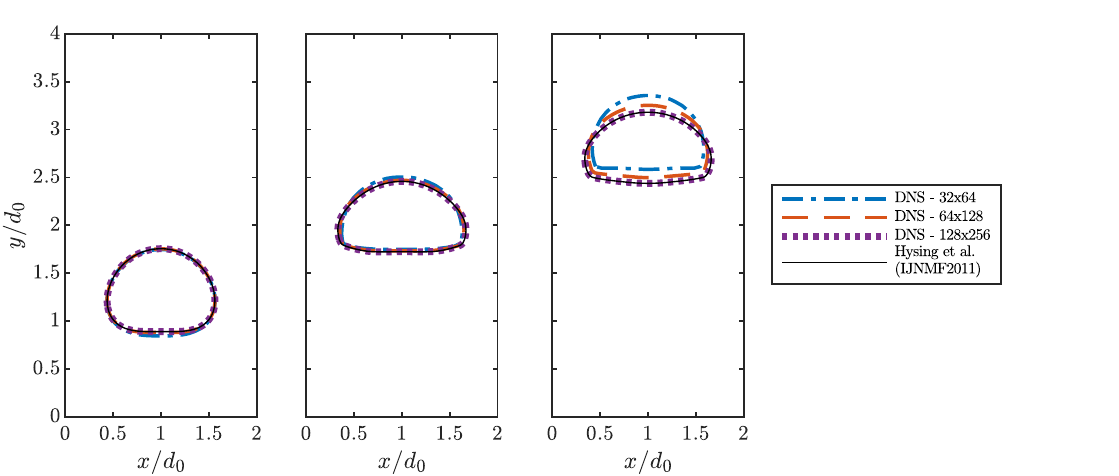}\hspace{-2.0 cm}
 \caption{Interface position for the first rising bubble test case at $t\sqrt{|\vec{g}|/d_0}=0.9,\ 2.7$ and $4.5$ (from the left to the right) and for different grid resolution $N_x\times N_y=32\times 64, 64\times 128$ and $128\times 256$.}
 \label{fig:bubble_shape}
 \end{figure}

\begin{figure}[t]
 \centering
 \includegraphics[width=6.0 cm, height=5.0 cm]{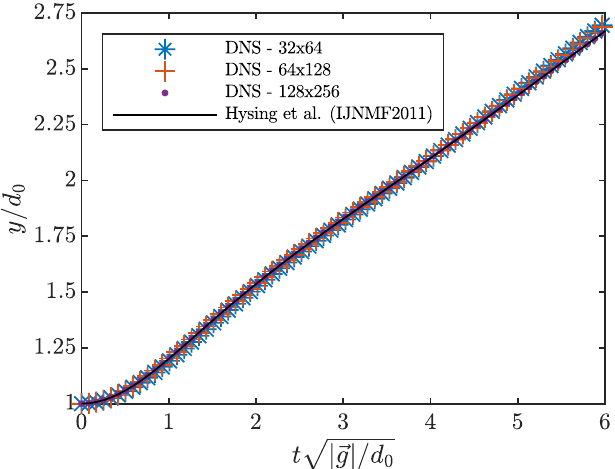}\hspace{+0.40 cm}
 \includegraphics[width=6.0 cm, height=5.0 cm]{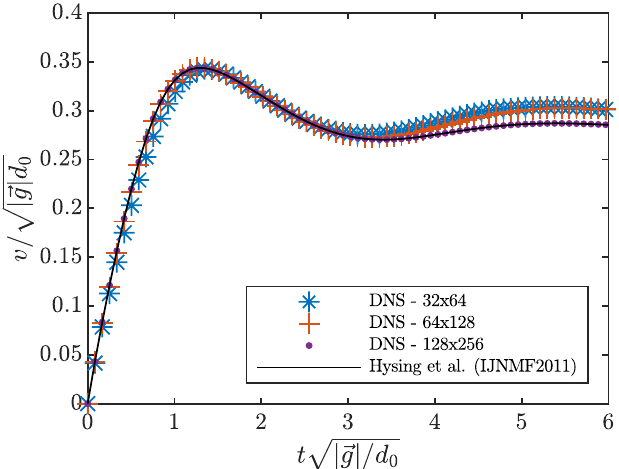}
    \put(-369,132){\small(\textit{a})}
    \put(-179,132){\small(\textit{b})}
 \caption{(a) Position of the center of mass of the rising bubble, (b) vertical velocity of the center of mass of the rising bubble. Results correspond to the first test case.}
 \label{fig:rising_bubble}
 \end{figure}

First, Fig.~\ref{fig:bubble_shape} shows the position of the interface at different time instances, providing a qualitative assessment of the numerical solution for the three grid resolutions. An excellent agreement is observed between the numerical solution on the most refined grid and the reference solution in all the analysed time instances. Similarly, an excellent agreement is confirmed by comparing the temporal evolution of the bubble's center of mass position and vertical velocity, reported in Fig.~\ref{fig:rising_bubble}. Note that the proposed benchmark is often employed for assessing the accuracy of incompressible two-phase codes without temperature variation in the bulk regions of the phases. In the present work, since the two phases exhibit compressible effects, local temperature variations cannot be avoided a-priori given the local variation of the pressure field. However, the maximum  temperature variations remain below $\Delta T/T_r\approx 10^{-5}$; confirming the excellent behaviour of the proposed numerical method in the incompressible limit.

%FFFFFFFFFFFFFFFFFFFFFFFFFFFFFFFFFFFFFFFFFFFF
%FFFFFFFFFFFFFFFFFFFFFFFFFFFFFFFFFFFFFFFFFFFF

%FFFFFFFFFFFFFFFFFFFFFFFFFFFFFFFFFFFFFFFFFFFF
%FFFFFFFFFFFFFFFFFFFFFFFFFFFFFFFFFFFFFFFFFFFF

{\SecondRef The second test case considers larger property variations and involves extreme events such as small bubble breakups. More specifically, the parameters considered in this second benchmark are $Re=3.5$, $We=0.125$, $\lambda_{\rho}=1000$ and $\lambda_{\mu}=100$, while all other simulation parameters remain the same as the first test case. As before, the EOS parameters reported in Table~\ref{table:ris_bub} are chosen to match the specific $\lambda_{\rho}$ considered here. Three uniform grid resolutions are used to simulate the flow: $128\times256$, $256\times512$ and $512\times1024$. First, the comparison of the interface position at dimensionless time $t\sqrt{|\vec{g}|/d_0}=3$ is presented in Fig.~\ref{fig:rising_bub_dr1000}, for the three different grid resolutions considered. In this case, we observe the formation of two trailing bubble tails instead of two small circular bubbles that break away from the original structure in the reference solution. As a result, the bubble rise velocity is lower than the reference results, as depicted in Fig.~\ref{fig:rising_vel_dr1000}. This deviation is reduced as the grid becomes finer, reaching a value of approximately $9\%$ at dimensionless time $t\sqrt{|\vec{g}|/d_0}=3$, for the $512\times1024$ grid. 

%FFFFFFFFFFFFFFFFFFFFFFFFFFFFFFFFFFFFFFFFFFFF
 \begin{figure}[h!]
 \centering
 %\hspace{3.5 cm}
    %\includegraphics[width=0.40\textwidth]{figures/rising_bubble/y_vel_rb_dr1000.pdf}
    \hspace{2.0 cm}
    \includegraphics[width=10.0 cm, height=4.680 cm]{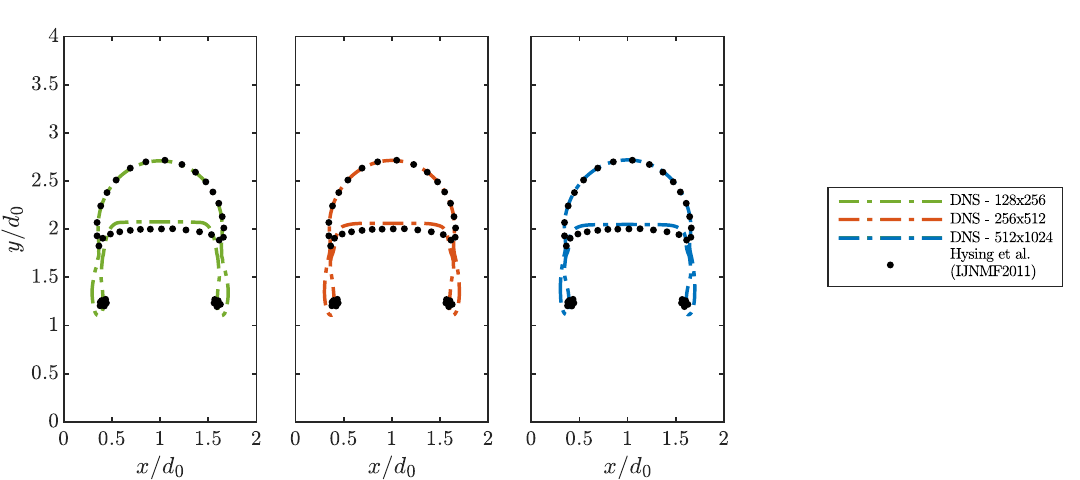}%\hspace{-2.0 cm}
    %\includegraphics[width=6.0 cm, height=5.0 cm]{figures/rising_bubble/y_pos_rb.pdf}\hspace{+0.40 cm}
    %\put(-369,132){\small(\textit{a})}
    %\put(-179,132){\small(\textit{b})}
    \caption{Interface position for the second rising bubble test case at $t\sqrt{|\vec{g}|/d_0}=3$ for different grid resolutions $N_x\times N_y=128\times 256, 256\times 512$ and $512\times 1024$.}
 \label{fig:rising_bub_dr1000}
 \end{figure}
%FFFFFFFFFFFFFFFFFFFFFFFFFFFFFFFFFFFFFFFFFFFF

%FFFFFFFFFFFFFFFFFFFFFFFFFFFFFFFFFFFFFFFFFFFF
 \begin{figure}[h!]
 \centering
 %\hspace{3.5 cm}
    \includegraphics[width=6.0 cm, height=5.0 cm]{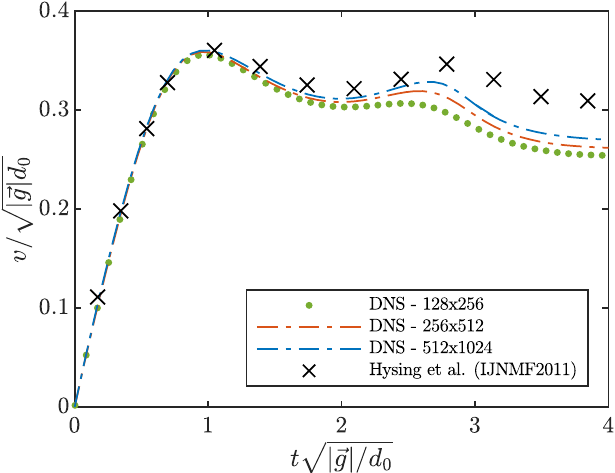}
    %\includegraphics[width=0.50\textwidth]{figures/rising_bubble/contour_vol_dr1000.pdf}%\hspace{-2.0 cm}
    %\includegraphics[width=6.0 cm, height=5.0 cm]{figures/rising_bubble/y_pos_rb.pdf}\hspace{+0.40 cm}
    %\put(-369,132){\small(\textit{a})}
    %\put(-179,132){\small(\textit{b})}
    \caption{Vertical velocity of the center of mass of the rising bubble for different grid resolutions $N_x\times N_y=128\times 256, 256\times 512$ and $512\times 1024$. Results correspond to the second test case.}
 \label{fig:rising_vel_dr1000}
 \end{figure}
%FFFFFFFFFFFFFFFFFFFFFFFFFFFFFFFFFFFFFFFFFFFF

Even though this method is able to accurately simulate flows with large property variations (as demonstrated in Sections~\ref{sec:cavtube} and~\ref{sec:tube_boil}) and capture large bubbles that break away from larger structures (demonstrated in Section~\ref{sec:demonstration}), it requires a fine grid resolution to capture the break away of small bubbles. This drawback is solely attributed to the inherent diffusive nature of the diffuse interface method. Small bubbles that break away cannot get fully detached from the bigger structures without additional treatment of the interface thickness. Section~\ref{sec:future} describes several approaches that can be incorporated in the proposed methodology to manage the interface thickness.}

%=============================
\subsection{Water liquid-vapor expansion tube}
%=============================
\label{sec:cavtube}

The one-dimensional water liquid-vapor expansion tube test case was first proposed in~\cite{saurel2008modelling} and was later adopted in other studies such as~\cite{zein2010modeling,pelanti2014mixture}. This test case is presented here to verify the code in a fully compressible flow and to demonstrate the accuracy of the phase transition solver. A tube of $L_x=1$~m length is filled with a two-phase mixture of liquid water with a uniformly distributed small amount of vapor $a_2=0.01$. The tube is open at both ends and the water is subject to atmospheric pressure, $p=10^5$~Pa, with a liquid density of $\rho_1=1150$~kg~m$^{-3}$. Initially, a velocity discontinuity is present at the centre of the cavity $x=0.5$~m, with $u=-u_0$ for $x<0.5$~m and $u=u_0$ for $x>0.5$~m, where $u_0=2$~m~s$^{-1}$. Viscosity, thermal conductivity and surface tension effects are not considered. The EOS parameters adopted for this test case are listed in Table~\ref{table:CAV_eos}. Using this parameters, the temperature of the two-phase mixture is calculated $T=354.728$~K and the vapor density  $\rho_2=0.63$~kg~m$^{-3}$. 

% TTTTTTTTTTTTTTTTTTTTTTTTTTTTTTTTTTTTTT
\begin{table}[b]
\centering
\begin{tabular}{lcccccccc}
\hline
 & $\gamma$ & $\eta$ & $\tilde{\eta}$ & $p_{\infty}$ & $b$ & $\kappa_v$\\
\hline
%&- & - & - & - & $\mathrm{N/m}$ & $\mathrm{J/(kg\cdot K)}$ & $\mathrm{Pa}$ & $\mathrm{m/s^2}$ \\
%\hline
 liquid (1) & 2.35 & -1.167$\times10^6$ & 0 & $10^9$ & 0 & 1816 \\
 vapor (2) & 1.43 & 2.030$\times10^6$ & -2.34$\times10^4$ & 0 & 0 & 1040 \\
\hline
\end{tabular}
\caption{EOS parameters adopted for the study of water liquid-vapor expansion.}
\label{table:CAV_eos}
\end{table}
%TTTTTTTTTTTTTTTTTTTTTTTTTTTTTTTTTTTTTT

Two cases were simulated: (i) phase transition is not activated and (ii) phase transition is activated when the condition $T>T_{sat}(p)$ is met. A uniform grid with 1024 points is used and a constant time step $\Delta t=3.2\times10^{-6}$~s is adopted. To avoid numerical instabilities, the initial velocity discontinuity used in the present simulations is approximated via a hyperbolic tangent function in the form,
\begin{equation}
u(x)=u_0\mathrm{tanh}((x-0.5)/\tilde{c}), \label{eq:tanh}
\end{equation}
where $u_0=2$~m~s$^{-1}$ and a value of $\tilde{c}=10^{-2}$~m is adopted for the sharpness parameter. The flow was allowed to develop for 1000 time steps, and the relevant fields at $t=0.032$~s are used for comparison against the reference results from~\cite{pelanti2014mixture}.

\begin{figure}[t]
 \centering
 \includegraphics[width=0.45\textwidth]{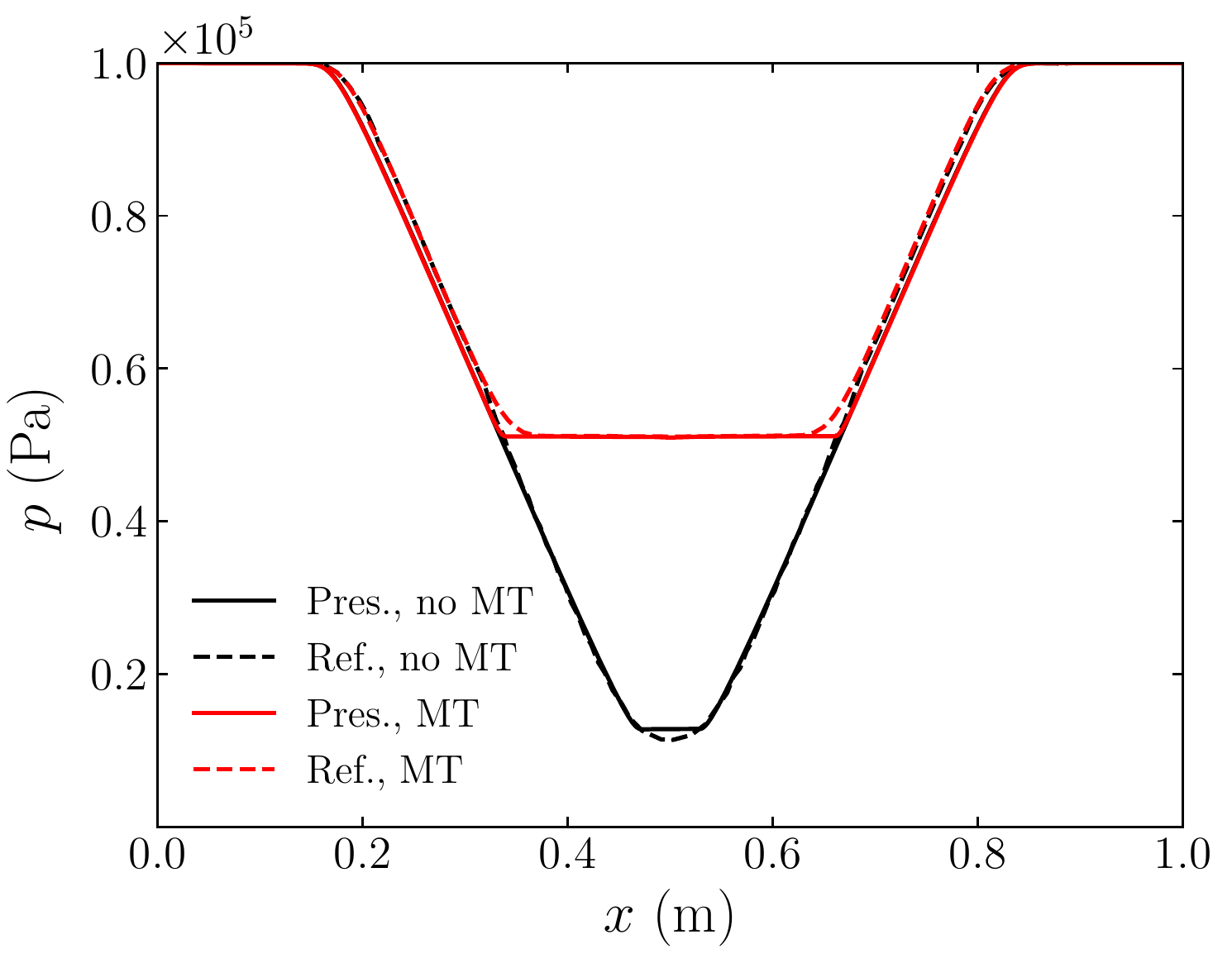}
 \includegraphics[width=0.45\textwidth]{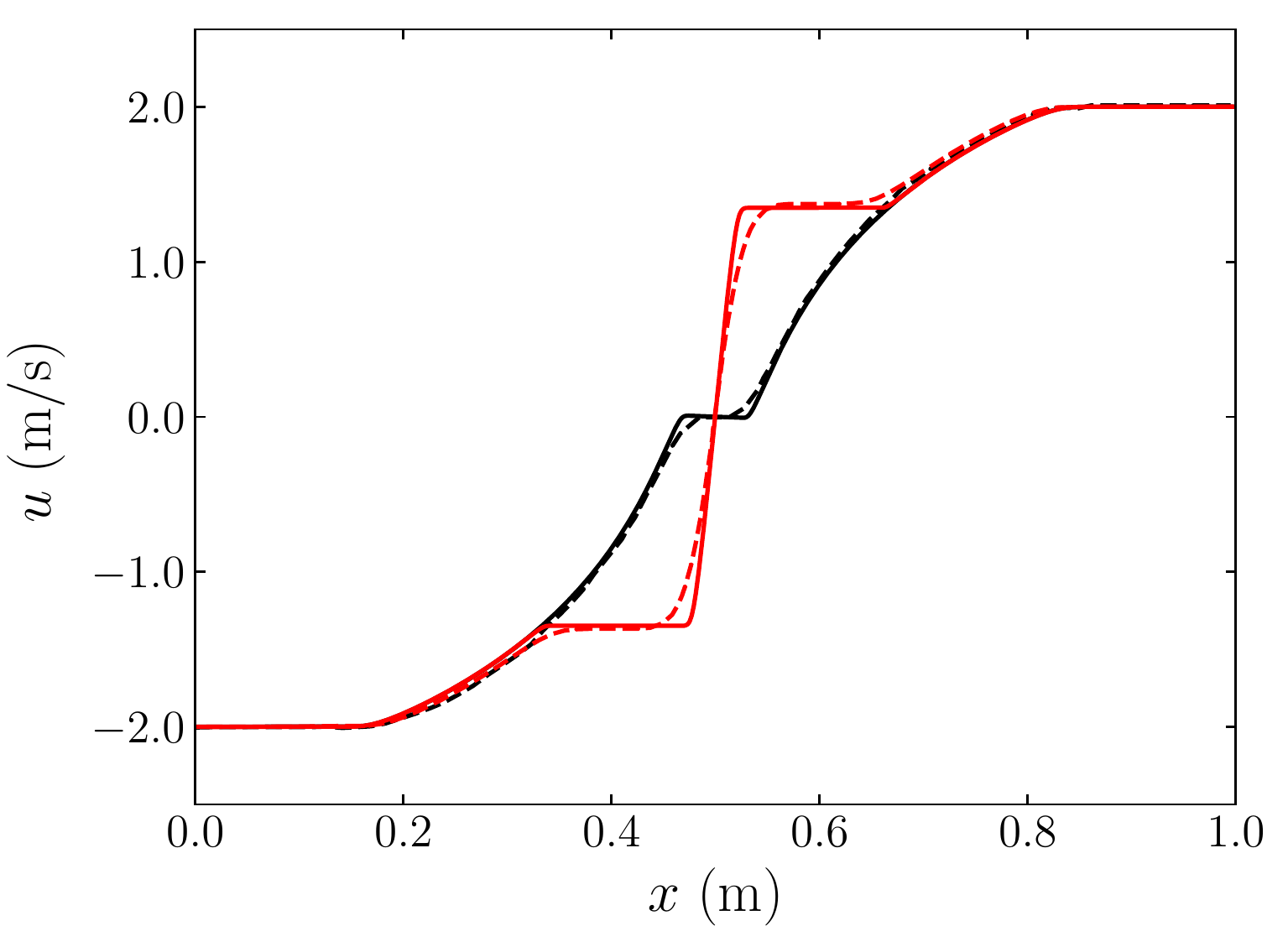}
 \includegraphics[width=0.45\textwidth]{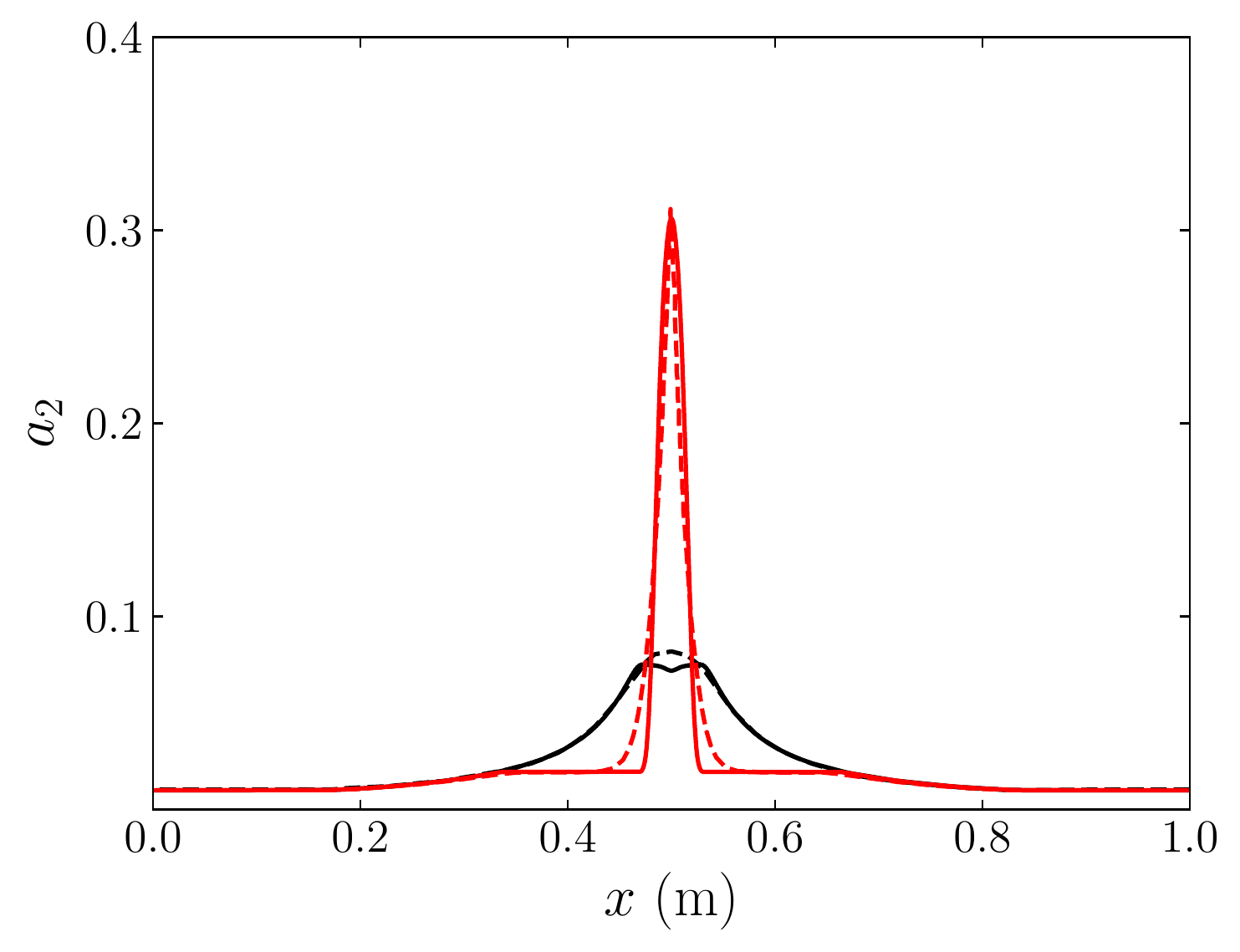}
 \includegraphics[width=0.45\textwidth]{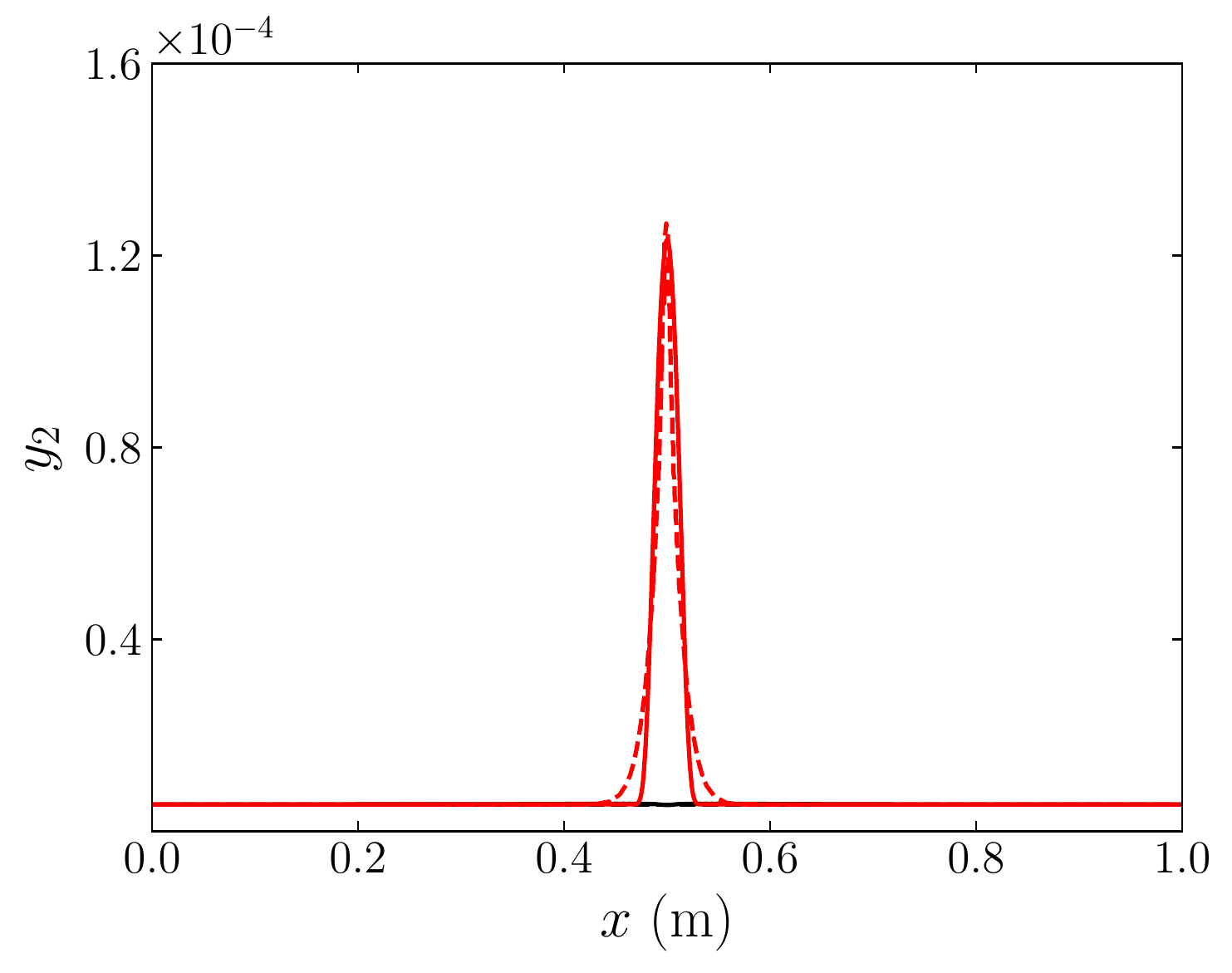}
 \caption{Solution of the water liquid-vapor expansion case with mass transfer (red) and without mass transfer (black), at $t=0.032$ s. Solid lines correspond to the present results while dashed lines correspond to the reference solutions of~\cite{pelanti2014mixture}.}
 \label{fig:cav_fields}
 \end{figure}

The comparison between the present and reference results is shown in Fig.~\ref{fig:cav_fields}. Starting from the case without mass transfer, two rarefaction waves appear, moving in opposite directions. The induced small mechanical expansion of the vapor phase causes an increase in vapor volume fraction at the center of the tube. Due to the presence of the rarefaction waves, the liquid phase becomes metastable and the inclusion of mass transfer influences the solution significantly. More specifically, in the presence of mass transfer, the expansion of liquid water causes the decrease of the pressure at the centre of the tube, reaching its saturation value. Consequently, a small amount of vapour is generated, as evidenced in the increase of the vapor volume and mass fractions. These characteristics provide the basis of the comparison depicted in Fig.~\ref{fig:cav_fields}, where the agreement between present and reference results is verified with only minor discrepancies. Overall, the proposed methodology accurately captures the solution of this compressible flow with and without mass transfer.

%In both cases, with and without mass transfer, two rarefactions are present, moving in opposite directions. This induces a pressure drop at the center of the tube, which consequently causes an increase in volume fraction for the vapor phase in this region, which for the case without mass transfer corresponds to a mechanical cavitation process. 
%When mass transfer is present, the increase in the volume fraction of the vapor phase is more pronounced while the pressure at the centre of the tube decreases until it reaches its saturation value. %{\lb not clear...can you just say that: the pressure drop at the centre of the tube decreases until it reaches its saturation value}. 
%The mass fraction of the vapor phase remains constant for the case without mass transfer, while it increases in the central region of the tube when mass transfer is activated. {\it These characteristics provide the basis of the comparison depicted in Fig.~\ref{fig:cav_fields}, where the quality of the present results is evident with only some minor discrepancies present. } {\lb I do not understand...do we need to describe shortly and not so clearly the physics? If you want to give osme physical insight, first talk about the case without mass transfer and then with. You should say more clearly that the code gives the correct results?}Overall, the proposed methodology captures the solution of this compressible flow with and without mass transfer accurately.

%FFFFFFFFFFFFFFFFFFFFFFFFFFFFFFFFFFFFFF
%FFFFFFFFFFFFFFFFFFFFFFFFFFFFFFFFFFFFFF

\begin{figure}[t]
 \centering
\includegraphics[width=0.45\textwidth]{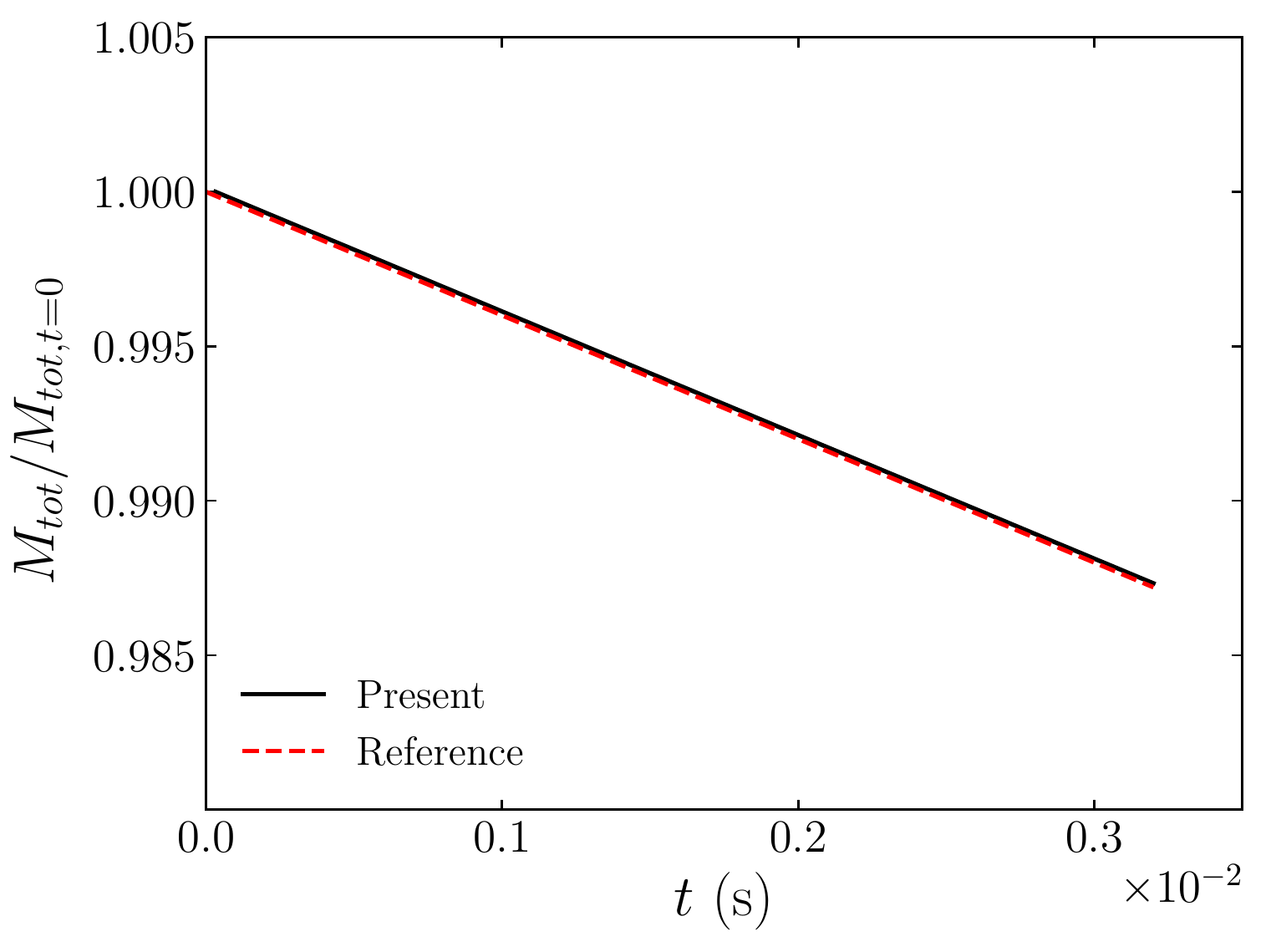}
 \includegraphics[width=0.45\textwidth]{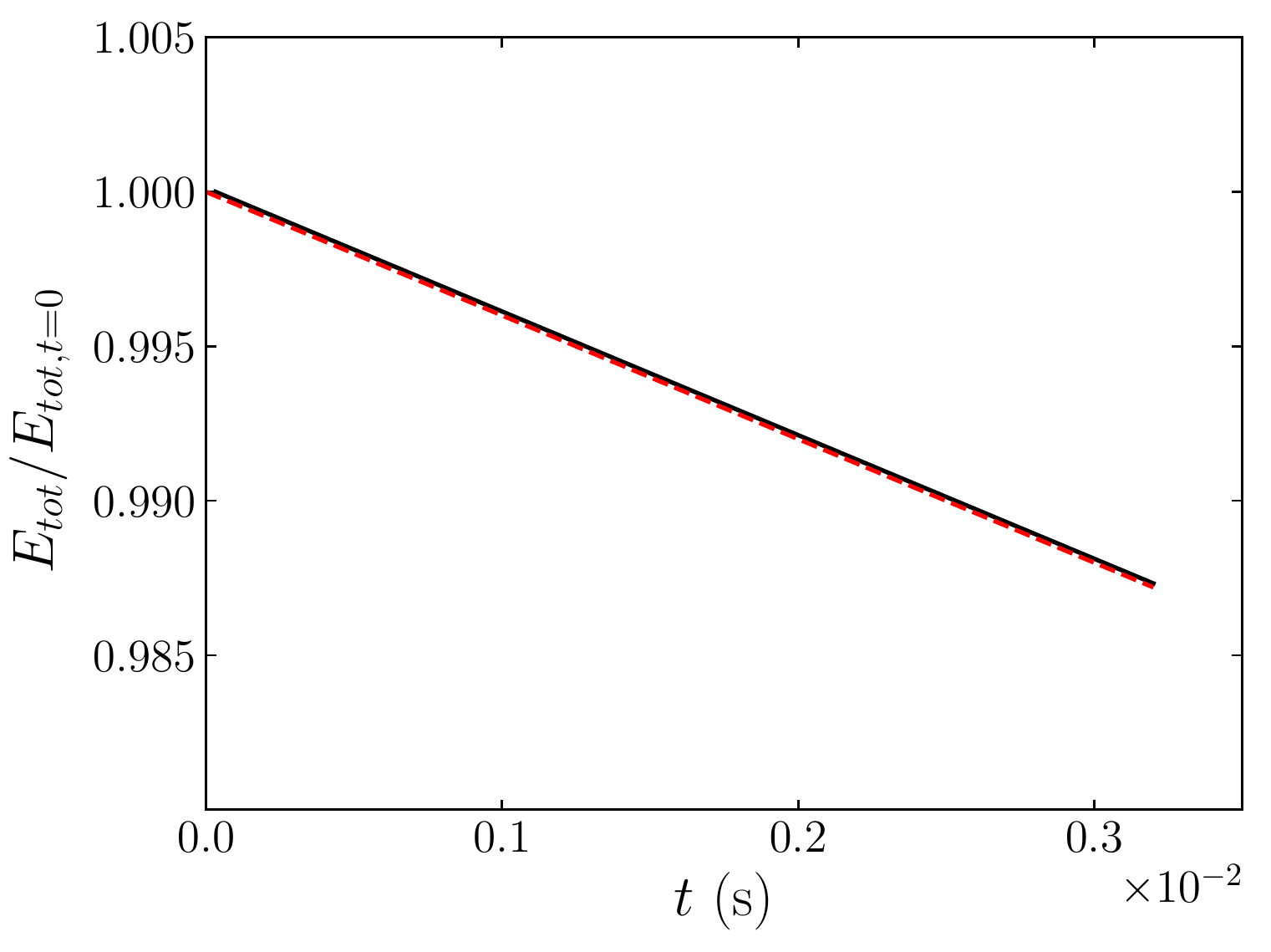}
 \caption{Temporal evolution of the total mass and total energy inside the tube for the water liquid-vapor expansion case with mass transfer. These quantities are divided by the corresponding initial values. Solid black lines correspond to the present results while dashed red lines correspond to the analytical solution described by Eqs.~\eqref{eq:tot_mass} and~\eqref{eq:tot_energy}. }
 \label{fig:cav_tot}
 \end{figure}

Since the equations for the conservation of mass and energy are not explicitly solved within the context of the proposed methodology, the correct calculation of these quantities needs further assessment. Assuming that the rarefactions do not reach the boundaries, the exact total mass $M_{tot}$ and total energy $E_{tot}$ in the domain as a function of time can be calculated as, 
\begin{eqnarray}
M_{tot}(t) &=& \left(1-2 u_0 t\right) \int_0^1 \rho_{t=0} dx , \label{eq:tot_mass} \\
E_{tot}(t) &=& \left(1-2 u_0 t\right) \int_0^1 \left(\mathcal{E} + \frac{1}{2}\rho u^2\right)_{t=0} dx \label{eq:tot_energy}.
\end{eqnarray}
The ratios between these quantities and their corresponding initial values are shown in Fig.~\ref{fig:cav_tot}, for the case where mass transfer is activated. An excellent agreement is observed between the computed and analytically derived quantities.
%, evidencing the ability of the methodology to conserve both mass and energy in the presence of mass transfer.
%FFFFFFFFFFFFFFFFFFFFFFFFFFFFFFFFFFFFFF
%FFFFFFFFFFFFFFFFFFFFFFFFFFFFFFFFFFFFFF

%=============================
\subsection{Water liquid-vapor filled tube with a superheated region}\label{sec:tube_boil}
%=============================

The test case presented in this section was specifically designed to verify the code in conditions that are more relevant to boiling flows. A 1.7m one-dimensional long tube is filled with a water mixture, with a liquid volume fraction of $a_1=0.9$ for $x<0.68$~m and $a_1=0.1$ for $x>0.68$~m. The tube is open at both ends and the mixture is subject to a pressure of $p=3.0104051\times10^6$~Pa. The gas-dominated region $x>0.68$~m is set at saturation conditions, while the liquid-dominated region is overheated by 2.0K. The EOS parameters adopted for this test case are listed in Table~\ref{table:tub_eos}. Using these parameters, the temperature for $x>0.68$~m is set at $T=T_{sat}=513.21628$~K, while for $x<0.68$~m is set at $T=515.21628$~K. 
Viscosity, thermal conductivity and surface tension are not considered.
% TTTTTTTTTTTTTTTTTTTTTTTTTTTTTTTTTTTTTT
\begin{table}[b]
\centering
\begin{tabular}{lcccccccc}
\hline
 & $\gamma$ & $\eta$ & $\tilde{\eta}$ & $p_{\infty}$ & $b$ & $\kappa_v$\\
\hline
%&- & - & - & - & $\mathrm{N/m}$ & $\mathrm{J/(kg\cdot K)}$ & $\mathrm{Pa}$ & $\mathrm{m/s^2}$ \\
%\hline
 liquid (1) & 1.3878826 & -1.244191$\times10^6$ & 0 & 8.899$\times 10^8$ & 4.78$\times 10^{-4}$ & 3202 \\
 vapor (2) & 1.9545455 & 2.287484$\times10^6$ & 6417 & 0 & 0 & 462 \\
\hline
\end{tabular}
\caption{EOS parameters adopted for the study of water liquid-vapor filled tube with a superheated region.}
\label{table:tub_eos}
\end{table}
%TTTTTTTTTTTTTTTTTTTTTTTTTTTTTTTTTTTTTT

To assess the accuracy of the results produced by the present methodology, the same test case was also simulated by using an established methodology for compressible two-phase flows with phase transition, documented in~\cite{pelanti2014mixture,de2019hyperbolic,pelanti2019numerical,pelanti2021arbitrary}. The reference methodology (formally second-order accurate) is based on a six-equation two-phase diffuse interface model, and it uses a HLLC-type Riemann solver for the homogeneous equations together with mechanical, and thermo-chemical relaxation procedures for inter-phase processes.  A grid of 1024 points and a constant time step of 6$\times 10^{-6}$~s was used to produce the present results, where the temperature and volume fraction discontinuities were approximated with a hyperbolic tangent function in the form of Eq.~\eqref{eq:tanh}. The reference results were obtained with a grid of 5000 points
and a varying time step with fixed Courant number equal to $0.5$, without considering any smoothing of the initial discontinuities. For this numerical test, chemical relaxation is activated everywhere in the domain.
%specific case, thermal relaxation was applied to all the domain, even in regions where the temperature %did not exceed the saturation value. {\color{red} (Why did we do that?)}.

\begin{figure}[ht!]
 \centering
 \includegraphics[width=0.42\textwidth]{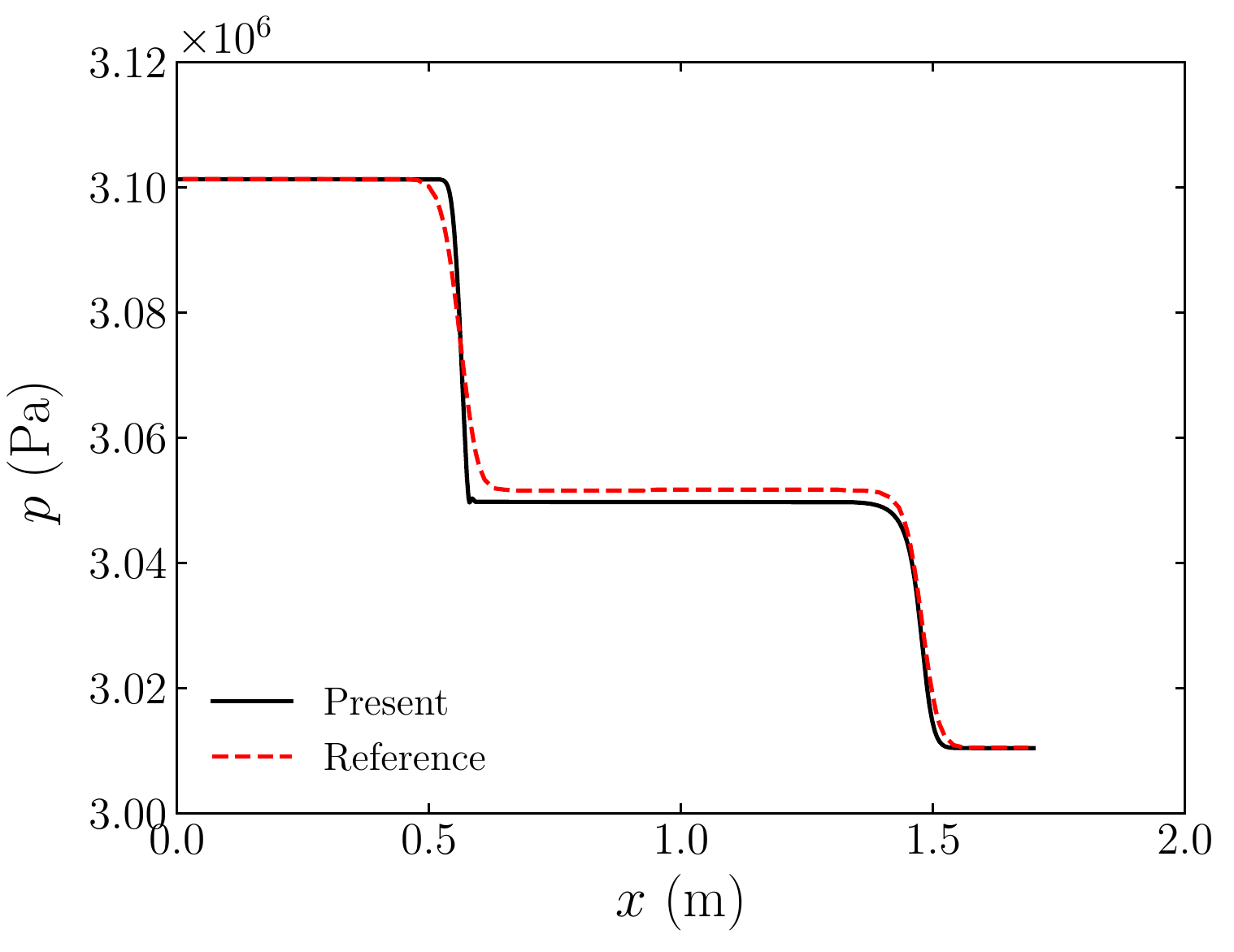}
 \includegraphics[width=0.42\textwidth]{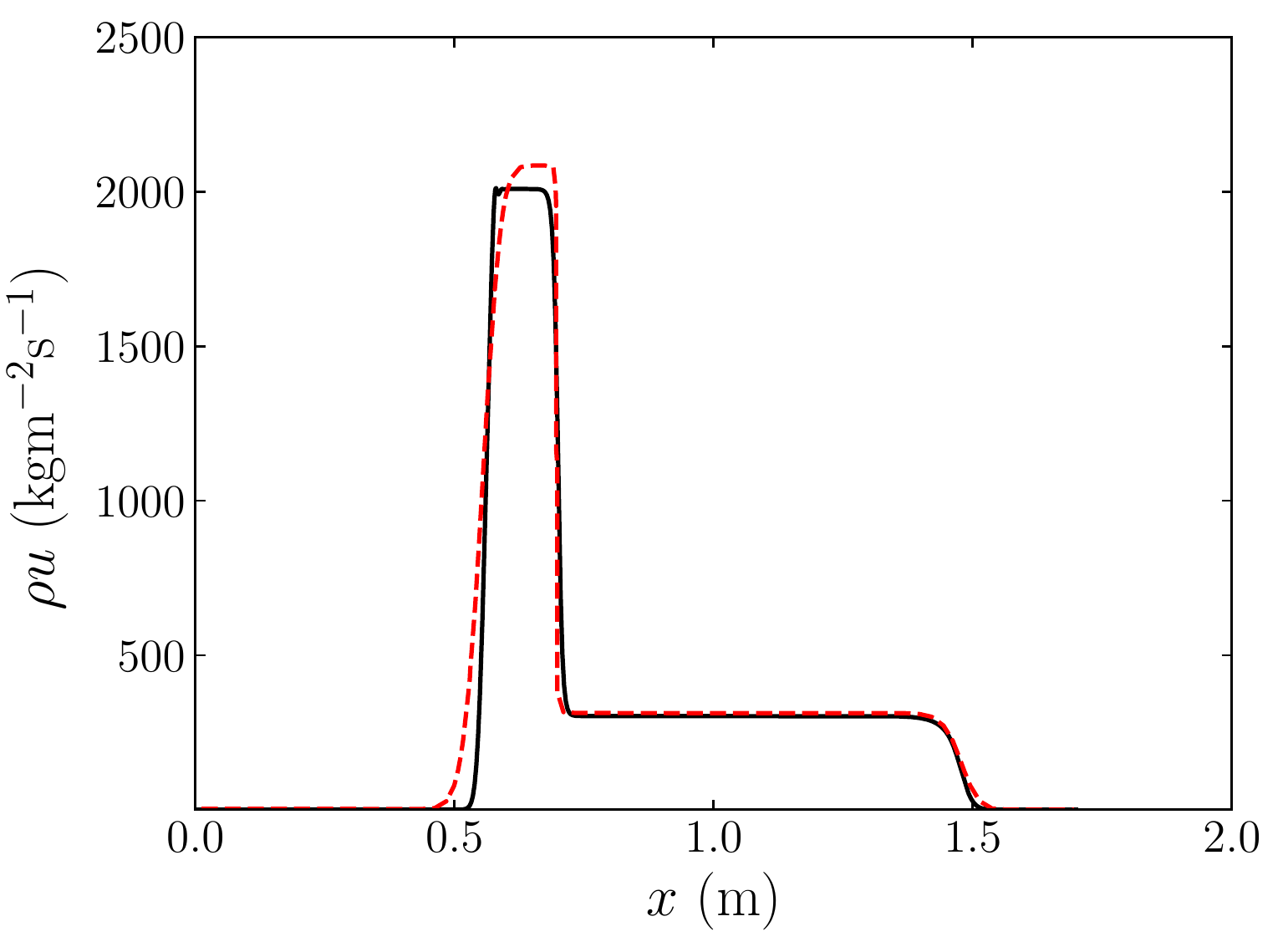}
 \includegraphics[width=0.42\textwidth]{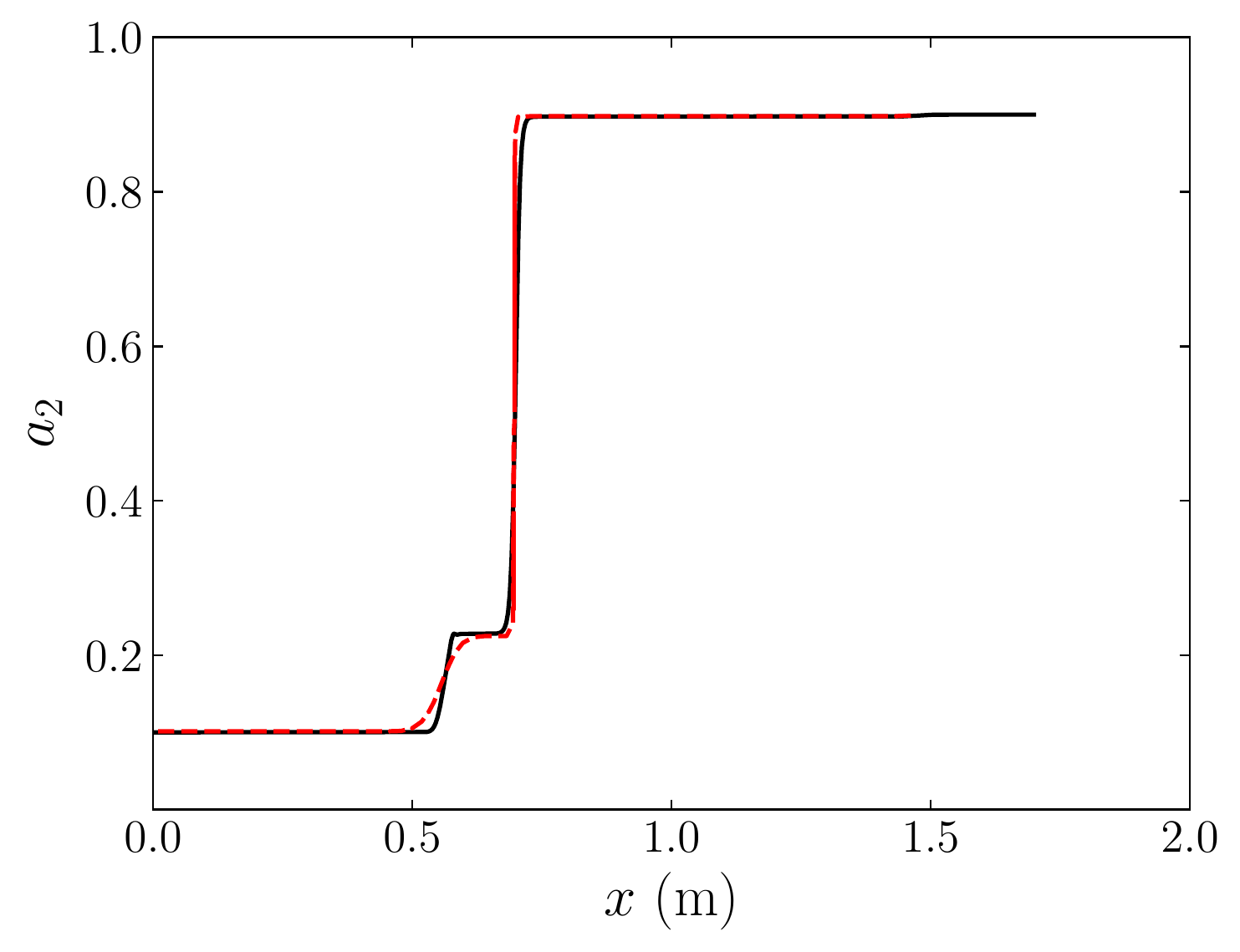}
 \includegraphics[width=0.42\textwidth]{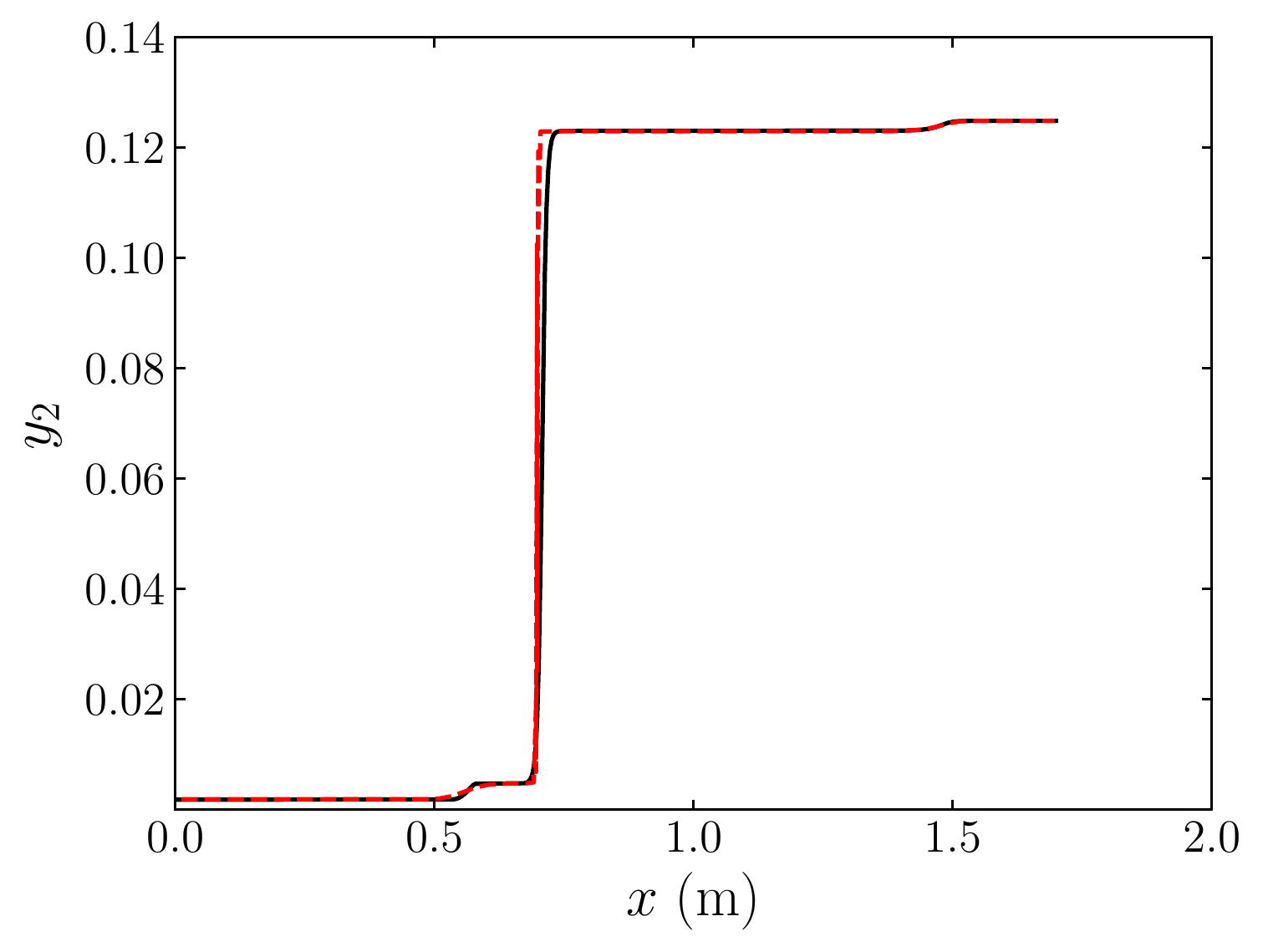}
 \includegraphics[width=0.42\textwidth]{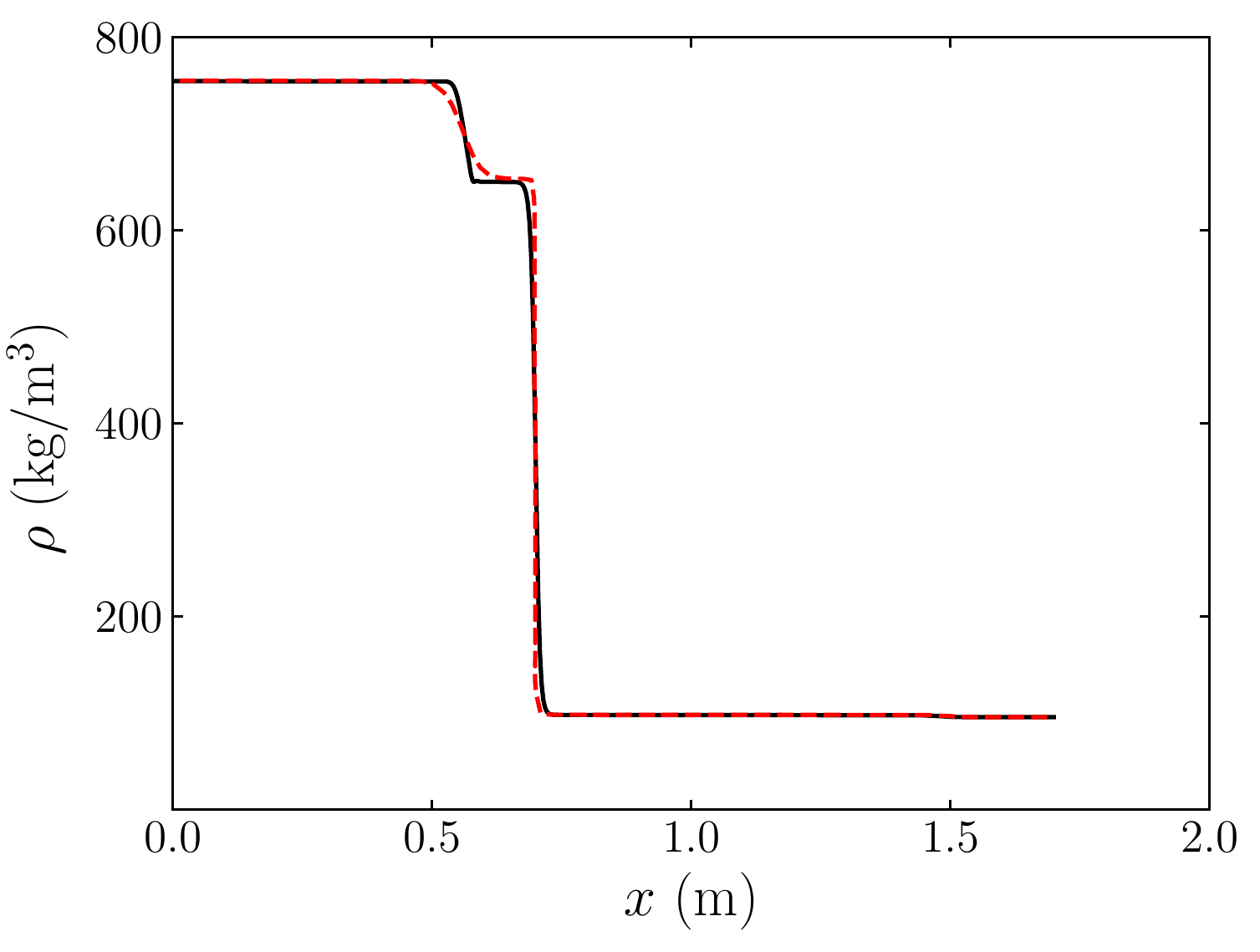}
 \includegraphics[width=0.42\textwidth]{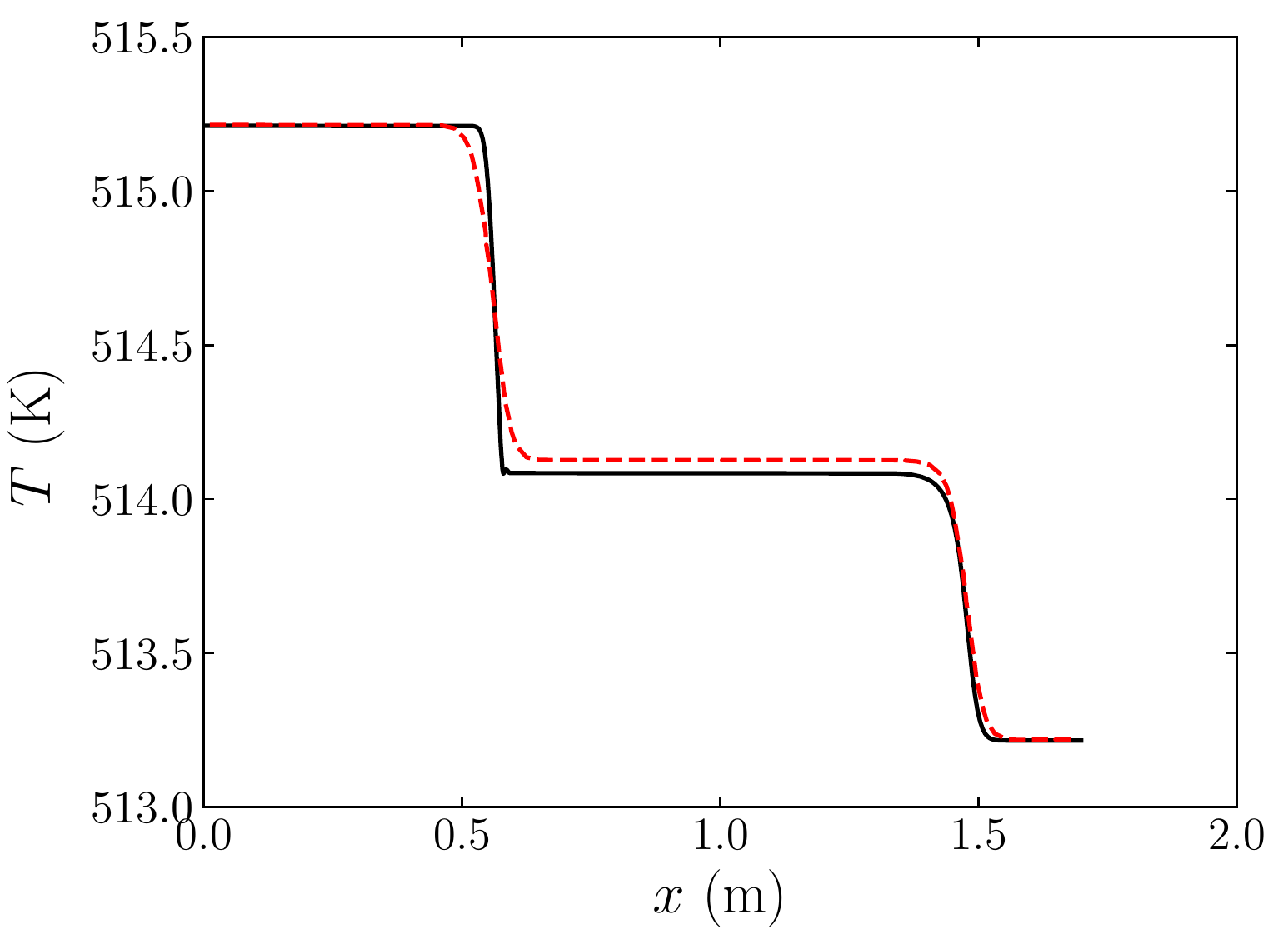}
 \caption{Solution of the water liquid-vapor filled tube with a superheated region at $t=0.006$s. Solid black lines correspond to the present results while dashed red lines correspond to the reference solution.}
 \label{fig:tube}
 \end{figure}

The comparison between the present and reference solution at $t=0.006$~s is shown in Fig.~\ref{fig:tube}. The pressure on the liquid side ($x<0.68$~m) almost instantaneously jumps to the saturation value. Therefore a pressure discontinuity is generated, giving rise to two waves that propagate to induce a thermodynamic equilibrium between the two regions. Overall, a good agreement is observed between the results of the two numerical methods.{\SecondRef The reference results display larger numerical diffusion, mainly due to the use of a dissipative HLLC-type Riemann solver compared to the Van-Leer flux limiter used in the present work.} Some numerical diffusion is also added due to the fact that the six-equation numerical model of~\cite{pelanti2014mixture} needs two additional relaxation steps (mechanical and thermal) in the fractional step algorithm after the solution of the homogeneous equations. The present method, therefore, provides a more accurate representation of the sharp variations, despite the smoothing applied to the initial temperature and volume fraction fields.

%FFFFFFFFFFFFFFFFFFFFFFFFFFFFFFFFFFFFFF
%FFFFFFFFFFFFFFFFFFFFFFFFFFFFFFFFFFFFFF
%
%===============================
\section{Three-dimensional nucleate boiling in water}\label{sec:demonstration}
\begin{figure}[t]
 \centering
 \includegraphics[width=0.45\textwidth]{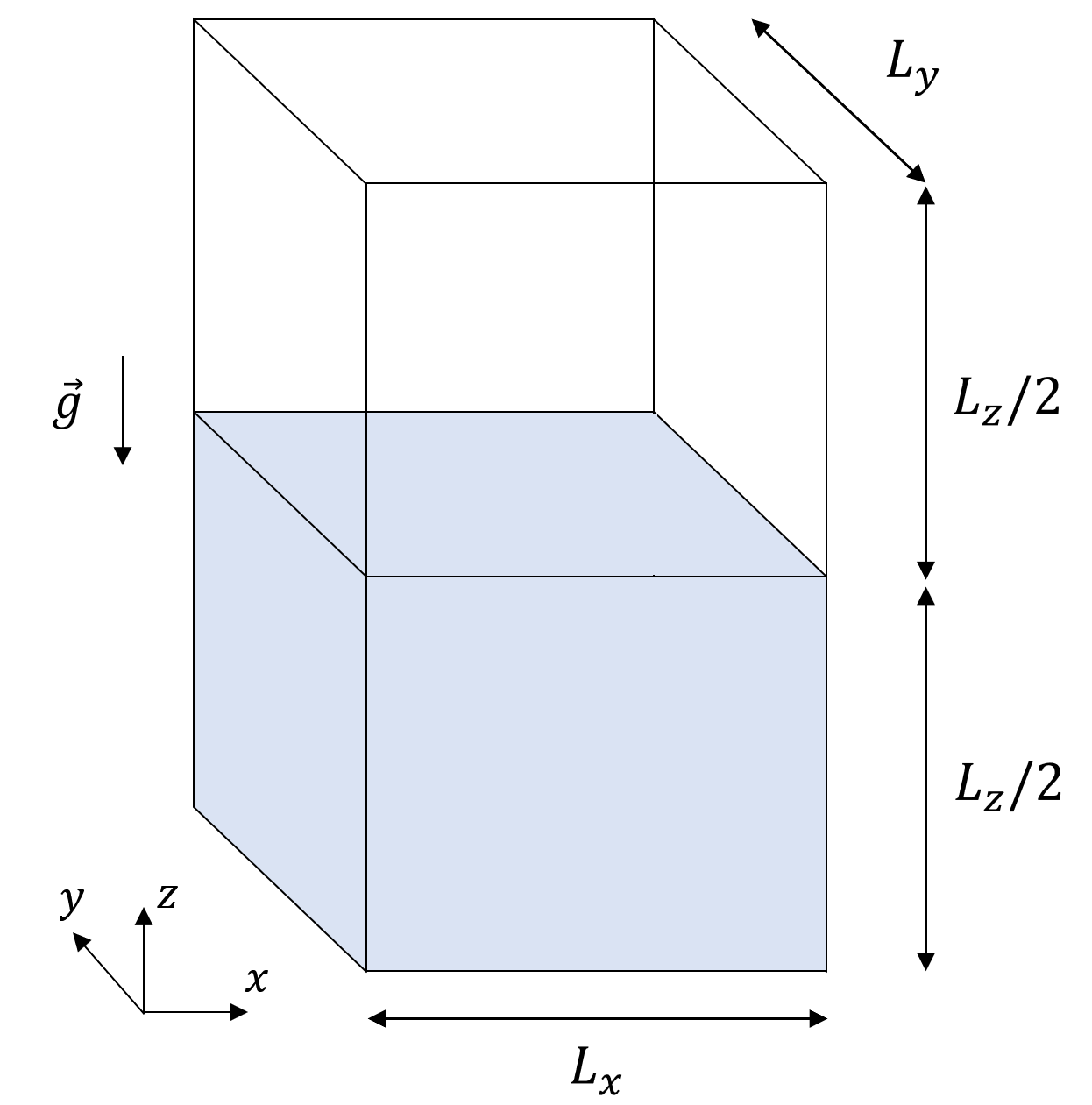}
 \caption{Schematic representation of the closed cuboid cavity used for the three-dimensional nucleate boiling in water. The dimensions of the cavity are $L_x\times L_y\times L_z=7\times 7\times 12$~cm, where the bottom half is filled with 99.99\% liquid and the upper half with 99.99\% vapor.}
 \label{fig:nucl_boil_geometry}
 \end{figure}

%===============================
To demonstrate the potential of the proposed methodology to simulate challenging boiling flows, a three-dimensional nucleate boiling simulation was carried out. The setup for this case was previously presented in~\cite{le2014towards}, albeit in two-dimensions. A closed cuboid box of dimensions $L_x\times L_y\times L_z=7\times 7\times 12$~cm is filled with a water liquid-vapor mixture, with $a_1=0.9999$ for $z<6$~cm and $a_1=0.0001$ for $z>6$~cm. A schematic representation of the domain is shown in Fig.~\ref{fig:nucl_boil_geometry}. All the domain boundaries are considered solid walls, therefore no-slip boundary conditions are applied for the velocity field. The vertical and the top walls are considered adiabatic ($\vec{\nabla}T|_w\cdot\hat{n}_w$=0), while a time-varying temperature is applied on the bottom wall, in the form, 
\begin{equation}
    T(t,x,y,z=0) = 
    \begin{cases}
      T_{sat} +\left(\frac{t}{0.15}\right)\Delta T &t\leq 0.15\mathrm{\ s}  \\
      T_{sat} +\Delta T &t > 0.15\mathrm{\ s},  
    \end{cases} 
 \end{equation}
where $T_{sat}=372.74$~K is the saturation temperature at $z=0$ and $\Delta T=15$~K. In this way, the bottom wall is slowly heated beyond the saturation temperature of the surrounding liquid, without causing too strong pressure waves. In addition to the three-dimensional representation, a more accurate EOS is used here in accordance to the NASG framework, as shown in Table~\ref{table:LeMart_eos}.

%FFFFFFFFFFFFFFFFFFFFFFFFFFFFFFFFFFFFFF
%FFFFFFFFFFFFFFFFFFFFFFFFFFFFFFFFFFFFFF

% TTTTTTTTTTTTTTTTTTTTTTTTTTTTTTTTTTTTTT
\begin{table}[b]
\centering
\begin{tabular}{lcccccccc}
\hline
 & $\gamma$ & $\eta$ & $\tilde{\eta}$ & $p_{\infty}$ & $b$ & $\kappa_v$\\
\hline
%&- & - & - & - & $\mathrm{N/m}$ & $\mathrm{J/(kg\cdot K)}$ & $\mathrm{Pa}$ & $\mathrm{m/s^2}$ \\
%\hline
 liquid (1) & 1.187 & -1.177788$\times10^6$ & 0 & 7.028$\times10^8$ & 6.61$\times 10^{-4}$ & 3610 \\
 vapor (2) & 1.467 & 2.077616$\times10^6$ & 1.4317$\times10^4$ & 0 & 0 & 955 \\
\hline
\end{tabular}
\caption{EOS parameters adopted for the three-dimensional study of nucleate boiling in water.}
\label{table:LeMart_eos}
\end{table}
%TTTTTTTTTTTTTTTTTTTTTTTTTTTTTTTTTTTTTT

The acceleration of gravity is set to $|\vec{g}|=9.81$~m~s$^{-2}$, the surface tension coefficient $\sigma=0.073$~N~m$^{-1}$ while the thermal conductivity in each phase  $\lambda_{c,1}=0.6788$~W~m$^{-1}$~K$^{-1}$ and $\lambda_{c,2}=0.0249$~W~m$^{-1}$~K$^{-1}$. Similarly to~\cite{le2014towards}, the effects of viscosity are not considered. Initially, the two-phase mixture is stagnant and the pressure in the cavity is  101325~Pa. The temperature is equal to the saturation value based on the local value of pressure. A uniform grid of $N_x\times N_y\times N_z = 256\times256\times512$ is used (approximately 33.6 million grid points), while the time step is dynamically adjusted in accordance to Eq.~\eqref{eqn:max_dt}, with $C_{\Delta t}=0.5$.

\begin{figure}[ht!]
 \centering
 \includegraphics[width=1.0\textwidth]{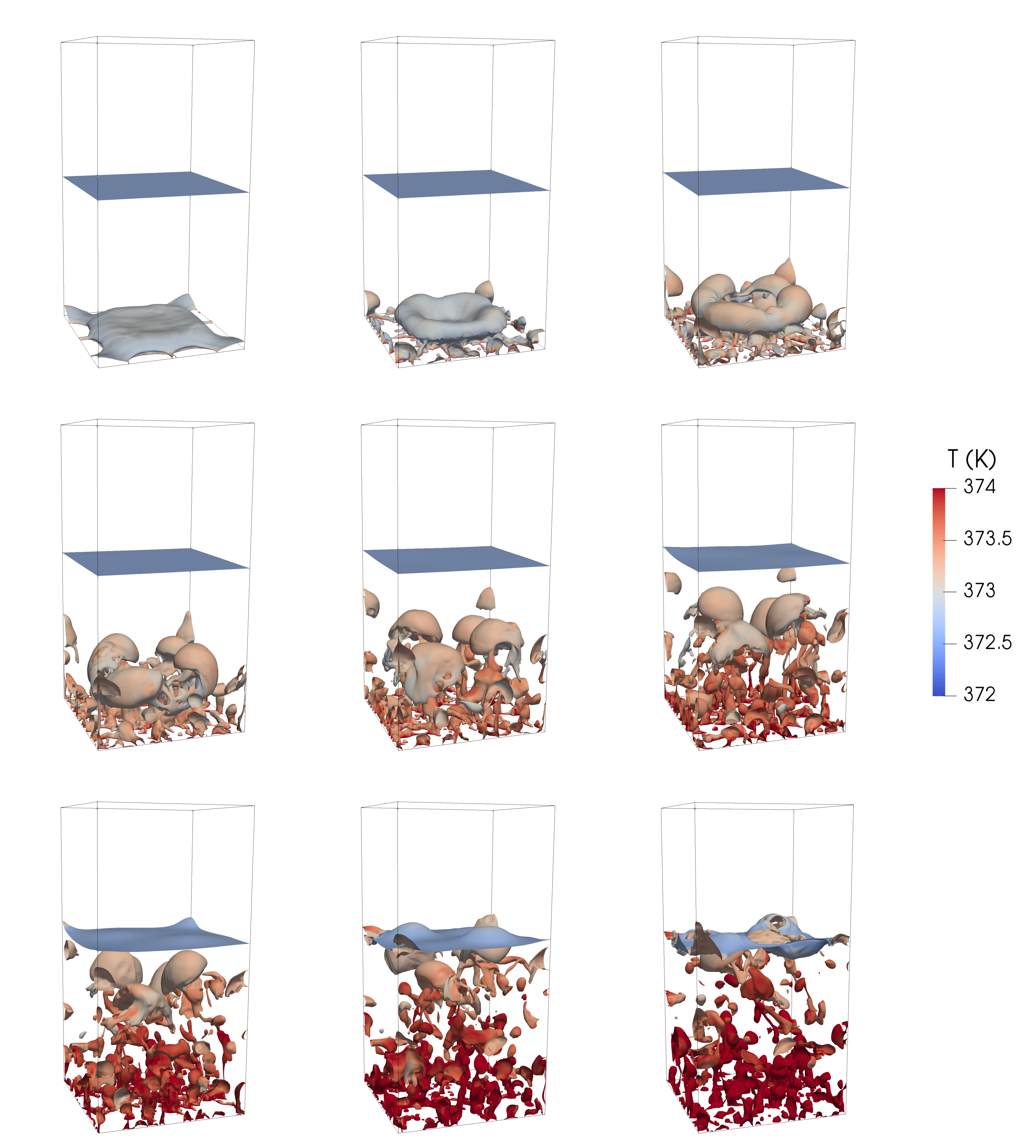}
 \caption{Instantaneous snapshots of $a_1=0.5$ iso-surfaces, coloured using the temperature field for the three-dimensional nucleate boiling in water. The first snapshot is at  $t \approx 0.262$~s, and each subsequent snapshot at intervals $\Delta t \approx 0.046$~s. }
 \label{fig:nucl_boil_inst}
 \end{figure}

At the very early stages of the flow development a vapor film starts forming at the bottom wall, which soon becomes unstable and breaks up. Instantaneous snapshots of $a_1=0.5$ iso-surfaces after the initial vapor film breakup are shown in Fig.~\ref{fig:nucl_boil_inst}. The film gives way to a torus-like structure centered around the vertical axis at the center of the cavity. In addition, bubbles are formed at the four bottom corners of the cavity and other smaller structures along the bottom edges of the cavity. As the flow develops further, the torus structure breaks up into four large bubbles along the $x-y$ diagonals. Eventually, bubbles of different sizes reach the interface and release their vapor content to the vapor-filled top half of the cavity. Overlooking the complexity of the three-dimensional structures, this phenomenological description is similar to what was reported in the reference two-dimensional study of~\cite{le2014towards}, where the initial vapor film breaks up in an elongated bubble at the center of the cavity and two smaller bubbles at the two bottom corners.

%FFFFFFFFFFFFFFFFFFFFFFFFFFFFFFFFFFFFFF
%FFFFFFFFFFFFFFFFFFFFFFFFFFFFFFFFFFFFFF

Even though an in-depth investigation of this specific nucleate boiling case is outside the scope of the present study, it is evident that the proposed method can provide reliable results for such a challenging physical phenomenon. Furthermore, the current implementation is efficient enough to solve problems on high-resolution numerical grids over long integration times. More specifically, the simulation presented in this section was carried out on 1024 processors on a system based on Intel Xeon Gold 6130 CPU's for ten days, consuming approximately 0.25 million core hours. 

%===============================
\section{Future improvements}\label{sec:future}
%===============================
In this section, a list of future modifications is compiled and motivated, with the potential to improve different aspects of the proposed methodology. The following improvements can help to enrich the physical description of the adopted model:
\begin{itemize}
    \item \underline{Wall treatment}: During nucleate boiling next to a heated surface, a number of physical mechanisms are responsible for transferring energy to the forming bubble. Depending on the conditions in which boiling takes place, energy can be transferred to a growing bubble through the micro-layer (thin liquid layer trapped between the bubble and the wall) and the three-phase contact line, amongst other mechanisms~\cite{kim2009review}. Both these mechanisms act on scales that are orders of magnitude smaller than the resolution requirements for the other physical mechanisms that affect the flow. Therefore, the appropriate modeling of these mechanisms (e.g.using the models presented in~\cite{cooper1969microlayer,stephan1992analysis,stephan1994new,sato2012new}) is very important for the accurate representation of flows involving bubble nucleation close to a wall.
    \item \underline{Realistic EOS}:
    Simple EOS are very convenient because they can be easily manipulated and included in the numerical method in an analytical form. On the other hand, realistic EOS for industrial applications such as the 
    IAPWS Industrial Formulation for Water and Steam \cite{iapws97} are very complex and their use presents significant challenges, both in terms of thermodynamical consistency of the numerical method and computational efficiency. For such complex equations of state the presented
    numerical model could be coupled to a table-look up method as proposed for instance in~\cite{delor-laf-pelanti-IJMF,de2021hyperbolic} to achieve fast and accurate thermodynamic calculations.
\end{itemize}

Furthermore, additional numerical techniques can be incorporated to improve the accuracy and efficiency of the computational methodology:
\begin{itemize}
    \item \underline{Managing interface thickness}: Even though the thermodynamic consistency is retained on the vapor-liquid interface, the interface typically becomes thicker with time in DI simulations.{\SecondRef This inherent drawback of diffuse interface methods hinders the detachment of small bubbles from larger structures, as discussed in Section~\ref{sec:rising_bubble}.} To treat this issue, two main approaches exist: (i) use of \textit{interface compression techniques}, where carefully constructed source terms (also called regularization terms) are introduced into the equations (e.g.~\cite{shukla2010interface,jain2020conservative}), and (ii) construct a sharper colour function from the more diffused mass or volume fractions and use this sharper function to calculate interface terms such as surface tension (e.g.~\cite{le2014towards}). Employing either approach can help retain the thickness of the interface at an appropriate size.
    \item \underline{Improved RK3 method}: The solution of the Helmholtz equation carries the biggest computational cost compared to the other algorithmic tasks. In its present form, the proposed methodology invokes the Helmholtz solver every RK3 sub-step, i.e. three times per time step. By attempting to extend the treatment of~\cite{le1991improvement} and~\cite{capuano2016approximate} to the pressure-based formulation of the present study, the Helmholtz problem will be solved  only once per time step, reducing the overall computational cost of the numerical solution significantly.
\end{itemize}

%===============================
\section{Conclusions}\label{sec:conclusions}
%===============================
In this study, a novel pressure-based methodology has been presented for the solution of a four-equation two-phase diffuse interface model, capable of solving 
%{\lb can we specify better here weakly? fully compressible flows at low speeds} 
low-Mach flows with mass transfer processes. The four-equation model results from the kinetic, mechanical and thermal relaxation of the general seven-equation Baer--Nunziato model, with the addition of extra terms to account for the effects of viscosity, surface tension, thermal conductivity and gravity. Mass transfer is modeled as a Gibbs free energy relaxation term.
%incorporated in the methodology using a Gibbs free energy relaxation procedure when the necessary %thermodynamic conditions are locally met.

The key characteristic that makes the proposed methodology stand out from the current state of the art is its pressure-based nature, which results in the solution of a Helmholtz equation for the pressure. This feature allows the utilisation of scalable and efficient solvers, able to exploit the full potential of high performance computing systems. With such capabilities, complex cases can be simulated with unprecedented resolution, giving a new insight into the underlying physical mechanisms.

The methodology was verified in a number of different cases, involving single- and two-phase configurations with large density ratios, under both compressible and incompressible conditions, with and without mass transfer. In addition to a very good agreement reached with relevant reference data, a second-order accurate solution was demonstrated in a range of Mach numbers. Moreover, the ability of the method to conserve mass and energy was demonstrated numerically in different scenarios.

Finally, the potential of the proposed methodology to simulate challenging  compressible two-phase flows with mass transfer was demonstrated with the three-dimensional nucleate boiling simulation in water. Initially, a vapor film was developed at the bottom heated wall, which in time broke up into a torus-like structure at the center and other smaller structures along the edges of the bottom wall. Eventually the torus structure generated four large bubbles, before releasing their vapor content to the top of the cavity.

%
%===============================
\section*{Acknowledgements}
%===============================
Andreas Demou, Nicol\`o Scapin and Luca Brandt acknowledge the support from the Swedish Research Council via the multidisciplinary research environment INTERFACE, Hybrid multiscale modelling of transport phenomena for energy efficient processes and the Grant No.\ 2016-06119. The computational resources were provided by SNIC (Swedish National Infrastructure for Computing) and by the National Infrastructure for High Performance Computing and Data Storage in Norway (project no. NN9561K). Marica Pelanti  was partially supported by the French Government Directorate for Armament (Direction G\'en\'erale de l'Armement, DGA) 
under grant N.~2018.60.0071.00.470.75.01. 
%
%===============================
\appendix
%===============================
\section{Derivation of the relaxed pressure and temperature equilibrium model}\label{BNrelax}
In the limit of instantaneous velocity, pressure, and temperature equilibrium, the 7--equation of Baer--Nunziato model ~\cite{baer1986two} is reduced to a 4--equation model. This relaxation procedure is presented here, following the asymptotic technique used by  Murrone and Guillard~\cite{murrone2005five} (see also \cite{chen1994hyperbolic}) to derive the 5--equation Kapila \textit{et al.}\ model \cite{kapila2001two} from
the 7--equation model. The notation followed is the same as the main body of the manuscript, except from symbols $\mu$, $\varphi$ and $\tau$ which are redefined in the context of this Appendix.
First, we write the Baer--Nunziato system in the variant of Saurel--Abgrall \cite{saurel1999multiphase} in terms of the vector of primitive variables
$w\in\mathbb{R}^{7}$ specified below as,
\begin{subequations}
\label{eq:sysrel}
\begin{equation}
\dts w + A(w)\dxs w = \frac{1}{\tau}\Psi(w) + \Phi(w), 
\end{equation} 
\begin{equation}
w=
\left[
\begin{array}{c}
\alpha_1\\
T_1\\
T_2\\
u_1\\
u_2\\
p_1\\
p_2
\end{array}
\right ],\,\,
A(w) =\left[
\begin{array}{ccccccc}
u_I & 0 & 0 & 0 & 0  & 0  & 0  \\
[1mm]
\textstyle -\left[\frac{\varphi_1}{C_{p1}}(p_1-p_I)-\frac{\Gamma_1T_1}{\alpha_1}\right] (u_1-u_I) 
& u_1 &  0 & \Gamma_1 T_1 & 0 & 0 & 0\\ 
[1mm]
\textstyle \left[\frac{\varphi_2}{C_{p2}}(p_2-p_I)-\frac{\Gamma_2T_2}{\alpha_2}\right] (u_2-u_I) 
& 0&  u_2 & 0 &\Gamma_2 T_2 & 0 & 0\\ 
[1mm]
\frac{p_1-p_I}{\alpha_1 \rho_1}  & 0 & 0 & u_1 & 0 & \frac{1}{\rho_1} & 0\\
[1mm]
-\frac{p_2-p_I}{\alpha_1 \rho_1}  & 0 & 0 & 0 & u_2 & 0& \frac{1}{\rho_2} \\
[1mm]
-\frac{\xi_1}{\alpha_1}(u_1-u_I) & 0 &  0 & \rho_1c_1^2 & 0 & u_1 &0\\
[1mm]
\frac{\xi_2}{\alpha_2}(u_2-u_I) & 0 &  0 & 0 & \rho_2 c_2^2 & 0 & u_2
\end{array}
\right],
\end{equation}
\begin{equation}
\label{eq:PsiPhi}
\Psi(w) = \left[
\begin{array}{c}
\tilde{\mu}(p_1-p_2)\\
[1mm]
\textstyle -\left[\frac{\varphi_1}{C_{p1}}(p_1-p_I)-\frac{\Gamma_1T_1}{\alpha_1}\right]\tilde{\mu}(p_1-p_2)
+\frac{\varphi_1}{C_{p1}}\tilde{\lambda}(u_I-u_1)(u_2-u_1)+\frac{\varphi_1}{C_{p1}}\tilde{\theta} (T_2-T_1)\\
[1mm]
\textstyle \left[\frac{\varphi_2}{C_{p2}}(p_2-p_I)-\frac{\Gamma_2T_2}{\alpha_2}\right]\tilde{\mu}(p_1-p_2)
-\frac{\varphi_2}{C_{p2}}\tilde{\lambda}(u_I-u_2)(u_2-u_1)-\frac{\varphi_2}{C_{p2}}\tilde{\theta} (T_2-T_1)\\
\frac{\tilde{\lambda}}{\alpha_1\rho_1}(u_2-u_1)\\
[1mm]
-\frac{\tilde{\lambda}}{\alpha_2\rho_2}(u_2-u_1)\\
[1mm]
\frac{\xi_1}{\alpha_1}\tilde{\mu}(p_1-p_2) +\frac{\Gamma_1}{\alpha_1}\tilde{\lambda}(u_I-u_1)(u_2-u_1)
+\frac{\Gamma_1}{\alpha_1}\tilde{\theta}(T_2-T_1)\\
[1mm]
-\frac{\xi_2}{\alpha_2}\tilde{\mu}(p_1-p_2) -\frac{\Gamma_2}{\alpha_2}\tilde{\lambda}(u_I-u_2)(u_2-u_1)
-\frac{\Gamma_1}{\alpha_1}\tilde{\theta}(T_2-T_1)
\end{array}
\right],
\end{equation}
where 
\begin{equation}
\varphi_k = 1+\Gamma_k^2\frac{\kappa_{pk}T_k}{c_k^2}=
1+\Gamma_k^2\frac{C_{pk}T_k}{\alpha_k \rho_k c_k^2}\,,\quad \xi_k = \Gamma_k (p_k-p_I)-\rho_k c_k^2\,.
\end{equation}
\end{subequations}
Here, $p_I$ and $u_I$ are the interface pressure and velocity. Vector $\Psi(w)$ contains all the relaxation source terms describing transfer processes that we consider in the limit of instantaneous equilibrium: the velocity relaxation term $\lambda(u_2-u_1)$, the pressure relaxation term $\mu(p_1-p_2)$, and the thermal relaxation term $\theta(T_2-T_1)$. The relaxation parameters are redefined as $\lambda=\frac{\tilde{\lambda}}{\tau}$, $\mu=\frac{\tilde{\mu}}{\tau}$ and $\theta=\frac{\tilde{\theta}}{\tau}$. Vector $\Phi(w)$ contains all the other source terms (e.g.\ gravity). The focus is on the behavior of the solutions of Eq.~\eqref{eq:sysrel} in the limit $\tau \rightarrow 0^+$. It is expected that these solutions are close to the set $\mathfrak{U} = \{w\in \mathbb{R}^{7};\Psi(w)=0\}$. Furthermore, it is assumed that the set of equations $\Psi(w)=0$ defines a smooth manifold of dimension $4$ and that for any $ w\in \mathfrak{U}$, a parameterization $\Xi$ (the Maxwellian) is known from an open subset $\Omega$ of $\mathbb{R}^4$ on a neighborhood of $w$ in $\mathfrak{U}$. For any $v \in \Omega \subset \mathbb{R}^4$ the Jacobian matrix $d\Xi_{v}$ is a full rank matrix, moreover, the column vectors of  $d\Xi_{v}$ form a basis of $\ker (\Psi'(\Xi(v)))$~\cite{murrone2005five}. 

Based on the above, matrix $C\in \mathbb{R}^{7 \times 7}$ can now be defined as, 
\begin{equation}
\label{eq:Cmatr}
C = [d\Xi_v^1\ldots d\Xi_v^4\,V^1 \, V^2 \, V^{3}] ,
\end{equation}
where $d\Xi_v^1,\ldots, d\Xi_v^4$ are the column vectors of $d\Xi_{v}$ and $\{V^1,V^2,V^{3}\}$
is a basis of the range of $\Psi'(\Xi(v))$. Based on the observations above, the matrix $C$ is invertible. Another $4 \times 7$ matrix $P$ can be defined, comprising of the first $4$ rows of the inverse $C^{-1}$. With the use of matrix $P$ the following results can be obtained (see~\cite{murrone2005five}),
\begin{equation}
\label{eq:ident}
P \,d\Xi_v =\mathbb{I}_{4} \qeq P \, \Psi'(\Xi(v)) = 0,
\end{equation}
where $\mathbb{I}_4$ denotes the $4\times 4$ identity matrix.
In order to obtain a reduced velocity, pressure and temperature equilibrium model, solutions in the form $w=\Xi(v)+\tau z$ are pursued, where $z$ is a small perturbation around the equilibrium state $\Xi(v)$. Using this form into the system \eqref{eq:sysrel} one obtains,
\begin{equation}
\dts (\Xi(v)) +A(\Xi(v))\dxs(\Xi(v)) -\Psi'(\Xi(v)) \,z=\Phi(\Xi(v)) +\mathcal{O}(\tau).
\end{equation}
By multiplying the above equation by $P$, using \eqref{eq:ident}, and neglecting terms of order $\tau$, the following reduced model system is obtained,
\begin{equation}
\label{eq:redsys}
\dts v + P A(\Xi(v))d\Xi_v\dxs v= P \Phi(\Xi(v)),
\end{equation} 
where $v \in \mathbb{R}^4$ and $A_r(v)\equiv P A(\Xi(v))d\Xi_v \in \mathbb{R}^{4\times4}$.
In the limit of instantaneous velocity, pressure and temperature relaxation, $u_1=u_2=u_I=u$, $p_1=p_2=p_I=p$, $T_1=T_2=T$, the vector of the variables of the reduced pressure-relaxed model can be expressed as,
\begin{equation}
v = [\alpha_1,T,u,p]\trasp \in \mathbb{R}^{4}.
\end{equation}   
The equilibrium state $\Xi(v)$ is defined by,
\begin{equation}
\label{eq:eqmaxw}
\Xi: v\rightarrow \Xi(v) = [\alpha_1,T, T,u,u,p,p]\trasp
\in \mathbb{R}^{7}.
\end{equation}  
The Jacobian $d\Xi_v \in \mathbb{R}^{7\times 7}$ of the Maxwellian is expressed as,
\begin{equation}
d\Xi_v = \left[
\begin{array}{cccc}
1 & 0 & 0 &0 \\
0 & 1 & 0 &0\\
0  & 1 & 0 &0\\
0  & 0 & 1 &0\\
0  & 0 & 1 &0\\
0  & 0 & 0 &1\\
0  & 0 & 0 &1
\end{array}
\right].
\end{equation}
A basis $\{V^1,V^2,V^3\}$, $V^k \in {R}^{7}$, $k=1,2,3$, for the range of 
$\Psi'(\Xi(v)) \in \mathbb{R}^{7\times 7}$ is found as,
\begin{equation}
V^1 = \left[
\begin{array}{c}
0\\
[1mm]
\frac{\varphi_1}{C_{p1}} \\
[1mm]
-\frac{\varphi_2}{C_{p2}}\\
[1mm]
0\\
[1mm]
0\\
[1mm]
\frac{\Gamma_1}{\alpha_1}\\
[1mm]
-\frac{\Gamma_2}{\alpha_2}
\end{array}
\right], \,\,
V^2 = \left[
\begin{array}{c}
0\\
[1mm]
0 \\
[1mm]
0\\
[1mm]
\frac{1}{\alpha_1\rho_1}\\
[1mm]
-\frac{1}{\alpha_2\rho_2}\\
[1mm]
0\\
[1mm]
0
\end{array}
\right], \,\,
V^{3} = \left[
\begin{array}{c}
1\\
[1mm]
-\frac{\Gamma_1 T_1}{\alpha_1} \\
[1mm]
\frac{\Gamma_2 T_2}{\alpha_2}\\
[1mm]
0\\
[1mm]
0\\
[1mm]
\frac{\xi_1}{\alpha_1}\\
[1mm]
-\frac{\xi_2}{\alpha_2}
\end{array}
\right].
\end{equation}
The matrix $C \in \mathbb{R}^{7\times 7}$ \eqref{eq:Cmatr} can then be constructed, inverted, and matrix $P \in \mathbb{R}^{4\times7}$ can be obtained by taking the first $4$ rows of $C^{-1}$. 
Finally, the reduced limit 4--equation model is obtained from \eqref{eq:redsys} by calculating the new matrix $A_r(v) \equiv PA(\Xi(v))d\Xi_v \in \mathbb{R}^{4\times4}$
and the new source term $\Phi_r(v) = P\Phi(\Xi(v)) \in \mathbb{R}^4$. Note that $A(w)$ and $\Phi(w)$ are evaluated in the equilibrium state $\Xi(v)$ in \eqref{eq:eqmaxw}. Following this procedure, the reduced limit 4--equation model takes the form (here assuming $\Phi=0$),
\begin{subequations}
\begin{eqnarray}
\partial_t a_1 + \vec{\nabla} \cdot \left( a_1 \vec{u} \right) + \left(S_{a}^{(3)}-a_1\right) \vec{\nabla}\cdot \vec{u} &=& 0, \\
\partial_t T + \vec{\nabla} \cdot \left( T \vec{u} \right) + \left(S_{T}^{(3)}-T\right) \vec{\nabla}\cdot \vec{u} &=& 0, \\
\partial_t \left(\rho \vec{u}\right) +  \vec{\nabla}\cdot \left(\rho \vec{u}\otimes \vec{u}\right) +  \vec{\nabla}p  &=& 0, \\
\partial_t p+ \vec{u} \cdot \vec{\nabla} p +\rho c^2 \vec{\nabla}\cdot \vec{u} &=& 0,
\end{eqnarray}
\end{subequations}
where,
\begin{subequations}
\begin{eqnarray}
S_{a}^{(3)} &=& 
\rho c^2 \left[\left(\frac{a_1 a_2}{\rho_2 c_2^2}-\frac{a_1 a_2}{\rho_1 c_1^2}\right)+\frac{T C_{p1}C_{p2}}{C_{p1}+C_{p2}}\left(\frac{\Gamma_2}{\rho_2 c_2^2}-\frac{\Gamma_1}{\rho_1 c_1^2}\right)\left(\frac{a_1\Gamma_2}{\rho_2 c_2^2}+\frac{a_2\Gamma_1}{\rho_1 c_1^2}\right)\right], \\
S_{T}^{(3)} &=& \frac{\rho c^2 T}{C_{p1}+C_{p2}}\left(\frac{C_{p1}\Gamma_1}{\rho_1 c_1^2}+\frac{C_{p2}\Gamma_2}{\rho_2 c_2^2}\right), \\
\dfrac{1}{c^2}&=& \rho\left(\dfrac{a_1}{\rho_1 c_1^2}+\dfrac{a_2}{\rho_2 c_2^2}\right)+\dfrac{\rho T C_{p1}C_{p2}}{C_{p1}+C_{p2}}\left( \dfrac{\Gamma_2}{\rho_2 c_2^2}-\dfrac{\Gamma_1}{\rho_1 c_1^2} \right)^2.
\end{eqnarray}
\end{subequations}

%=============================================
%=============================================
%=============================================
\section{Source terms for the enriched physical description}\label{extraPhysics}
%=============================================
%=============================================
%=============================================
The effects of viscosity, thermal conductivity and mass transfer are described by source terms 
in the governing equations. The formulation of the four-equation model with
this enriched physical description in terms of the conserved variables
is given by Eq. (\ref{eq:sysTboilc}). Here we show how to derive from (\ref{eq:sysTboilc}) the expressions the source terms in the formulation (\ref{eq:sysprim}). Based on (\ref{eq:sysTboilc}), we can write: 
\begin{eqnarray}
\partial_t(a_1\rho_1)+\vec{\nabla}\cdot(a_1\rho_1) &=& M \\
\partial_t(a_2\rho_2)+\vec{\nabla}\cdot(a_2\rho_2) &=& -M \\
\partial_t(\mathcal{E})+\vec{\nabla}\cdot(\mathcal{E}\vec{u}) +p\vec{\nabla}\cdot\vec{u} &=& D_{\mathcal{E}}+K,
\end{eqnarray}
where $M$, $D_{\mathcal{E}}$, $K$ are defined in Eqs.~\eqref{eq:M_define},~\eqref{eq:DE_define} and~\eqref{eq:K_define} respectively. The corresponding source terms in the volume fraction, temperature and pressure equations, are determined after the transformation from equations for $\tilde{q}$ to $w$, defined as,
\begin{equation}
    \tilde{q}= \left[
        \begin{array}{c} 
            a_1 \rho_1 \\
            a_2 \rho_2 \\
            \mathcal{E}
        \end{array} \right],
\mathrm{\hspace{0.5cm} and \hspace{0.5cm}}
     w= \left[
        \begin{array}{c} 
            a_1  \\
            T \\
            p
\end{array} \right],
\end{equation}
with a transformation matrix,
\begin{equation}
    \frac{\partial \tilde{q}}{\partial w}= \left[
        \begin{array}{ccc} 
            \rho_1 & a_1\phi_1 & a_1\zeta_1\\
           -\rho_2 & a_2\phi_2 & a_2\zeta_2\\
            \mathcal{E}_1-\mathcal{E}_2 & C_{\mathcal{E}p} & C_{\mathcal{E}T}
        \end{array} \right],
\end{equation}
where $\phi_k$ and $\zeta_k$ are defined in Eqs.~\eqref{eq:phi} and~\eqref{eq:zeta} respectively and,
\begin{eqnarray}
C_{\mathcal{E}p}&=&a_1\left(\frac{\partial \mathcal{E}_1}{\partial T}\right)_p + a_2\left(\frac{\partial \mathcal{E}_2}{\partial T}\right)_p ,\\
C_{\mathcal{E}T}&=&a_1\left(\frac{\partial \mathcal{E}_1}{\partial p}\right)_T + a_2\left(\frac{\partial \mathcal{E}_2}{\partial p}\right)_T.
\end{eqnarray}
Therefore, the source terms of $w$ (RHS$(w)$) are determined as,
\begin{equation}
    \mathrm{RHS}(w)=\left(\frac{d\tilde{q}}{dw} \right)^{-1}\left[
    \begin{array}{c} 
            M  \\
            -M \\
            D_{\mathcal{E}}+K
    \end{array}
                \right]=\left[
    \begin{array}{c} 
            S_a^{(1)}M+ S_a^{(2)}(D_{\mathcal{E}}+K) \\
            S_T^{(1)}M+ S_T^{(2)}(D_{\mathcal{E}}+K) \\
            S_p^{(1)}M+ S_p^{(2)}(D_{\mathcal{E}}+K)
    \end{array}
                \right].
\end{equation}
Taking into account the expressions for the derivatives of $\mathcal{E}_k$,
\begin{equation}
    \left(\frac{\partial \mathcal{E}_k}{\partial T} \right)_p = -\frac{\chi_k}{\Gamma_k}\phi_k, \mathrm{\hspace{0.5cm} and \hspace{0.5cm}}
    \left(\frac{\partial \mathcal{E}_k}{\partial p} \right)_T = \frac{1}{\Gamma_k}(1-\chi_k\zeta_k),
\end{equation}
along with $\mathcal{E}_1-\mathcal{E}_2=\rho_1 h_1-\rho_2 h_2$ and $h_k=(c_k^2-\chi_k)/\Gamma_k$, the coefficients of the source terms are obtained as,
\begin{subequations}
\begin{eqnarray}
S_{a}^{(1)} &=&\frac{1}{\bar{D}}\left[
\left(\frac{\chi_1}{\Gamma_1}-\frac{\chi_2}{\Gamma_2}\right)(\phi\zeta)^T+\left(\frac{a_1}{\Gamma_1}+\frac{a_2}{\Gamma_2}\right)\phi_v
\right], \\ 
S_{T}^{(1)} &=&\frac{1}{\bar{D}}\left[
\left(\frac{\chi_2}{\Gamma_2}-\frac{\chi_1}{\Gamma_1}\right)(\zeta\rho)^T+\left(\frac{\rho_1c_1^2}{\Gamma_1}-\frac{\rho_2c_2^2}{\Gamma_2}\right)\zeta_v+\left(\frac{a_1}{\Gamma_1}+\frac{a_2}{\Gamma_2}\right)\Delta\rho
\right], \\ 
S_{p}^{(1)} &=&\frac{1}{\bar{D}}\left[\left(\frac{\chi_1}{\Gamma_1}-\frac{\chi_2}{\Gamma_2}\right)(\phi\rho)^T +
\left(\frac{\rho_2 c^{2}_2}{\Gamma_2}-\frac{\rho_1 c^{2}_1}{\Gamma_1}\right)\phi_v\right],  \\
S_{a}^{(2)}&=&\frac{1}{\bar{D}}(\phi \zeta)^T, \\
S_{T}^{(2)}&=&-\frac{1}{\bar{D}}(\zeta\rho)^T, \\
S_{p}^{(2)}&=&\frac{1}{\bar{D}}(\phi\rho)^T,
\end{eqnarray}
\end{subequations}
where, 
\begin{subequations}
\begin{eqnarray}
(\phi\rho)^T&=& a_1\phi_1\rho_2+a_2\phi_2\rho_1\mathrm{,} \hspace{0.5 cm} \phi_v= a_1\phi_1+a_2\phi_2\mathrm{,} \\
(\zeta\rho)^T &=& a_1\zeta_1\rho_2+a_2\zeta_2\rho_1\mathrm{,} \hspace{0.65 cm} \zeta_v = a_1\zeta_1+a_2\zeta_2\mathrm{,}\\
(\phi \zeta)^T&=&a_1 a_2(\phi_1\zeta_2-\phi_2\zeta_1)\mathrm{,} \hspace{0.25 cm} \Delta\rho = \rho_2-\rho_1\mathrm{,} \\
\bar{D}&=& \left(\frac{\rho_1 c_1^2}{\Gamma_1}-\frac{\rho_2 c_2^2 }{\Gamma_2}\right)(\phi\zeta)^T+\left(\frac{a_1}{\Gamma_1}+\frac{a_2}{\Gamma_2}\right)(\phi\rho)^T\mathrm{,}\\
\chi_k&=& \left(\frac{\partial P}{\partial \rho_k}\right)_{\mathcal{E}_k}= c^2_k-\Gamma_k h_k\mathrm{.}
\end{eqnarray}
\end{subequations}
Note that the same results could be obtained by writing the source appropriate term
$\Phi$ in the Baer--Nunziato equations before the asymptotic procedure described in the previous Appendix.
\bibliography{main}

\begin{thebibliography}{95}
\expandafter\ifx\csname natexlab\endcsname\relax\def\natexlab#1{#1}\fi
\providecommand{\url}[1]{\texttt{#1}}
\providecommand{\href}[2]{#2}
\providecommand{\path}[1]{#1}
\providecommand{\DOIprefix}{doi:}
\providecommand{\ArXivprefix}{arXiv:}
\providecommand{\URLprefix}{URL: }
\providecommand{\Pubmedprefix}{pmid:}
\providecommand{\doi}[1]{\href{http://dx.doi.org/#1}{\path{#1}}}
\providecommand{\Pubmed}[1]{\href{pmid:#1}{\path{#1}}}
\providecommand{\bibinfo}[2]{#2}
\ifx\xfnm\relax \def\xfnm[#1]{\unskip,\space#1}\fi
%Type = Inproceedings
\bibitem[{Van~Doormaal et~al.(1986)Van~Doormaal, Raithby, and
  McDonald}]{van1986segregated}
\bibinfo{author}{J.~Van~Doormaal}, \bibinfo{author}{G.~Raithby},
  \bibinfo{author}{B.~McDonald},
\newblock \bibinfo{title}{The segregated approach to predicting viscous
  compressible fluid flows},
\newblock in: \bibinfo{booktitle}{ASME 1986 International Gas Turbine
  Conference and Exhibit}, \bibinfo{organization}{American Society of
  Mechanical Engineers Digital Collection}, \bibinfo{year}{1986}, pp.
  \bibinfo{pages}{268--277}.
%Type = Article
\bibitem[{Dhir(1998)}]{dhir1998boiling}
\bibinfo{author}{V.~Dhir},
\newblock \bibinfo{title}{Boiling heat transfer},
\newblock \bibinfo{journal}{Annual review of fluid mechanics}
  \bibinfo{volume}{30} (\bibinfo{year}{1998}) \bibinfo{pages}{365--401}.
%Type = Article
\bibitem[{Zhao et~al.(2003)Zhao, Guo, Bai, Hou, and Zhang}]{zhao2003convective}
\bibinfo{author}{L.~Zhao}, \bibinfo{author}{L.~Guo}, \bibinfo{author}{B.~Bai},
  \bibinfo{author}{Y.~Hou}, \bibinfo{author}{X.~Zhang},
\newblock \bibinfo{title}{Convective boiling heat transfer and two-phase flow
  characteristics inside a small horizontal helically coiled tubing
  once-through steam generator},
\newblock \bibinfo{journal}{International journal of heat and mass transfer}
  \bibinfo{volume}{46} (\bibinfo{year}{2003}) \bibinfo{pages}{4779--4788}.
%Type = Article
\bibitem[{Amalfi et~al.(2016)Amalfi, Vakili-Farahani, and
  Thome}]{amalfi2016flow}
\bibinfo{author}{R.~L. Amalfi}, \bibinfo{author}{F.~Vakili-Farahani},
  \bibinfo{author}{J.~R. Thome},
\newblock \bibinfo{title}{Flow boiling and frictional pressure gradients in
  plate heat exchangers. part 1: Review and experimental database},
\newblock \bibinfo{journal}{International Journal of Refrigeration}
  \bibinfo{volume}{61} (\bibinfo{year}{2016}) \bibinfo{pages}{166--184}.
%Type = Article
\bibitem[{Narumanchi et~al.(2008)Narumanchi, Troshko, Bharathan, and
  Hassani}]{narumanchi2008numerical}
\bibinfo{author}{S.~Narumanchi}, \bibinfo{author}{A.~Troshko},
  \bibinfo{author}{D.~Bharathan}, \bibinfo{author}{V.~Hassani},
\newblock \bibinfo{title}{Numerical simulations of nucleate boiling in
  impinging jets: Applications in power electronics cooling},
\newblock \bibinfo{journal}{International Journal of Heat and Mass Transfer}
  \bibinfo{volume}{51} (\bibinfo{year}{2008}) \bibinfo{pages}{1--12}.
%Type = Article
\bibitem[{Saurel et~al.(2016)Saurel, Boivin, and
  Le~M{\'e}tayer}]{saurel2016general}
\bibinfo{author}{R.~Saurel}, \bibinfo{author}{P.~Boivin},
  \bibinfo{author}{O.~Le~M{\'e}tayer},
\newblock \bibinfo{title}{A general formulation for cavitating, boiling and
  evaporating flows},
\newblock \bibinfo{journal}{Computers \& Fluids} \bibinfo{volume}{128}
  (\bibinfo{year}{2016}) \bibinfo{pages}{53--64}.
%Type = Book
\bibitem[{Tryggvason et~al.(2011)Tryggvason, Scardovelli, and
  Zaleski}]{tryggvason2011direct}
\bibinfo{author}{G.~Tryggvason}, \bibinfo{author}{R.~Scardovelli},
  \bibinfo{author}{S.~Zaleski}, \bibinfo{title}{Direct numerical simulations of
  gas--liquid multiphase flows}, \bibinfo{publisher}{Cambridge University
  Press}, \bibinfo{year}{2011}.
%Type = Article
\bibitem[{Unverdi and Tryggvason(1992)}]{unverdi1992front}
\bibinfo{author}{S.~O. Unverdi}, \bibinfo{author}{G.~Tryggvason},
\newblock \bibinfo{title}{A front-tracking method for viscous, incompressible,
  multi-fluid flows},
\newblock \bibinfo{journal}{Journal of computational physics}
  \bibinfo{volume}{100} (\bibinfo{year}{1992}) \bibinfo{pages}{25--37}.
%Type = Article
\bibitem[{Tryggvason et~al.(2001)Tryggvason, Bunner, Esmaeeli, Juric,
  Al-Rawahi, Tauber, Han, Nas, and Jan}]{tryggvason2001front}
\bibinfo{author}{G.~Tryggvason}, \bibinfo{author}{B.~Bunner},
  \bibinfo{author}{A.~Esmaeeli}, \bibinfo{author}{D.~Juric},
  \bibinfo{author}{N.~Al-Rawahi}, \bibinfo{author}{W.~Tauber},
  \bibinfo{author}{J.~Han}, \bibinfo{author}{S.~Nas}, \bibinfo{author}{Y.-J.
  Jan},
\newblock \bibinfo{title}{A front-tracking method for the computations of
  multiphase flow},
\newblock \bibinfo{journal}{Journal of computational physics}
  \bibinfo{volume}{169} (\bibinfo{year}{2001}) \bibinfo{pages}{708--759}.
%Type = Incollection
\bibitem[{Dervieux and Thomasset(1980)}]{dervieux1980finite}
\bibinfo{author}{A.~Dervieux}, \bibinfo{author}{F.~Thomasset},
\newblock \bibinfo{title}{A finite element method for the simulation of a
  rayleigh-taylor instability},
\newblock in: \bibinfo{booktitle}{Approximation methods for Navier-Stokes
  problems}, \bibinfo{publisher}{Springer}, \bibinfo{year}{1980}, pp.
  \bibinfo{pages}{145--158}.
%Type = Article
\bibitem[{Hirt and Nichols(1981)}]{hirt1981volume}
\bibinfo{author}{C.~W. Hirt}, \bibinfo{author}{B.~D. Nichols},
\newblock \bibinfo{title}{Volume of fluid (vof) method for the dynamics of free
  boundaries},
\newblock \bibinfo{journal}{Journal of computational physics}
  \bibinfo{volume}{39} (\bibinfo{year}{1981}) \bibinfo{pages}{201--225}.
%Type = Article
\bibitem[{Cahn and Hilliard(1958)}]{cahn1958free}
\bibinfo{author}{J.~W. Cahn}, \bibinfo{author}{J.~E. Hilliard},
\newblock \bibinfo{title}{Free energy of a nonuniform system. i. interfacial
  free energy},
\newblock \bibinfo{journal}{The Journal of chemical physics}
  \bibinfo{volume}{28} (\bibinfo{year}{1958}) \bibinfo{pages}{258--267}.
%Type = Article
\bibitem[{Saurel and Abgrall(1999)}]{saurel1999multiphase}
\bibinfo{author}{R.~Saurel}, \bibinfo{author}{R.~Abgrall},
\newblock \bibinfo{title}{A multiphase {Godunov} method for compressible
  multifluid and multiphase flows},
\newblock \bibinfo{journal}{Journal of Computational Physics}
  \bibinfo{volume}{150} (\bibinfo{year}{1999}) \bibinfo{pages}{425--467}.
%Type = Article
\bibitem[{LeMartelot et~al.(2013)LeMartelot, Nkonga, and
  Saurel}]{lemartelot2013liquid}
\bibinfo{author}{S.~LeMartelot}, \bibinfo{author}{B.~Nkonga},
  \bibinfo{author}{R.~Saurel},
\newblock \bibinfo{title}{Liquid and liquid--gas flows at all speeds},
\newblock \bibinfo{journal}{Journal of Computational Physics}
  \bibinfo{volume}{255} (\bibinfo{year}{2013}) \bibinfo{pages}{53--82}.
%Type = Article
\bibitem[{Baer and Nunziato(1986)}]{baer1986two}
\bibinfo{author}{M.~Baer}, \bibinfo{author}{J.~Nunziato},
\newblock \bibinfo{title}{A two-phase mixture theory for the
  deflagration-to-detonation transition (ddt) in reactive granular materials},
\newblock \bibinfo{journal}{International journal of multiphase flow}
  \bibinfo{volume}{12} (\bibinfo{year}{1986}) \bibinfo{pages}{861--889}.
%Type = Article
\bibitem[{Linga and Fl{\aa}tten(2019)}]{linga2019hierarchy}
\bibinfo{author}{G.~Linga}, \bibinfo{author}{T.~Fl{\aa}tten},
\newblock \bibinfo{title}{A hierarchy of non-equilibrium two-phase flow
  models},
\newblock \bibinfo{journal}{ESAIM: Proceedings and Surveys}
  \bibinfo{volume}{66} (\bibinfo{year}{2019}) \bibinfo{pages}{109--143}.
%Type = Article
\bibitem[{Saurel et~al.(2009)Saurel, Petitpas, and Berry}]{saurel2009simple}
\bibinfo{author}{R.~Saurel}, \bibinfo{author}{F.~Petitpas},
  \bibinfo{author}{R.~A. Berry},
\newblock \bibinfo{title}{Simple and efficient relaxation methods for
  interfaces separating compressible fluids, cavitating flows and shocks in
  multiphase mixtures},
\newblock \bibinfo{journal}{journal of Computational Physics}
  \bibinfo{volume}{228} (\bibinfo{year}{2009}) \bibinfo{pages}{1678--1712}.
%Type = Article
\bibitem[{Yeom and Chang(2013)}]{yeom2013modified}
\bibinfo{author}{G.-S. Yeom}, \bibinfo{author}{K.-S. Chang},
\newblock \bibinfo{title}{A modified {HLLC-type} riemann solver for the
  compressible six-equation two-fluid model},
\newblock \bibinfo{journal}{Computers \& Fluids} \bibinfo{volume}{76}
  (\bibinfo{year}{2013}) \bibinfo{pages}{86--104}.
%Type = Article
\bibitem[{Pelanti and Shyue(2014)}]{pelanti2014mixture}
\bibinfo{author}{M.~Pelanti}, \bibinfo{author}{K.-M. Shyue},
\newblock \bibinfo{title}{A mixture-energy-consistent six-equation two-phase
  numerical model for fluids with interfaces, cavitation and evaporation
  waves},
\newblock \bibinfo{journal}{Journal of Computational Physics}
  \bibinfo{volume}{259} (\bibinfo{year}{2014}) \bibinfo{pages}{331--357}.
%Type = Article
\bibitem[{Kapila et~al.(2001)Kapila, Menikoff, Bdzil, Son, and
  Stewart}]{kapila2001two}
\bibinfo{author}{A.~Kapila}, \bibinfo{author}{R.~Menikoff},
  \bibinfo{author}{J.~Bdzil}, \bibinfo{author}{S.~Son}, \bibinfo{author}{D.~S.
  Stewart},
\newblock \bibinfo{title}{Two-phase modeling of deflagration-to-detonation
  transition in granular materials: Reduced equations},
\newblock \bibinfo{journal}{Physics of fluids} \bibinfo{volume}{13}
  (\bibinfo{year}{2001}) \bibinfo{pages}{3002--3024}.
%Type = Article
\bibitem[{Allaire et~al.(2002)Allaire, Clerc, and Kokh}]{allaire2002five}
\bibinfo{author}{G.~Allaire}, \bibinfo{author}{S.~Clerc},
  \bibinfo{author}{S.~Kokh},
\newblock \bibinfo{title}{A five-equation model for the simulation of
  interfaces between compressible fluids},
\newblock \bibinfo{journal}{Journal of Computational Physics}
  \bibinfo{volume}{181} (\bibinfo{year}{2002}) \bibinfo{pages}{577--616}.
%Type = Article
\bibitem[{Murrone and Guillard(2005)}]{murrone2005five}
\bibinfo{author}{A.~Murrone}, \bibinfo{author}{H.~Guillard},
\newblock \bibinfo{title}{A five equation reduced model for compressible two
  phase flow problems},
\newblock \bibinfo{journal}{Journal of Computational Physics}
  \bibinfo{volume}{202} (\bibinfo{year}{2005}) \bibinfo{pages}{664--698}.
%Type = Article
\bibitem[{Perigaud and Saurel(2005)}]{perigaud2005compressible}
\bibinfo{author}{G.~Perigaud}, \bibinfo{author}{R.~Saurel},
\newblock \bibinfo{title}{A compressible flow model with capillary effects},
\newblock \bibinfo{journal}{Journal of Computational Physics}
  \bibinfo{volume}{209} (\bibinfo{year}{2005}) \bibinfo{pages}{139--178}.
%Type = Article
\bibitem[{Shukla et~al.(2010)Shukla, Pantano, and Freund}]{shukla2010interface}
\bibinfo{author}{R.~K. Shukla}, \bibinfo{author}{C.~Pantano},
  \bibinfo{author}{J.~B. Freund},
\newblock \bibinfo{title}{An interface capturing method for the simulation of
  multi-phase compressible flows},
\newblock \bibinfo{journal}{Journal of Computational Physics}
  \bibinfo{volume}{229} (\bibinfo{year}{2010}) \bibinfo{pages}{7411--7439}.
%Type = Article
\bibitem[{Jain et~al.(2020)Jain, Mani, and Moin}]{jain2020conservative}
\bibinfo{author}{S.~S. Jain}, \bibinfo{author}{A.~Mani},
  \bibinfo{author}{P.~Moin},
\newblock \bibinfo{title}{A conservative diffuse-interface method for
  compressible two-phase flows},
\newblock \bibinfo{journal}{Journal of Computational Physics}
  \bibinfo{volume}{418} (\bibinfo{year}{2020}) \bibinfo{pages}{109606}.
%Type = Article
\bibitem[{Abgrall(1996)}]{abgrall1996prevent}
\bibinfo{author}{R.~Abgrall},
\newblock \bibinfo{title}{How to prevent pressure oscillations in
  multicomponent flow calculations: a quasi conservative approach},
\newblock \bibinfo{journal}{Journal of Computational Physics}
  \bibinfo{volume}{125} (\bibinfo{year}{1996}) \bibinfo{pages}{150--160}.
%Type = Article
\bibitem[{Saurel and Abgrall(1999)}]{saurel1999simple}
\bibinfo{author}{R.~Saurel}, \bibinfo{author}{R.~Abgrall},
\newblock \bibinfo{title}{A simple method for compressible multifluid flows},
\newblock \bibinfo{journal}{SIAM Journal on Scientific Computing}
  \bibinfo{volume}{21} (\bibinfo{year}{1999}) \bibinfo{pages}{1115--1145}.
%Type = Article
\bibitem[{Johnsen and Ham(2012)}]{johnsen2012preventing}
\bibinfo{author}{E.~Johnsen}, \bibinfo{author}{F.~Ham},
\newblock \bibinfo{title}{Preventing numerical errors generated by
  interface-capturing schemes in compressible multi-material flows},
\newblock \bibinfo{journal}{Journal of Computational Physics}
  \bibinfo{volume}{231} (\bibinfo{year}{2012}) \bibinfo{pages}{5705--5717}.
%Type = Article
\bibitem[{Lund and Aursand(2012)}]{lund_proc}
\bibinfo{author}{H.~Lund}, \bibinfo{author}{P.~Aursand},
\newblock \bibinfo{title}{Two-phase flow of {${\rm CO}_2$} with phase
  transfer},
\newblock \bibinfo{journal}{Energy Procedia} \bibinfo{volume}{23}
  (\bibinfo{year}{2012}) \bibinfo{pages}{246--255}.
%Type = Article
\bibitem[{Le~Martelot et~al.(2014)Le~Martelot, Saurel, and
  Nkonga}]{le2014towards}
\bibinfo{author}{S.~Le~Martelot}, \bibinfo{author}{R.~Saurel},
  \bibinfo{author}{B.~Nkonga},
\newblock \bibinfo{title}{Towards the direct numerical simulation of nucleate
  boiling flows},
\newblock \bibinfo{journal}{International Journal of Multiphase Flow}
  \bibinfo{volume}{66} (\bibinfo{year}{2014}) \bibinfo{pages}{62--78}.
%Type = Article
\bibitem[{Murrone and Guillard(2008)}]{murrone2008behavior}
\bibinfo{author}{A.~Murrone}, \bibinfo{author}{H.~Guillard},
\newblock \bibinfo{title}{Behavior of upwind scheme in the low {Mach} number
  limit: Iii. preconditioned dissipation for a five equation two phase model},
\newblock \bibinfo{journal}{Computers \& fluids} \bibinfo{volume}{37}
  (\bibinfo{year}{2008}) \bibinfo{pages}{1209--1224}.
%Type = Article
\bibitem[{Pelanti(2017)}]{pelanti-amc}
\bibinfo{author}{M.~Pelanti},
\newblock \bibinfo{title}{Low {M}ach number preconditioning techniques for
  {Roe}-type and {HLLC}-type methods for a two-phase compressible flow model},
\newblock \bibinfo{journal}{Appl. Math. Comp.} \bibinfo{volume}{310}
  (\bibinfo{year}{2017}) \bibinfo{pages}{112--133}.
%Type = Article
\bibitem[{Jemison et~al.(2014)Jemison, Sussman, and
  Arienti}]{jemison2014compressible}
\bibinfo{author}{M.~Jemison}, \bibinfo{author}{M.~Sussman},
  \bibinfo{author}{M.~Arienti},
\newblock \bibinfo{title}{Compressible, multiphase semi-implicit method with
  moment of fluid interface representation},
\newblock \bibinfo{journal}{Journal of Computational Physics}
  \bibinfo{volume}{279} (\bibinfo{year}{2014}) \bibinfo{pages}{182--217}.
%Type = Article
\bibitem[{Denner et~al.(2018)Denner, Xiao, and van Wachem}]{denner2018pressure}
\bibinfo{author}{F.~Denner}, \bibinfo{author}{C.-N. Xiao},
  \bibinfo{author}{B.~G. van Wachem},
\newblock \bibinfo{title}{Pressure-based algorithm for compressible interfacial
  flows with acoustically-conservative interface discretisation},
\newblock \bibinfo{journal}{Journal of Computational Physics}
  \bibinfo{volume}{367} (\bibinfo{year}{2018}) \bibinfo{pages}{192--234}.
%Type = Article
\bibitem[{Weiss and Smith(1995)}]{weiss1995preconditioning}
\bibinfo{author}{J.~M. Weiss}, \bibinfo{author}{W.~A. Smith},
\newblock \bibinfo{title}{Preconditioning applied to variable and constant
  density flows},
\newblock \bibinfo{journal}{AIAA journal} \bibinfo{volume}{33}
  (\bibinfo{year}{1995}) \bibinfo{pages}{2050--2057}.
%Type = Article
\bibitem[{Turkel and Vatsa(2005)}]{turkel2005local}
\bibinfo{author}{E.~Turkel}, \bibinfo{author}{V.~N. Vatsa},
\newblock \bibinfo{title}{Local preconditioners for steady and unsteady flow
  applications},
\newblock \bibinfo{journal}{ESAIM: Mathematical Modelling and Numerical
  Analysis-Mod{\'e}lisation Math{\'e}matique et Analyse Num{\'e}rique}
  \bibinfo{volume}{39} (\bibinfo{year}{2005}) \bibinfo{pages}{515--535}.
%Type = Article
\bibitem[{Saurel and Pantano(2018)}]{saurel2018diffuse}
\bibinfo{author}{R.~Saurel}, \bibinfo{author}{C.~Pantano},
\newblock \bibinfo{title}{Diffuse-interface capturing methods for compressible
  two-phase flows},
\newblock \bibinfo{journal}{Annual Review of Fluid Mechanics}
  \bibinfo{volume}{50} (\bibinfo{year}{2018}) \bibinfo{pages}{105--130}.
%Type = Article
\bibitem[{Park and Munz(2005)}]{park2005multiple}
\bibinfo{author}{J.~Park}, \bibinfo{author}{C.-D. Munz},
\newblock \bibinfo{title}{Multiple pressure variables methods for fluid flow at
  all mach numbers},
\newblock \bibinfo{journal}{International journal for numerical methods in
  fluids} \bibinfo{volume}{49} (\bibinfo{year}{2005})
  \bibinfo{pages}{905--931}.
%Type = Article
\bibitem[{Klein(1995)}]{klein1995semi}
\bibinfo{author}{R.~Klein},
\newblock \bibinfo{title}{Semi-implicit extension of a godunov-type scheme
  based on low mach number asymptotics i: One-dimensional flow},
\newblock \bibinfo{journal}{Journal of Computational Physics}
  \bibinfo{volume}{121} (\bibinfo{year}{1995}) \bibinfo{pages}{213--237}.
%Type = Article
\bibitem[{Klein et~al.(2001)Klein, Botta, Schneider, Munz, Roller, Meister,
  Hoffmann, and Sonar}]{klein2001asymptotic}
\bibinfo{author}{R.~Klein}, \bibinfo{author}{N.~Botta},
  \bibinfo{author}{T.~Schneider}, \bibinfo{author}{C.-D. Munz},
  \bibinfo{author}{S.~Roller}, \bibinfo{author}{A.~Meister},
  \bibinfo{author}{L.~Hoffmann}, \bibinfo{author}{T.~Sonar},
\newblock \bibinfo{title}{Asymptotic adaptive methods for multi-scale problems
  in fluid mechanics},
\newblock \bibinfo{journal}{Journal of Engineering Mathematics}
  \bibinfo{volume}{39} (\bibinfo{year}{2001}) \bibinfo{pages}{261--343}.
%Type = Article
\bibitem[{Munz et~al.(2003)Munz, Roller, Klein, and Geratz}]{munz2003extension}
\bibinfo{author}{C.-D. Munz}, \bibinfo{author}{S.~Roller},
  \bibinfo{author}{R.~Klein}, \bibinfo{author}{K.~J. Geratz},
\newblock \bibinfo{title}{The extension of incompressible flow solvers to the
  weakly compressible regime},
\newblock \bibinfo{journal}{Computers \& Fluids} \bibinfo{volume}{32}
  (\bibinfo{year}{2003}) \bibinfo{pages}{173--196}.
%Type = Article
\bibitem[{Dumbser and Casulli(2016)}]{dumbser2016conservative}
\bibinfo{author}{M.~Dumbser}, \bibinfo{author}{V.~Casulli},
\newblock \bibinfo{title}{A conservative, weakly nonlinear semi-implicit finite
  volume scheme for the compressible {Navier- Stokes} equations with general
  equation of state},
\newblock \bibinfo{journal}{Applied Mathematics and Computation}
  \bibinfo{volume}{272} (\bibinfo{year}{2016}) \bibinfo{pages}{479--497}.
%Type = Article
\bibitem[{Berm{\'u}dez et~al.(2020)Berm{\'u}dez, Busto, Dumbser, Ferr{\'\i}n,
  Saavedra, and V{\'a}zquez-Cend{\'o}n}]{bermudez2020staggered}
\bibinfo{author}{A.~Berm{\'u}dez}, \bibinfo{author}{S.~Busto},
  \bibinfo{author}{M.~Dumbser}, \bibinfo{author}{J.~L. Ferr{\'\i}n},
  \bibinfo{author}{L.~Saavedra}, \bibinfo{author}{M.~E.
  V{\'a}zquez-Cend{\'o}n},
\newblock \bibinfo{title}{A staggered semi-implicit hybrid fv/fe projection
  method for weakly compressible flows},
\newblock \bibinfo{journal}{Journal of Computational Physics}
  \bibinfo{volume}{421} (\bibinfo{year}{2020}) \bibinfo{pages}{109743}.
%Type = Article
\bibitem[{Busto et~al.(2021)Busto, R{\'\i}o-Mart{\'\i}n,
  V{\'a}zquez-Cend{\'o}n, and Dumbser}]{busto2021semi}
\bibinfo{author}{S.~Busto}, \bibinfo{author}{L.~R{\'\i}o-Mart{\'\i}n},
  \bibinfo{author}{M.~E. V{\'a}zquez-Cend{\'o}n}, \bibinfo{author}{M.~Dumbser},
\newblock \bibinfo{title}{A semi-implicit hybrid finite volume/finite element
  scheme for all mach number flows on staggered unstructured meshes},
\newblock \bibinfo{journal}{Applied Mathematics and Computation}
  \bibinfo{volume}{402} (\bibinfo{year}{2021}) \bibinfo{pages}{126117}.
%Type = Inproceedings
\bibitem[{Re and Abgrall(2018)}]{re2018non}
\bibinfo{author}{B.~Re}, \bibinfo{author}{R.~Abgrall},
\newblock \bibinfo{title}{Non-equilibrium model for weakly compressible
  multi-component flows: the hyperbolic operator},
\newblock in: \bibinfo{booktitle}{International Seminar on Non-Ideal
  Compressible-Fluid Dynamics for Propulsion \& Power},
  \bibinfo{organization}{Springer}, \bibinfo{year}{2018}, pp.
  \bibinfo{pages}{33--45}.
%Type = Article
\bibitem[{Re and Abgrall(2021)}]{re2021pressure}
\bibinfo{author}{B.~Re}, \bibinfo{author}{R.~Abgrall},
\newblock \bibinfo{title}{A pressure-based method for weakly compressible
  two-phase flows under a baer-nunziato type model with generic equations of
  state and pressure and velocity disequilibrium},
\newblock \bibinfo{journal}{arXiv preprint arXiv:2107.12408}
  (\bibinfo{year}{2021}).
%Type = Article
\bibitem[{Kuhn and Desjardins(2021)}]{kuhn2021all}
\bibinfo{author}{M.~Kuhn}, \bibinfo{author}{O.~Desjardins},
\newblock \bibinfo{title}{An all-mach, low-dissipation strategy for simulating
  multiphase flows},
\newblock \bibinfo{journal}{Journal of Computational Physics}
  (\bibinfo{year}{2021}) \bibinfo{pages}{110602}.
%Type = Article
\bibitem[{Fuster and Popinet(2018)}]{fuster2018all}
\bibinfo{author}{D.~Fuster}, \bibinfo{author}{S.~Popinet},
\newblock \bibinfo{title}{An {all-Mach} method for the simulation of bubble
  dynamics problems in the presence of surface tension},
\newblock \bibinfo{journal}{Journal of Computational Physics}
  \bibinfo{volume}{374} (\bibinfo{year}{2018}) \bibinfo{pages}{752--768}.
%Type = Article
\bibitem[{Dalla~Barba et~al.(2021)Dalla~Barba, Scapin, Demou, Rosti, Picano,
  and Brandt}]{dalla2021interface}
\bibinfo{author}{F.~Dalla~Barba}, \bibinfo{author}{N.~Scapin},
  \bibinfo{author}{A.~D. Demou}, \bibinfo{author}{M.~E. Rosti},
  \bibinfo{author}{F.~Picano}, \bibinfo{author}{L.~Brandt},
\newblock \bibinfo{title}{An interface capturing method for liquid-gas flows at
  {low-Mach} number},
\newblock \bibinfo{journal}{Computers \& Fluids} \bibinfo{volume}{216}
  (\bibinfo{year}{2021}) \bibinfo{pages}{104789}.
%Type = Article
\bibitem[{Juric and Tryggvason(1998)}]{juric1998computations}
\bibinfo{author}{D.~Juric}, \bibinfo{author}{G.~Tryggvason},
\newblock \bibinfo{title}{Computations of boiling flows},
\newblock \bibinfo{journal}{International journal of multiphase flow}
  \bibinfo{volume}{24} (\bibinfo{year}{1998}) \bibinfo{pages}{387--410}.
%Type = Article
\bibitem[{Sato and Ni{\v{c}}eno(2013)}]{sato2013sharp}
\bibinfo{author}{Y.~Sato}, \bibinfo{author}{B.~Ni{\v{c}}eno},
\newblock \bibinfo{title}{A sharp-interface phase change model for a
  mass-conservative interface tracking method},
\newblock \bibinfo{journal}{Journal of Computational Physics}
  \bibinfo{volume}{249} (\bibinfo{year}{2013}) \bibinfo{pages}{127--161}.
%Type = Article
\bibitem[{Tanguy et~al.(2014)Tanguy, Sagan, Lalanne, Couderc, and
  Colin}]{tanguy2014benchmarks}
\bibinfo{author}{S.~Tanguy}, \bibinfo{author}{M.~Sagan},
  \bibinfo{author}{B.~Lalanne}, \bibinfo{author}{F.~Couderc},
  \bibinfo{author}{C.~Colin},
\newblock \bibinfo{title}{Benchmarks and numerical methods for the simulation
  of boiling flows},
\newblock \bibinfo{journal}{Journal of Computational Physics}
  \bibinfo{volume}{264} (\bibinfo{year}{2014}) \bibinfo{pages}{1--22}.
%Type = Article
\bibitem[{Scapin et~al.(2020)Scapin, Costa, and Brandt}]{scapin2020volume}
\bibinfo{author}{N.~Scapin}, \bibinfo{author}{P.~Costa},
  \bibinfo{author}{L.~Brandt},
\newblock \bibinfo{title}{A volume-of-fluid method for interface-resolved
  simulations of phase-changing two-fluid flows},
\newblock \bibinfo{journal}{Journal of Computational Physics}
  \bibinfo{volume}{407} (\bibinfo{year}{2020}) \bibinfo{pages}{109251}.
%Type = Article
\bibitem[{Jafari and Okutucu-{\"O}zyurt(2016)}]{jafari2016numerical}
\bibinfo{author}{R.~Jafari}, \bibinfo{author}{T.~Okutucu-{\"O}zyurt},
\newblock \bibinfo{title}{Numerical simulation of flow boiling from an
  artificial cavity in a microchannel},
\newblock \bibinfo{journal}{International Journal of Heat and Mass Transfer}
  \bibinfo{volume}{97} (\bibinfo{year}{2016}) \bibinfo{pages}{270--278}.
%Type = Article
\bibitem[{Wang et~al.(2021)Wang, Zheng, Chryssostomidis, and
  Karniadakis}]{wang2021phase}
\bibinfo{author}{Z.~Wang}, \bibinfo{author}{X.~Zheng},
  \bibinfo{author}{C.~Chryssostomidis}, \bibinfo{author}{G.~E. Karniadakis},
\newblock \bibinfo{title}{A phase-field method for boiling heat transfer},
\newblock \bibinfo{journal}{Journal of Computational Physics}
  (\bibinfo{year}{2021}) \bibinfo{pages}{110239}.
%Type = Article
\bibitem[{Brackbill et~al.(1992)Brackbill, Kothe, and
  Zemach}]{brackbill1992continuum}
\bibinfo{author}{J.~U. Brackbill}, \bibinfo{author}{D.~B. Kothe},
  \bibinfo{author}{C.~Zemach},
\newblock \bibinfo{title}{A continuum method for modeling surface tension},
\newblock \bibinfo{journal}{Journal of computational physics}
  \bibinfo{volume}{100} (\bibinfo{year}{1992}) \bibinfo{pages}{335--354}.
%Type = Article
\bibitem[{Fl{\aa}tten and Lund(2011)}]{flatten-lund:rel}
\bibinfo{author}{T.~Fl{\aa}tten}, \bibinfo{author}{H.~Lund},
\newblock \bibinfo{title}{Relaxation two-phase models and the subcharacteristic
  condition},
\newblock \bibinfo{journal}{Math. Models Methods Appl. Sci.}
  \bibinfo{volume}{21} (\bibinfo{year}{2011}) \bibinfo{pages}{2379--2407}.
%Type = Article
\bibitem[{Le~M{\'e}tayer and Saurel(2016)}]{le2016noble}
\bibinfo{author}{O.~Le~M{\'e}tayer}, \bibinfo{author}{R.~Saurel},
\newblock \bibinfo{title}{The noble-abel stiffened-gas equation of state},
\newblock \bibinfo{journal}{Physics of Fluids} \bibinfo{volume}{28}
  (\bibinfo{year}{2016}) \bibinfo{pages}{046102}.
%Type = Article
\bibitem[{Saurel and {Le~M\'etayer}(2001)}]{sa-lem:multi}
\bibinfo{author}{R.~Saurel}, \bibinfo{author}{O.~{Le~M\'etayer}},
\newblock \bibinfo{title}{A multiphase model for compressible flows with
  interfaces, shocks, detonation waves and cavitation},
\newblock \bibinfo{journal}{J. Fluid Mech.} \bibinfo{volume}{431}
  (\bibinfo{year}{2001}) \bibinfo{pages}{239--271}.
%Type = Book
\bibitem[{Wesseling(2009)}]{wesseling2009principles}
\bibinfo{author}{P.~Wesseling}, \bibinfo{title}{Principles of computational
  fluid dynamics}, volume~\bibinfo{volume}{29}, \bibinfo{publisher}{Springer
  Science \& Business Media}, \bibinfo{year}{2009}.
%Type = Article
\bibitem[{Van~Leer(1977)}]{van1977towards}
\bibinfo{author}{B.~Van~Leer},
\newblock \bibinfo{title}{Towards the ultimate conservative difference scheme.
  iv. a new approach to numerical convection},
\newblock \bibinfo{journal}{Journal of computational physics}
  \bibinfo{volume}{23} (\bibinfo{year}{1977}) \bibinfo{pages}{276--299}.
%Type = Book
\bibitem[{Prosperetti and Tryggvason(2009)}]{prosperetti2009computational}
\bibinfo{author}{A.~Prosperetti}, \bibinfo{author}{G.~Tryggvason},
  \bibinfo{title}{Computational methods for multiphase flow},
  \bibinfo{publisher}{Cambridge university press}, \bibinfo{year}{2009}.
%Type = Article
\bibitem[{Amsden and Harlow(1970)}]{amsden1970simplified}
\bibinfo{author}{A.~A. Amsden}, \bibinfo{author}{F.~H. Harlow},
\newblock \bibinfo{title}{A simplified {MAC} technique for incompressible fluid
  flow calculations},
\newblock \bibinfo{journal}{Journal of computational physics}
  \bibinfo{volume}{6} (\bibinfo{year}{1970}) \bibinfo{pages}{322--325}.
%Type = Inproceedings
\bibitem[{Falgout and Yang(2002)}]{falgout2002hypre}
\bibinfo{author}{R.~D. Falgout}, \bibinfo{author}{U.~M. Yang},
\newblock \bibinfo{title}{hypre: A library of high performance
  preconditioners},
\newblock in: \bibinfo{booktitle}{International Conference on Computational
  Science}, \bibinfo{organization}{Springer}, \bibinfo{year}{2002}, pp.
  \bibinfo{pages}{632--641}.
%Type = Article
\bibitem[{De~Lorenzo et~al.(2019)De~Lorenzo, Lafon, and
  Pelanti}]{de2019hyperbolic}
\bibinfo{author}{M.~De~Lorenzo}, \bibinfo{author}{P.~Lafon},
  \bibinfo{author}{M.~Pelanti},
\newblock \bibinfo{title}{A hyperbolic phase-transition model with
  non-instantaneous eos-independent relaxation procedures},
\newblock \bibinfo{journal}{Journal of Computational Physics}
  \bibinfo{volume}{379} (\bibinfo{year}{2019}) \bibinfo{pages}{279--308}.
%Type = Inproceedings
\bibitem[{Pelanti et~al.(2019)Pelanti, De~Lorenzo, and
  Lafon}]{pelanti2019numerical}
\bibinfo{author}{M.~Pelanti}, \bibinfo{author}{M.~De~Lorenzo},
  \bibinfo{author}{P.~Lafon},
\newblock \bibinfo{title}{A numerical model for liquid-vapor flows with
  arbitrary heat and mass transfer relaxation times and general equation of
  state},
\newblock in: \bibinfo{booktitle}{APS Division of Fluid Dynamics Meeting
  Abstracts}, \bibinfo{year}{2019}, pp. \bibinfo{pages}{G24--008}.
%Type = Article
\bibitem[{Pelanti(2021)}]{pelanti2021arbitrary}
\bibinfo{author}{M.~Pelanti},
\newblock \bibinfo{title}{Arbitrary-rate relaxation techniques for the
  numerical modeling of compressible two-phase flows with heat and mass
  transfer},
\newblock \bibinfo{journal}{arXiv preprint arXiv:2108.00556}
  (\bibinfo{year}{2021}).
%Type = Article
\bibitem[{Saurel et~al.(2008)Saurel, Petitpas, and
  Abgrall}]{saurel2008modelling}
\bibinfo{author}{R.~Saurel}, \bibinfo{author}{F.~Petitpas},
  \bibinfo{author}{R.~Abgrall},
\newblock \bibinfo{title}{Modelling phase transition in metastable liquids:
  Application to cavitating and flashing flows.},
\newblock \bibinfo{journal}{Journal of Fluid Mechanics} \bibinfo{volume}{607}
  (\bibinfo{year}{2008}) \bibinfo{pages}{313--350}.
%Type = Article
\bibitem[{Zein et~al.(2010)Zein, Hantke, and Warnecke}]{zein2010modeling}
\bibinfo{author}{A.~Zein}, \bibinfo{author}{M.~Hantke},
  \bibinfo{author}{G.~Warnecke},
\newblock \bibinfo{title}{Modeling phase transition for compressible two-phase
  flows applied to metastable liquids},
\newblock \bibinfo{journal}{Journal of Computational Physics}
  \bibinfo{volume}{229} (\bibinfo{year}{2010}) \bibinfo{pages}{2964--2998}.
%Type = Article
\bibitem[{Costa(2018)}]{costa2018fft}
\bibinfo{author}{P.~Costa},
\newblock \bibinfo{title}{A {FFT-based} finite-difference solver for
  massively-parallel direct numerical simulations of turbulent flows},
\newblock \bibinfo{journal}{Computers \& Mathematics with Applications}
  \bibinfo{volume}{76} (\bibinfo{year}{2018}) \bibinfo{pages}{1853--1862}.
%Type = Article
\bibitem[{Kang et~al.(2000)Kang, Fedkiw, and Liu}]{kang2000boundary}
\bibinfo{author}{M.~Kang}, \bibinfo{author}{R.~P. Fedkiw},
  \bibinfo{author}{X.-D. Liu},
\newblock \bibinfo{title}{A boundary condition capturing method for multiphase
  incompressible flow},
\newblock \bibinfo{journal}{Journal of Scientific Computing}
  \bibinfo{volume}{15} (\bibinfo{year}{2000}) \bibinfo{pages}{323--360}.
%Type = Article
\bibitem[{Liska and Wendroff(2003)}]{liska2003comparison}
\bibinfo{author}{R.~Liska}, \bibinfo{author}{B.~Wendroff},
\newblock \bibinfo{title}{Comparison of several difference schemes on 1d and 2d
  test problems for the euler equations},
\newblock \bibinfo{journal}{SIAM Journal on Scientific Computing}
  \bibinfo{volume}{25} (\bibinfo{year}{2003}) \bibinfo{pages}{995--1017}.
%Type = Phdthesis
\bibitem[{Miczek(2013)}]{miczek2013simulation}
\bibinfo{author}{F.~Miczek}, \bibinfo{title}{Simulation of {low Mach} number
  astrophysical flows}, Ph.D. thesis, Technische Universit{\"a}t M{\"u}nchen,
  \bibinfo{year}{2013}.
%Type = Article
\bibitem[{Thomann et~al.(2020)Thomann, Puppo, and Klingenberg}]{thomann2020all}
\bibinfo{author}{A.~Thomann}, \bibinfo{author}{G.~Puppo},
  \bibinfo{author}{C.~Klingenberg},
\newblock \bibinfo{title}{An all speed second order well-balanced {IMEX}
  relaxation scheme for the euler equations with gravity},
\newblock \bibinfo{journal}{Journal of Computational Physics}
  \bibinfo{volume}{420} (\bibinfo{year}{2020}) \bibinfo{pages}{109723}.
%Type = Article
\bibitem[{Kwatra et~al.(2009)Kwatra, Su, Gr{\'e}tarsson, and
  Fedkiw}]{kwatra2009method}
\bibinfo{author}{N.~Kwatra}, \bibinfo{author}{J.~Su}, \bibinfo{author}{J.~T.
  Gr{\'e}tarsson}, \bibinfo{author}{R.~Fedkiw},
\newblock \bibinfo{title}{A method for avoiding the acoustic time step
  restriction in compressible flow},
\newblock \bibinfo{journal}{Journal of Computational Physics}
  \bibinfo{volume}{228} (\bibinfo{year}{2009}) \bibinfo{pages}{4146--4161}.
%Type = Article
\bibitem[{Gray and Giorgini(1976)}]{gray1976validity}
\bibinfo{author}{D.~D. Gray}, \bibinfo{author}{A.~Giorgini},
\newblock \bibinfo{title}{The validity of the boussinesq approximation for
  liquids and gases},
\newblock \bibinfo{journal}{International Journal of Heat and Mass Transfer}
  \bibinfo{volume}{19} (\bibinfo{year}{1976}) \bibinfo{pages}{545--551}.
%Type = Article
\bibitem[{de~Vahl~Davis(1983)}]{de1983natural}
\bibinfo{author}{G.~de~Vahl~Davis},
\newblock \bibinfo{title}{Natural convection of air in a square cavity: a bench
  mark numerical solution},
\newblock \bibinfo{journal}{International Journal for numerical methods in
  fluids} \bibinfo{volume}{3} (\bibinfo{year}{1983}) \bibinfo{pages}{249--264}.
%Type = Article
\bibitem[{Hortmann et~al.(1990)Hortmann, Peri{\'c}, and
  Scheuerer}]{hortmann1990finite}
\bibinfo{author}{M.~Hortmann}, \bibinfo{author}{M.~Peri{\'c}},
  \bibinfo{author}{G.~Scheuerer},
\newblock \bibinfo{title}{Finite volume multigrid prediction of laminar natural
  convection: bench-mark solutions},
\newblock \bibinfo{journal}{International journal for numerical methods in
  fluids} \bibinfo{volume}{11} (\bibinfo{year}{1990})
  \bibinfo{pages}{189--207}.
%Type = Article
\bibitem[{Le~Qu{\'e}r{\'e}(1991)}]{le1991accurate}
\bibinfo{author}{P.~Le~Qu{\'e}r{\'e}},
\newblock \bibinfo{title}{Accurate solutions to the square thermally driven
  cavity at high rayleigh number},
\newblock \bibinfo{journal}{Computers \& Fluids} \bibinfo{volume}{20}
  (\bibinfo{year}{1991}) \bibinfo{pages}{29--41}.
%Type = Article
\bibitem[{Le~Qu{\'e}r{\'e} et~al.(2005)Le~Qu{\'e}r{\'e}, Weisman, Paill{\`e}re,
  Vierendeels, Dick, Becker, Braack, and Locke}]{le2005modelling}
\bibinfo{author}{P.~Le~Qu{\'e}r{\'e}}, \bibinfo{author}{C.~Weisman},
  \bibinfo{author}{H.~Paill{\`e}re}, \bibinfo{author}{J.~Vierendeels},
  \bibinfo{author}{E.~Dick}, \bibinfo{author}{R.~Becker},
  \bibinfo{author}{M.~Braack}, \bibinfo{author}{J.~Locke},
\newblock \bibinfo{title}{Modelling of natural convection flows with large
  temperature differences: a benchmark problem for low mach number solvers.
  part 1. reference solutions},
\newblock \bibinfo{journal}{ESAIM: Mathematical Modelling and Numerical
  Analysis} \bibinfo{volume}{39} (\bibinfo{year}{2005})
  \bibinfo{pages}{609--616}.
%Type = Article
\bibitem[{Armengol et~al.(2017)Armengol, Bannwart, Xaman, and
  Santos}]{armengol2017effects}
\bibinfo{author}{J.~Armengol}, \bibinfo{author}{F.~Bannwart},
  \bibinfo{author}{J.~Xaman}, \bibinfo{author}{R.~Santos},
\newblock \bibinfo{title}{Effects of variable air properties on transient
  natural convection for large temperature differences},
\newblock \bibinfo{journal}{International Journal of Thermal Sciences}
  \bibinfo{volume}{120} (\bibinfo{year}{2017}) \bibinfo{pages}{63--79}.
%Type = Article
\bibitem[{Demou et~al.(2019)Demou, Frantzis, and Grigoriadis}]{demou2019low}
\bibinfo{author}{A.~Demou}, \bibinfo{author}{C.~Frantzis},
  \bibinfo{author}{D.~G. Grigoriadis},
\newblock \bibinfo{title}{A low-mach methodology for efficient direct numerical
  simulations of variable property thermally driven flows},
\newblock \bibinfo{journal}{International Journal of Heat and Mass Transfer}
  \bibinfo{volume}{132} (\bibinfo{year}{2019}) \bibinfo{pages}{539--549}.
%Type = Article
\bibitem[{Demou and Grigoriadis(2020)}]{demou2020variable}
\bibinfo{author}{A.~D. Demou}, \bibinfo{author}{D.~G. Grigoriadis},
\newblock \bibinfo{title}{Variable property dns of differentially heated
  cavities filled with air},
\newblock \bibinfo{journal}{International Journal of Heat and Mass Transfer}
  \bibinfo{volume}{149} (\bibinfo{year}{2020}) \bibinfo{pages}{119259}.
%Type = Article
\bibitem[{Hysing et~al.(2009)Hysing, Turek, Kuzmin, Parolini, Burman, Ganesan,
  and Tobiska}]{hysing2009quantitative}
\bibinfo{author}{S.-R. Hysing}, \bibinfo{author}{S.~Turek},
  \bibinfo{author}{D.~Kuzmin}, \bibinfo{author}{N.~Parolini},
  \bibinfo{author}{E.~Burman}, \bibinfo{author}{S.~Ganesan},
  \bibinfo{author}{L.~Tobiska},
\newblock \bibinfo{title}{Quantitative benchmark computations of
  two-dimensional bubble dynamics},
\newblock \bibinfo{journal}{International Journal for Numerical Methods in
  Fluids} \bibinfo{volume}{60} (\bibinfo{year}{2009})
  \bibinfo{pages}{1259--1288}.
%Type = Article
\bibitem[{Kim(2009)}]{kim2009review}
\bibinfo{author}{J.~Kim},
\newblock \bibinfo{title}{Review of nucleate pool boiling bubble heat transfer
  mechanisms},
\newblock \bibinfo{journal}{International Journal of Multiphase Flow}
  \bibinfo{volume}{35} (\bibinfo{year}{2009}) \bibinfo{pages}{1067--1076}.
%Type = Article
\bibitem[{Cooper and Lloyd(1969)}]{cooper1969microlayer}
\bibinfo{author}{M.~Cooper}, \bibinfo{author}{A.~Lloyd},
\newblock \bibinfo{title}{The microlayer in nucleate pool boiling},
\newblock \bibinfo{journal}{International Journal of Heat and Mass Transfer}
  \bibinfo{volume}{12} (\bibinfo{year}{1969}) \bibinfo{pages}{895--913}.
%Type = Article
\bibitem[{Stephan and Busse(1992)}]{stephan1992analysis}
\bibinfo{author}{P.~Stephan}, \bibinfo{author}{C.~Busse},
\newblock \bibinfo{title}{Analysis of the heat transfer coefficient of grooved
  heat pipe evaporator walls},
\newblock \bibinfo{journal}{International Journal of heat and mass transfer}
  \bibinfo{volume}{35} (\bibinfo{year}{1992}) \bibinfo{pages}{383--391}.
%Type = Article
\bibitem[{Stephan and Hammer(1994)}]{stephan1994new}
\bibinfo{author}{P.~Stephan}, \bibinfo{author}{J.~Hammer},
\newblock \bibinfo{title}{A new model for nucleate boiling heat transfer},
\newblock \bibinfo{journal}{Heat and Mass Transfer} \bibinfo{volume}{30}
  (\bibinfo{year}{1994}) \bibinfo{pages}{119--125}.
%Type = Article
\bibitem[{Sato and Niceno(2012)}]{sato2012new}
\bibinfo{author}{Y.~Sato}, \bibinfo{author}{B.~Niceno},
\newblock \bibinfo{title}{A new contact line treatment for a conservative level
  set method},
\newblock \bibinfo{journal}{Journal of computational physics (Print)}
  \bibinfo{volume}{231} (\bibinfo{year}{2012}) \bibinfo{pages}{3887--3895}.
%Type = Article
\bibitem[{Wagner et~al.(2000)Wagner, Cooper, Dittmann, Kijima, Kretzschmar,
  Kruse, {Mare\v{s}}, Oguchi, Sato, {St\"ocker}, {\v{S}ifner}, Takaishi,
  Tanishita, {Tr\"ubenbach}, and Willkommen}]{iapws97}
\bibinfo{author}{W.~Wagner}, \bibinfo{author}{J.~R. Cooper},
  \bibinfo{author}{A.~Dittmann}, \bibinfo{author}{J.~Kijima},
  \bibinfo{author}{H.-J. Kretzschmar}, \bibinfo{author}{A.~Kruse},
  \bibinfo{author}{R.~{Mare\v{s}}}, \bibinfo{author}{K.~Oguchi},
  \bibinfo{author}{H.~Sato}, \bibinfo{author}{I.~{St\"ocker}},
  \bibinfo{author}{O.~{\v{S}ifner}}, \bibinfo{author}{Y.~Takaishi},
  \bibinfo{author}{I.~Tanishita}, \bibinfo{author}{J.~{Tr\"ubenbach}},
  \bibinfo{author}{T.~Willkommen},
\newblock \bibinfo{title}{The {IAPWS} {Industrial} {Formulation} 1997 for the
  {Thermodynamic} {Properties} of {Water} and {Steam}},
\newblock \bibinfo{journal}{Transactions of the ASME} \bibinfo{volume}{122}
  (\bibinfo{year}{2000}) \bibinfo{pages}{150--182}.
%Type = Article
\bibitem[{{De~Lorenzo} et~al.(2017){De~Lorenzo}, Lafon, Matteo, Pelanti,
  Seynhaeve, and Bartosiewicz}]{delor-laf-pelanti-IJMF}
\bibinfo{author}{M.~{De~Lorenzo}}, \bibinfo{author}{P.~Lafon},
  \bibinfo{author}{M.~D. Matteo}, \bibinfo{author}{M.~Pelanti},
  \bibinfo{author}{J.-M. Seynhaeve}, \bibinfo{author}{Y.~Bartosiewicz},
\newblock \bibinfo{title}{Homogeneous two-phase flow models and accurate
  steam-water table look-up method for fast transient simulations},
\newblock \bibinfo{journal}{Int. J. Multiphase Flow} \bibinfo{volume}{95}
  (\bibinfo{year}{2017}) \bibinfo{pages}{199--219}.
%Type = Article
\bibitem[{De~Lorenzo et~al.(2021)De~Lorenzo, Lafon, Pelanti, Pantano,
  Di~Matteo, Bartosiewicz, and Seynhaeve}]{de2021hyperbolic}
\bibinfo{author}{M.~De~Lorenzo}, \bibinfo{author}{P.~Lafon},
  \bibinfo{author}{M.~Pelanti}, \bibinfo{author}{A.~Pantano},
  \bibinfo{author}{M.~Di~Matteo}, \bibinfo{author}{Y.~Bartosiewicz},
  \bibinfo{author}{J.-M. Seynhaeve},
\newblock \bibinfo{title}{A hyperbolic phase-transition model coupled to
  tabulated {EoS} for two-phase flows in fast depressurizations},
\newblock \bibinfo{journal}{Nuclear Engineering and Design}
  \bibinfo{volume}{371} (\bibinfo{year}{2021}) \bibinfo{pages}{110954}.
%Type = Article
\bibitem[{Le and Moin(1991)}]{le1991improvement}
\bibinfo{author}{H.~Le}, \bibinfo{author}{P.~Moin},
\newblock \bibinfo{title}{An improvement of fractional step methods for the
  incompressible navier-stokes equations},
\newblock \bibinfo{journal}{Journal of computational physics}
  \bibinfo{volume}{92} (\bibinfo{year}{1991}) \bibinfo{pages}{369--379}.
%Type = Article
\bibitem[{Capuano et~al.(2016)Capuano, Coppola, Chiatto, and
  de~Luca}]{capuano2016approximate}
\bibinfo{author}{F.~Capuano}, \bibinfo{author}{G.~Coppola},
  \bibinfo{author}{M.~Chiatto}, \bibinfo{author}{L.~de~Luca},
\newblock \bibinfo{title}{Approximate projection method for the incompressible
  {Navier--Stokes} equations},
\newblock \bibinfo{journal}{AIAA Journal} \bibinfo{volume}{54}
  (\bibinfo{year}{2016}) \bibinfo{pages}{2179--2182}.
%Type = Article
\bibitem[{Chen et~al.(1994)Chen, Levermore, and Liu}]{chen1994hyperbolic}
\bibinfo{author}{G.-Q. Chen}, \bibinfo{author}{C.~D. Levermore},
  \bibinfo{author}{T.-P. Liu},
\newblock \bibinfo{title}{Hyperbolic conservation laws with stiff relaxation
  terms and entropy},
\newblock \bibinfo{journal}{Communications on Pure and Applied Mathematics}
  \bibinfo{volume}{47} (\bibinfo{year}{1994}) \bibinfo{pages}{787--830}.

\end{thebibliography}

\end{document}